\documentclass[useAMS,usenatbib]{mn2e}
\usepackage{graphicx}
\usepackage{amsmath}
\numberwithin{equation}{section}

\title[Radiatively heated, protoplanetary discs with dead zones. I.]{Radiatively heated, protoplanetary discs with dead zones. I. 
    Dust settling and thermal structure of discs around M stars}
\author[Y. Hasegawa and R. E. Pudritz]{Yasuhiro Hasegawa$^{1}$\thanks{E-mail:
hasegay@physics.mcmaster.ca (YH); pudritz@physics.mcmaster.ca (REP)} and Ralph E. Pudritz$^{1,2}$\footnotemark[1]\\
$^{1}$Department of Physics and Astronomy, McMaster University, Hamilton, ON L8S 4M1, Canada\\
$^{2}$Origins Institute, McMaster University, Hamilton, ON L8S 4M1, Canada}

\begin{document}

\date{}

\pagerange{\pageref{firstpage}--\pageref{lastpage}} \pubyear{2009}

\maketitle

\label{firstpage}

\begin{abstract}
The irradiation of protoplanetary discs by central stars is the main heating mechanism for discs, 
resulting in their flared geometric structure. In a series of papers, we investigate the deep links 
between 2D self-consistent disc structure and planetary migration in irradiated discs, focusing 
particularly on those around M stars. In this first paper, we analyse the thermal structure of discs 
that are irradiated by an M star by solving the radiative transfer equation by means of a Monte Carlo 
code. Our simulations of irradiated hydrostatic discs are realistic and self-consistent in that they include 
dust settling with multiple grain sizes (N=15), the gravitational force of an embedded planet on the disc, 
and the presence of a dead zone (a region with very low levels of turbulence) within it. We show that dust 
settling drives the temperature of the 
mid-plane from an $r^{-3/5}$ distribution (well mixed dust models) toward an $r^{-3/4}$. The dead zone, 
meanwhile, leaves a dusty wall at its outer edge because dust settling in this region is enhanced compared 
to the active turbulent disc at larger disc radii. The disc heating produced by this irradiated wall 
provides a positive gradient region of the temperature in the dead zone in front of the wall. This is 
crucially important for slowing planetary migration because Lindblad torques are inversely proportional 
to the disc temperature. Furthermore, we show that low turbulence of the dead zone is self-consistently 
induced by dust settling, resulting in the Kelvin-Helmholtz instability (KHI). We show that the strength 
of turbulence arising from the KHI in the dead zone is $\alpha=10^{-5}$.     
\end{abstract}

\begin{keywords}
accretion, accretion discs -- radiative transfer-- turbulence -- methods:numerical -- 
planetary systems:protoplanetary discs
\end{keywords}

\section{Introduction}

The discovery of a gas giant planet with a small orbital radius around 51 Peg \citep{mq95} motivated 
the proposal that such planets must form far from their central star, and then migrate \citep{gt79,gt80} 
inward to their observed orbits. The total number of detected exoplanets has increased to more than 
300,\footnote{See the website http://exoplanet.eu/} providing abundant supportive evidence for this idea. 
In addition to Hot Jupiters, massive Earth-like planets (1-10 $M_{\oplus }$) called Super-Earths, have 
recently been discovered around M stars \citep{bbfw06}. The detected number of such low mass exoplanets 
has also been increasing. These very different planetary populations both require migration as an 
explanation of their observed mass - period relation \citep{us07}. 

It is well established that the thermal structure of protoplanetary discs plays a central role in 
determining planetary migration. This is because Lindblad resonances which control the strength of 
the disc-planet interaction are strongly influenced by the thermal gas pressure \citep{a93,ward97}. 
The dominant heat source for discs is stellar irradiation, although viscous heating dominates within 
the central 1 au for classical T Tauri star (CTTS) systems (\citealt{cg97}, hereafter CG97; \citealt{dccl98}). 
The irradiation of the disc under the assumption of vertical hydrostatic equilibrium results in a flared 
disc shape \citep{kh87} that reproduces the observed infrared excess in the spectral energy distributions 
(SEDs). 

Dust in protoplanetary discs is the main absorber of stellar radiation \citep[e.g.,][]{dhkd07}. An 
important ingredient in how discs are heated is the grain size distribution. \citet{mrn77} found that 
the size distribution of dust in the interstellar medium (ISM) is well represented by a power-law, the 
so-called MRN distribution. A related, but somewhat shallower power-law is likely to be applicable to 
dust in discs \citep[e.g.,][hereafter DCH01]{dch01} due to grain growth, as discussed below. The 
composition of dust in discs is different from that in the 
ISM mainly with respect to the contribution of carbon \citep[hereafter P94]{dl84,phbs94}. For discs, 
carbon is included as organics while they are graphite in the ISM. Recently, polycyclic aromatic 
hydrocarbons (PAHs) have received a lot of attention, and found that they play an important role in 
the SEDs \citep[e.g.,][]{dl07} although their formation mechanisms remain to be investigated.     

Another important aspect of dust in discs is its time evolution, that is, grain growth and dust settling. 
The sizes of dust grains grow with time, resulting in ever increasing dust settling (\citealt{dms95}, 
hereafter DMS95; \citealt{sh04}; \citealt{fp06}). More millimeter and less far-infrared emission in the 
observed SEDs are a good indication of grain growth and the resultant dust settling. The data have been 
recently confronted by comparisons with theoretical models \citep{cjc01,dd04b,dch06}. Grain growth and 
dust settling are robust features of discs around various masses of stars from Herbig Ae/Be 
\citep{aad04}, CTTSs \citep{fcdh05} to brown dwarfs \citep[e.g.,][hereafter S07]{apbn05,sjw07}. 
Grain growth is also thought to be the trigger of planetary formation in discs \citep{yg05,jomk07}.  

The structure of protoplanetary discs is far more interesting than just pure power-law behaviour, 
however. A robust feature in discs is the presence of regions of very low amplitude turbulent regions, 
called dead zones \citep{g96}. Turbulence is most likely the outcome of the magnetorotational 
instability (MRI) \citep{bh91a,bh91b}. The MRI requires good coupling between magnetic fields and 
the weakly conducting ionized molecules produced by the X-rays from the central star and cosmic-rays. 
The high density mid-plane region within 0.01 - 10 au of a star has difficulty in being ionized, 
resulting in a dead zone. Extensive studies so far have shown that dead zones are initially extended 
over roughly 10 au from the central star \citep{smun00,is05,smp03,mp06}. They subsequently shrink in time 
due to viscous evolution \citep{mpt09}.

Dead zones play a significant role in dust settling and the thermal structure of discs. This is because 
dust settling is determined by the balance between the gravity and turbulence (e.g., DMS95). Dust 
settling is enhanced in dead zones because turbulence is significantly reduced there. As we show below, this leaves 
a dusty wall at the boundary between active and dead regions for turbulence. We will show that the 
{\it direct} irradiation of this high scale height of dust in the active region in turn creates a 
positive gradient of the disc temperature. We also demonstrate that dust settling in the dead zone 
drives a low level of turbulence by the Kelvin-Helmholtz instability with a turbulent amplitude of 
$\alpha\simeq 10^{-5}$.

The presence of planets also distorts the distribution of gas and dust in discs. 
\citet[hereafter JS03, JS04, respectively]{js03,js04} investigated the effect of the gravitational 
force of a planet on the thermal structure of discs by using models of \citet{cpmd91}. They assumed that 
stellar irradiation and viscous heating were the main heat sources where dust was assumed to be well-mixed 
with the gas. They found that the gravity of the planet produces a self-shadowing (on dayside of planet) 
and a resulting illumination region (on nightside of planet) around the planet. This causes temperature 
variations compared with the unperturbed case. The maximum variation is about 30 per cent for planets 
which have the threshold masses to open up a gap. Recently, \citet[hereafter J08]{jc08} improved the 
calculations of JS03, JS04 by considering the full 3D discs and including self-consistent treatments for 
the density and temperature structures, and found that the self-consistency is very important to 
determine the temperature structure accurately. This is because the temperature is very sensitive to 
the density distribution of dust.  

Thus, the dust distribution is crucial for the thermal structure of discs. A major consequence is its 
role in controlling planetary migration. Analytical and numerical simulations of this process so far 
assume discs to have isothermal or a simple power-law temperature profile \citep{ttw02,npmk00,dkh03}. 
More recently, non-isothermal discs have been treated in numerical simulations although they cannot 
include stellar radiation \citep{dhk03,kk06,pm06}. Analytically, \citet{mg04} have taken into account 
the radiation of stars, and \citet{js05} have also included viscous heating and the gravity of planets 
as well as stellar radiation although their disc models assume dust to be well-mixed with the gas, which 
is still far from realistic disc models. We have recently shown that dust settling has a very important 
effect on planetary migration in a companion paper \citep[submitted]{hp09}. 

In this paper, we present detailed radiative transfer calculations of the effects of dust settling and 
dead zones on the thermal structure of discs. We performed numerical simulations, solving the radiative 
transfer equation by means of a Monte Carlo method \citep{dd04a}. We included dust settling with multiple 
grains sizes, a planet, and a dead zone in our 2D disc models. We especially focus on discs around M stars 
whose properties are similar to discs observed around brown dwarfs (S07). This is physically interesting 
because the recent primary target for observations is low mass planets like Super-Earths and the detection 
probability is the highest around low mass stars and also because viscous heating can be safely ignored. 
Also, the computational time for the Monte Carlo method increases for massive discs. Low mass discs around 
M dwarfs are therefore an ideal target. We show that the presence of dead zones can have a very significant 
impact on the thermal structure of protoplanetary discs.     

Our plan of this paper is the following. In $\S$ \ref{diskmodel}, we describe our numerical methods, 
disc models, and dust properties.  In $\S$ \ref{results}, we present our results and describe how each 
of component (dust settling, a planet, and a dead zone) and the combination of them affect the density 
and thermal structure of discs. In $\S$ \ref{discu}, we verify our findings by performing a parameter 
study for disc properties and discuss other effects on the temperature, such as viscous heating and 
planetary accretion heating that are ignored in our calculations. In $\S$ \ref{khi}, we discuss the 
Kelvin-Helmholtz instability that can be active in the dead zone. Finally, we present 
our conclusions in $\S$ \ref{conc}.  

\section{Numerical methods and disc models} \label{diskmodel}

\subsection{Numerical methods}

We performed numerical simulations by using a 2D radiative transfer code, called 
RADMC\footnote{See the website: http://www.mpia-hd.mpg.de/homes/dullemon/radmc/index.html} \citep{dd04a}. 
RADMC is a versatile and highly reliable code based on the Monte Carlo technique in which the full 
radiative transfer equation is solved \citep{pws04}. The basic principle of the Monte Carlo method is 
very simple, and is well discussed in the literature \citep[e.g.,][references herein]{ww02}. The Monte 
Carlo method traces photons as they scatter and perform random walks through discs. The implementation 
of absorption processes is crucial for determining the thermal structure of discs. In this code, an 
improved version of \citet{bw01} is adopted. Their method assumes discs to be in the local thermodynamic 
equilibrium, so that photons absorbed by the dust grains are forced to re-emit immediately. 
Furthermore, the re-emission of photons is adjusted so that they have the corrected temperature 
distribution. This makes it possible to avoid time-consuming iterative calculations.   

RADMC uses spherical coordinate systems in order to increase the efficiency of following random walks 
of photons \citep{dt00}. However, cylindrical coordinate systems are generally used in analytical 
theories of planetary migration \citep[e.g.,][]{ward97}. In order to estimate the migration time 
accurately, we performed a coordinate transformation from spherical to cylindrical ones (see Appendix A). 
This is non-trivial because the path of photons projected into 2D planes are curved due to this projection 
effect, and because it depends on projected coordinate systems \citep{nhl02}. Our implementation of 
this coordinate transform was tested against a standard benchmark \citep{pws04}, and proves successful 
\citep{h08}.

\subsection{Stellar \& disc models}

We adopt a disc model that can reproduce the observed SEDs for M stars (S07). These models assume that 
the density distribution of discs (gas + dust) can be represented by 
\begin{equation}
\rho=\frac{\Sigma}{\sqrt{2\pi}h} \exp \left( -\frac{z^2}{2h^2}  \right),
\label{rho_wo_pl}
\end{equation}
with $\Sigma \propto r^{-1}$ and $h=h_0(r/R_*)^{\beta}$ (see Table \ref{param_disk} for the meaning of 
symbols). By adjusting $h_0$ and $\beta$ with the fixed total disc mass and size, they reproduce the SEDs 
of very low mass stars such as M and brown dwarfs observed by {\it Spitzer}. Although they also took into 
account the effect of dust settling by considering two kind scale heights $h$ for two kinds of grain size 
distributions: one of which has large $h_0$ for a ISM-like grain size distribution (identical to that of gas), 
the other of which has small $h_0$ for a size distribution extending to larger grain sizes, we use only 
large $h_0$ since we use much more detailed dust settling models, as discussed below. This large scale 
height is employed for gas and all sizes of dust when dust is assumed to be well-mixed with the gas, while 
it is used for gas and only the smallest dust grain when dust settling is included in disc models. For 
the star, we assume that it is described by blackbody radiation with the characteristic temperature $T_*$.

Table \ref{param_disk} summarises the stellar and disc parameters we use (S07). Note that we increase 
the total disc mass by a factor of 10 relative to that of S07 following the literature \citep{ki98,ambw05}. 
This increment allows discs to form gas giants which have been mainly observed around many intermediate 
massive stars like our Sun \citep{ufq07}. We run simulations only of the inner region of the discs, which 
is typically half the standard size, in order to shorten computational time. This requires the disc mass 
to be adjusted, so that the surface density is kept 10 times more massive compared with that of S07. We 
call this configuration our fiducial disc model. In $\S$ \ref{discu}, we perform parameter studies showing 
that our findings are model-independent.    

\begin{table*}
\begin{minipage}{17cm}
\begin{center}
\caption{Summary of parameters \& symbols}
\label{param_disk}
\begin{tabular}{ccc}
\hline
Symbol     & Meaning                        & Value                \\ \hline
$M_{* }$   & Stellar mass                   & 0.1 M$_{\odot }$       \\
$R_{*}$    & Stellar radius                 & 0.4R$_{\odot }$       \\
$T_{*}$    & Stellar effective temperature  & $2850$ K              \\
$R_{out}$  & Outer disc radius              & $50$ au              \\
$R_{in}$   & Inner disc radius              & $6R_*$                \\
$r_o$      & Characteristic disc radius     & $R_*$                 \\
$h_o$      & Characteristic disc height     & 0.02 $r_o $           \\
$\beta$    & The exponent of disc height    & 1.1                   \\
$\Sigma$   & The surface density of (gas + dust)     & $\propto (r/r_o)^{-1}$ \\
$M_d^1$    & The total disc mass (gas + dust) & $4.5\times 10^{-3}$ M$_{\odot}$   \\
           & gas-to-dust ratio              & 100                   \\
$\alpha$   & Parameter for turbulence (active/dead) & $10^{-2}/10^{-5}$ \\
\hline
\end{tabular}
\medskip
\\
M$_{\odot }$ is solar mass, and R$_{\odot }$ is solar radius. \ 
$^1$ $M_d$ is increased by a factor of 10 relative to S07.
\end{center}
\end{minipage}
\end{table*}

\subsection{Heat sources}

We assume the main heat source for discs to be stellar irradiation and neglect any other possible heat 
sources such as viscous heating.

Viscous heating arises from the accretion processes of discs by stars. The temperature due to viscous 
heating under the assumption of gray atmosphere is
\begin{equation}
T_v^4=\frac{3F_v}{4\sigma_B} \left( \tau_d+ \frac{2}{3} \right),
\label{t_v}
\end{equation}
where $\sigma_B$ is the Stefan-Boltzmann constant, $\tau_d$ is the optical depth of the thermal emission 
of dust, the viscous flux $F_v$ is
\begin{equation}
F_v=\frac{3GM_*\dot{M}_a}{4\pi r^3} \left[ 1- \left( \frac{R_*}{r}\right)^{1/2} \right],
\label{f_v}
\end{equation}
and $\dot{M}_a$ is the accretion rate of discs by the star \citep[also see Table \ref{param_disk}]{p81}. 
For discs around the CTTSs, viscous heating dominates the heating by the star within about 1 au (e.g., JS04). 
This is because the temperature due to stellar irradiation is roughly $r^{-1/2}$ since the inverse square 
law while the temperature due to viscous heating is about $r^{-3/4}$ (see equation (\ref{t_v}) and 
(\ref{f_v})). Thus, the viscous temperature decreases rapidly. 

For discs around M stars, however, the turnover point is at a much smaller disc radius than the case of 
CTTSs because the accretion rate $\dot{M}_a$ of M stars is a few orders of magnitude smaller than that of 
CTTSs. We adopt $\dot{M}_a=10^{-10}$ M$_{\odot }$ yr$^{-1}$ for M stars \citep{mjb05}. Under the Eddington 
approximation, $\tau_d=2/3$ or higher corresponds to the optically thick regions. Thus, we consider 
$\tau_{d}=1$ as the optically thick region although this choice does not change $T_v$ very much since 
$T_v \propto \tau_d^{1/4}$ (see equation (\ref{t_v})). Our preliminary simulations showed that viscous 
heating is dominant only within 0.1 au for discs around M stars. Thus, our neglect of viscous heating for 
low mass systems is valid, especially for M star systems.

\subsection{Dust properties}

The thermal structure of discs is controlled by two main properties of dust, its composition and 
size distribution. For the composition of dust, we adopt the model of P94 in which real and imaginary 
refractive indices of dust are derived from available laboratory and astronomical data, theory, and 
the chemical composition of the primitive bodies in the solar system.
\footnote{See the website : http://www.mpia.de/homes/henning/\\Dust{\_}opacities/Opacities/opacities.html} 
Table \ref{dust_model} summarises the composition, abundance, and density of dust. The opacity of dust is 
calculated based on the Mie theory using these refractive indices.

For the composition, we assume that water ice exists over the entire extent of discs. This is because 
DCH01 found that mass absorption coefficients with and without the water ice are very similar 
(see their fig. 1), and because JS04 found that the resultant density structures are also similar 
(see their fig. 3). Thus, our assumption is appropriate. Note that the so-called ice line that defines 
the region in which water ice can exist is important for planet formation. \citet{il08v} found that 
planet formation is accelerated at disc radii beyond the ice line, due to the formation of water ice and 
its retention arising from the pressure maximum \citep{kl07}. This retention is likely to arise at the 
inner edge of the dead zones because the pressure maxima require a positive surface density gradient 
while we focus on the outer edge of the dead zones in this paper.      

\begin{figure}
\begin{center}
\includegraphics[width=8.4cm]{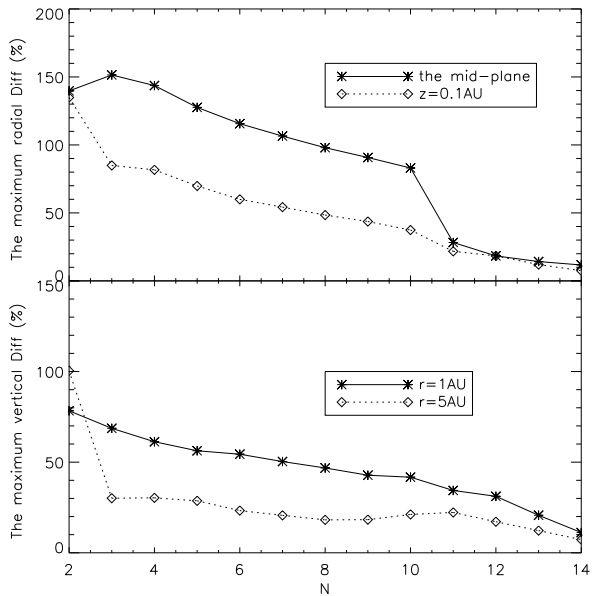}
\caption{The maximum temperature difference as a function of the total number of sizes of dust grains. 
Top: the radial temperature difference. The solid line denotes the difference at the mid-plane. The dotted 
line denotes the difference at $z=0.1$ au from the mid-plane. Bottom: the vertical temperature difference. 
The solid line denotes the difference at $r=1$ au from the central star. The dotted line denotes the 
difference at $r=5$ au. For both cases, the temperature with 15 sizes of dust included is used as the 
reference. Both panels show convergence as more sizes of dust are added.}
\label{fig1}
\end{center}
\end{figure}

For the size distribution of dust, we adopt discretised version of a power-law similar to the MRN 
distribution \citep{mrn77} because dust settling causes different sizes of dust to have different scale 
heights. Furthermore, as shown by \citet{wolf03}, the treatment of a real size distribution is important for calculating the dust 
temperature accurately compared with the approximate treatment in which the mean opacity for the 
ensemble size distribution is adopted. In contrast, we adopt the mean opacity for the 
composition. Thus the size distribution function $n(a)$ is generally written as
\begin{equation}
n(a)=n_0 \sum_{a_i=a_{min}}^{a_{max}} a^p\delta (a_i-a),
\end{equation}
where $n_0$ is a constant independent of grain size $a$, $\delta (a)$ is the Dirac's delta function, 
$a_i$ is the grain size we pick up, $a_{min}$ is the minimum size of dust, and $a_{max}$ is the maximum 
size of dust. We set $a_{min}=0.01$ $\umu$m and $a_{max}=1000$ $\umu$m \citep{cjc01}. The total dust mass
 $m_{dust}$ over the size distribution above becomes
\begin{equation} 
m_{dust}=\int da \frac{4\pi}{3} a^3 \bar{\rho} n(a)\propto a^{3+p},
\end{equation}
where dust is assumed to be spherical and $\bar{\rho}$ is the mean density for the ensemble of 
dust in the composition. In order for $m_{dust}$ to be an increasing function of $a$, 
we set $p=-2.5$. This shallower slope does not affect mass absorption coefficients very much and is 
likely to be preferred in discs in order to take into account grain growth (DCH01). 
The intermediate sizes of dust are logarithmically assigned between 
$a_{min}$ and $a_{max}$. Fig. \ref{fig1} depicts the maximum difference defined as 
\begin{equation}
\mbox{The maximum Diff [\%]}=\mbox{Max} \left[ \frac{T(x_0,N)-T(x_0,N=15)}{T(x_0,N=15)}\times 100 \right],
\end{equation}
where $T$ is the disc temperature, $x_0$ is the fixed position, $N$ is the total number of sampled 
grain sizes. We define the disc temperature $T$ as below
\begin{equation}
 T=\frac{\sum_{a} \frac{4\pi}{3} a^3 \bar{\rho} n(a)\bar{T}(a)}{\sum_a \frac{4\pi}{3} a^3 \bar{\rho} n(a)},
\label{def_diskT}
\end{equation}
where $\bar{T}(a)$ is the dust temperature for size $a$. The temperature at $N=15$ is used as the reference 
value. The top panel shows the radial differences at the mid-plane and $z=0.1$ au ($=x_0$) by the solid 
and dotted lines, respectively, and the bottom panel shows that the vertical differences at $r=1$ and 
5 au ($=x_0$) by the solid and dotted lines, respectively. Both panels show the sign of convergence as 
more sizes of dust are included and that the maximum differences decrease to about or less 10 per cent. 
Our numerical convergence studies show that 15 sizes of dust are sufficient and we adopt this in all 
models shown the paper. We emphasise that our calculations using a real size distribution enables one to 
provide more accurate dust temperatures than using the mean opacity for the ensemble of many dust grains 
done by S07.

\begin{table}
\caption{The composition and related quantities of dust}
\label{dust_model}
\begin{tabular}{ccc}
\hline
Composition    & Abundance & Density [g cm$^{-3}$] \\ \hline
Olivine        & 2.6       & 3.49                 \\
Orthopyroxene  & 0.8       & 3.4                  \\
Iron           & 0.1       & 7.87                 \\
Troilite       & 0.8       & 4.83                 \\
Organics       & 4.1       & 1.5                  \\
Water ice       & 5.6       & 0.92                 \\
\hline
\end{tabular} 
\end{table}

\subsection{Dust settling}

The motion of dust is different from that of gas. This is because dust is in Keplerian motion while gas 
is in slightly sub-Keplerian since it is affected by its thermal pressure. This difference implies 
collisions or friction between them \citep{w77}, resulting in an identical density distribution for them 
when the friction is efficient. Dust in protoplanetary discs is generally in the so-called Epstein 
regime wherein $\tau_f/\Omega_{Kep}^{-1} < 1$, where the friction timescale $\tau_f$ is
\begin{equation}
\tau_{f}=\frac{\rho_s}{\rho_g}\frac{a}{c_s},
\label{friction_time}
\end{equation}
$\rho_s$ is the material bulk density of dust, $\rho_g$ is the gas density, $a$ is the radius of dust 
grains, and $c_s$ is the sound speed. In this regime, the collisional efficiency ($\propto 1/\tau _f$) 
is mainly dependent on $a$. Thus, larger dust grains feel the friction from the gas less than smaller 
ones, resulting in decoupling from the gas. For laminar flows, the dust subsequently sinks into the 
mid-plane. Consequently, different sizes of dust have different scale heights, all of which are 
less than that of gas. This settling motion is prevented in turbulent flows - such as the MRI-active 
outer regions of discs. Dust settling is diffusive in a turbulent environment. Thus, dust settling is 
controlled by the balance between the gravitational force of the star and the amplitude of the turbulence 
in the discs for a given size of dust.

We adopt the analytical model of DMS95 for dust settling in which turbulence in discs is described 
by $\alpha$-prescription \citep{ss73}. Also, its energy spectrum is assumed to be represented by 
the Kolmogorov type, that is, $E(k) \propto k^{-\gamma_{turb}} $, where $k$ is the wavenumber, 
and $\gamma_{turb}$ is dependent on the nature of the turbulence. Since $\gamma_{turb}$ generally has 
the value between 5/3 and 3 (DMS95), we set $\gamma_{turb}=2$. This choice is not significant for 
our results (see equation (\ref{bar_h})). The reduced scale height of dust $h_d$ due to the dust settling 
for its size $a$ is 
\begin{equation}
\frac{h_d(a)}{h}=\bar H / \sqrt{ 1+ \bar H ^2},
\label{h_d}
\end{equation}  
where 
\begin{equation}
\bar H=\left(\frac{1}{1+\gamma_{turb}} \right)^{1/4} \sqrt{\frac{\alpha \Sigma}{\sqrt{2\pi} \rho_s a}},
\label{bar_h}
\end{equation}
and $h$ is the scale height of gas. Equation (\ref{bar_h}) was derived from the advection diffusion 
equation with the diffusion coefficient calculated from the properties of the turbulence, and confirmed 
by their numerical calculations. This approach has been recently verified by magnetohydrodynamical 
simulations \citep{fp06}. 

Thus, the density distribution of the size $a$ of dust with settling is described by equation 
(\ref{rho_wo_pl}) with the replacement of $h$ with $h_d(a)$.

\subsection{The gravity of planets}

The gravitational force of a planet distorts the distribution of gas in the disc (JS03; JS04). This 
distortion is taken into account by extending the original vertical hydrostatic balance equation to one 
that includes the gravitational field of the planet as well as the central star. Consequently, the 
density distribution of gas becomes
\begin{equation}
\rho=\rho_0 \exp \left( -\frac{z^2}{2h^2} + \frac{\mu }{h^2} \left[\frac{r}{\sqrt{(r-r_p)^2+z^2}}-\frac{r}{|r-r_p|} \right]  \right),
\label{rho_w_pl}
\end{equation}
where $\mu=M_p/M_*$, $M_p$ is the planetary mass, $r_p$ is the location of the planet from the host star, 
and the normalization constant $\rho_0$ is chosen so that the density at $z=0$ corresponds to the 
unperturbed density (following JS03). Note that the hydrostatic assumption breaks down within the Hill 
radius $r_H\approx r_p(M_p/M_*)^{1/3}$, so that we interpolate the density in this region \citep{js05}.

Thus, for a disc model without dust settling, but with a planet, the density distribution of dust for 
any size $a$ is identical to that of gas (equation (\ref{rho_w_pl})) while the distribution of the size 
$a$ of dust for the case of dust settling is represented by the same equation above with the replacement 
of $h$ with $h_d(a)$ that is calculated by equation (\ref{h_d}). Note that the density distribution of 
gas with dust settling as well as a planet is still represented by equation (\ref{rho_w_pl}).

\subsection{Dead zones with dust settling}

Our disc models include dead zones. In this paper, we assume a characteristic radius for the dead zone 
to be 4 au in our fiducial disc model. Many studies have shown that it is roughly 10 au in the early 
phases of disc evolution. \citet{mpt09} have recently shown that dead zones evolve with time. The outer 
edge of the dead zone moves inward following the accretion of disc material onto the star. In order to 
take time-evolution into account, we perform parameter studies in $\S$ \ref{discu}, confirming the 
same results. We adopt the $\alpha$-prescription for turbulence, so that whether discs are active or 
"dead", they have a characteristic value of  $\alpha$. For dead zones ($r\leq 4$ au), 
$\alpha_{DZ}=10^{-5}$ while $\alpha_{active}=10^{-2}$ for active regions ($r> 4$ au). We show later 
in the paper that $\alpha=10^{-5}$ in dead zones is self-consistent.  

Dead zones have higher degrees of dust settling than active regions. Since $\alpha$ in equation 
(\ref{bar_h}) takes two different values, depending on either active or dead zones, the density 
distribution of dust for this case is described by equation (\ref{rho_wo_pl}) with the replacement 
of $h$ with $h_d(a)$ and two values of $\alpha$. If a planet is present, it is represented by 
equation (\ref{rho_w_pl}) with the replacement of $h$ with $h_d(a)$ and two values of $\alpha$. 
The distribution of gas is denoted by equation (\ref{rho_wo_pl}) and (\ref{rho_w_pl}) with no and a 
planet, respectively.

\subsection{Self-consistency \& the disc temperature}

Self-consistency is important for the disc temperature as J08 discussed. In order to achieve it, 
iterative calculations are essential.  Our Monte Carlo calculations are time-consuming. About a week 
is needed to run a single disc model. This is because the mid-plane region in a disc, which is 
important for torque calculations, is optically thick so that a large number of photons ($\sim 10^8$) 
is required to avoid the random noise. Therefore, we perform an iteration for each disc model above. 
The temperature of each size of dust is first calculated for an initial distribution of dust. The 
distribution of gas is not important for the temperature calculations since its heat capacity is much 
less than that of dust. Assuming the temperature of gas to be equal to the mass-averaged dust 
temperature (see equation (\ref{def_diskT})), the new scale height of gas is calculated, allowing us to 
calculate the new scale height of dust. Consequently, the new density distributions of gas and dust are 
obtained.

\section{Results} \label{results}

\subsection{Dust settling}

\begin{figure*}
\begin{minipage}{17cm}
\begin{center}
\includegraphics[width=8.4cm]{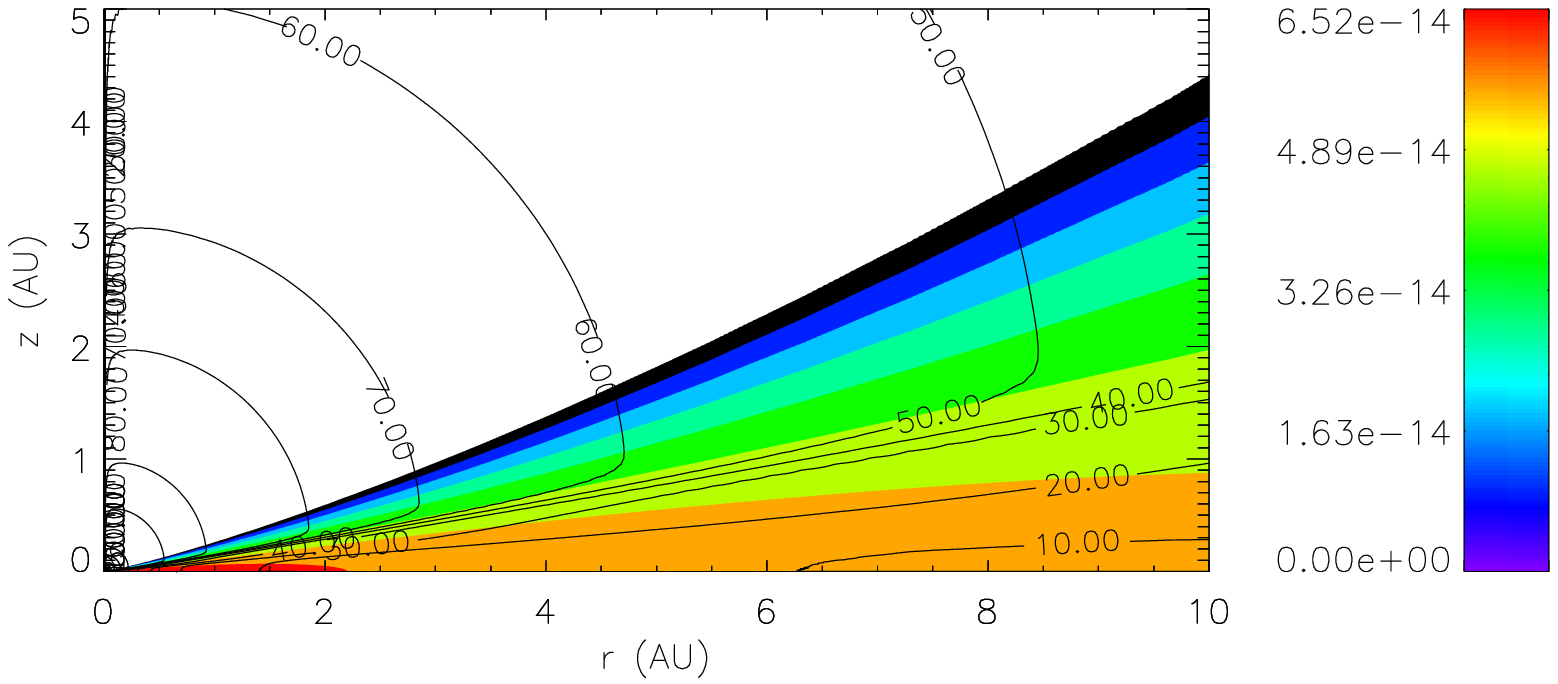}
\includegraphics[width=8.4cm]{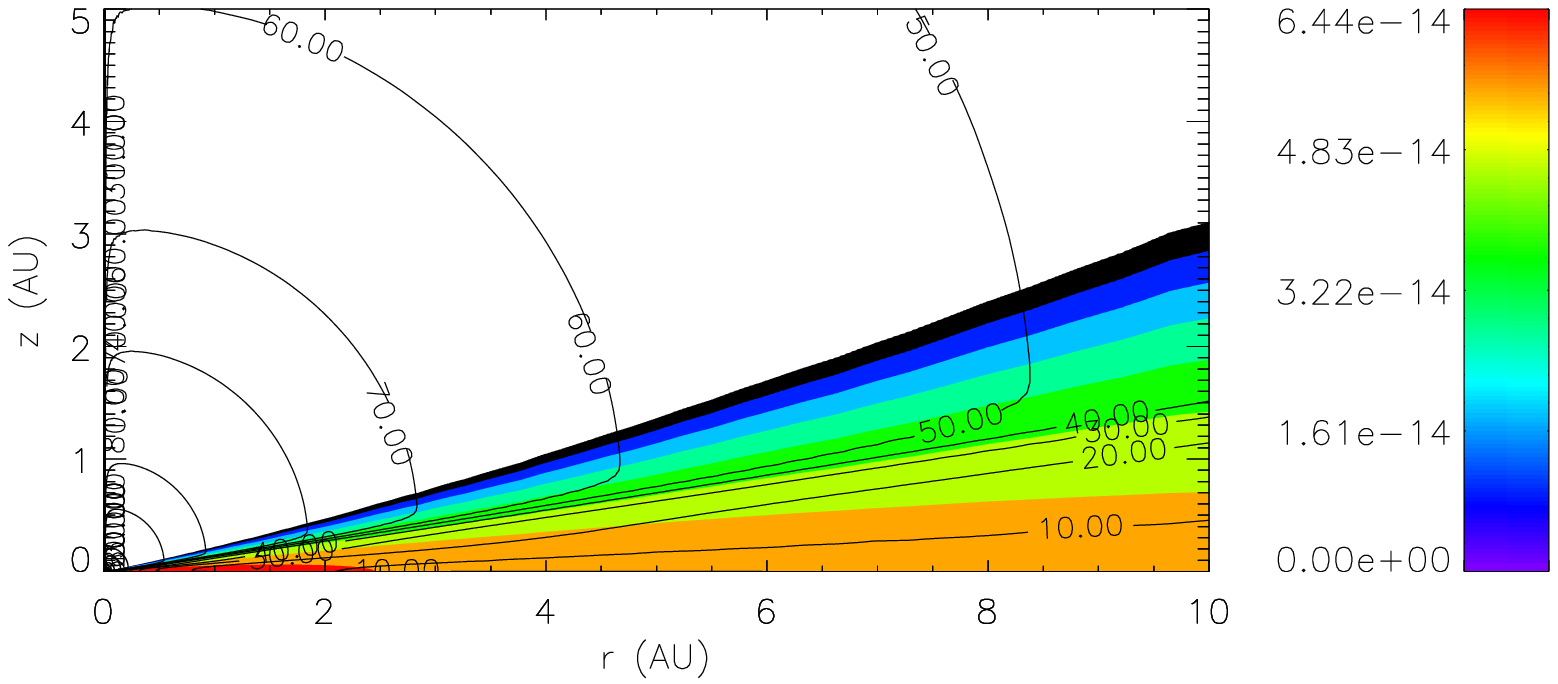}
\includegraphics[width=8.4cm]{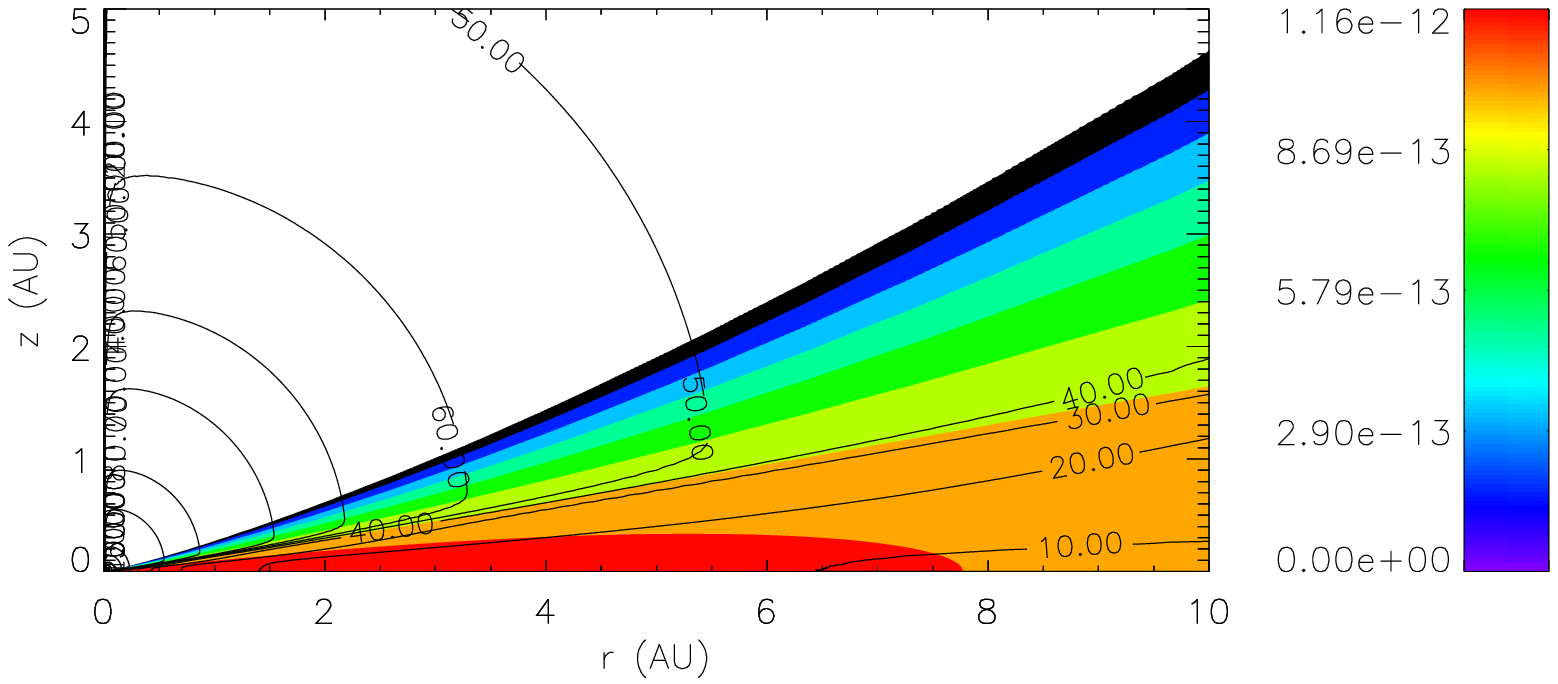}
\includegraphics[width=8.4cm]{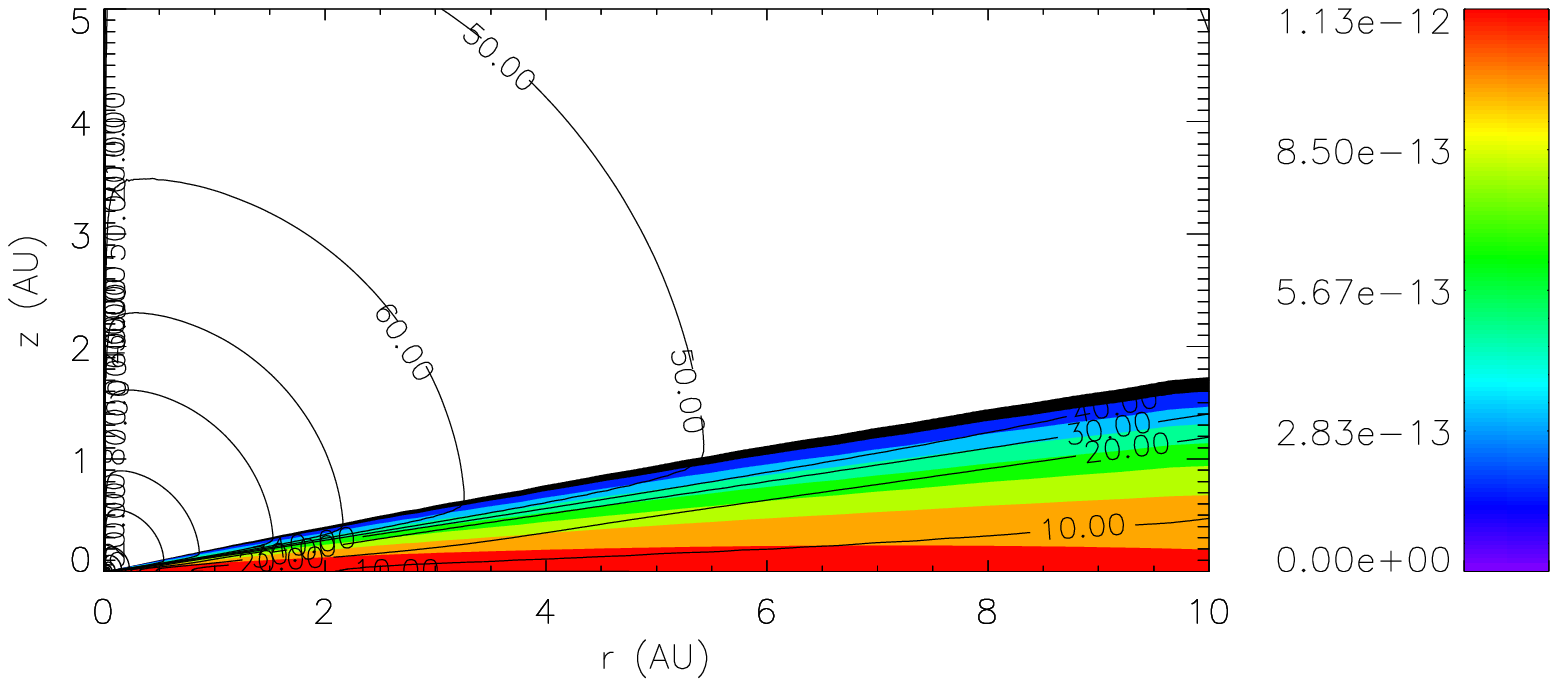}
\includegraphics[width=8.4cm]{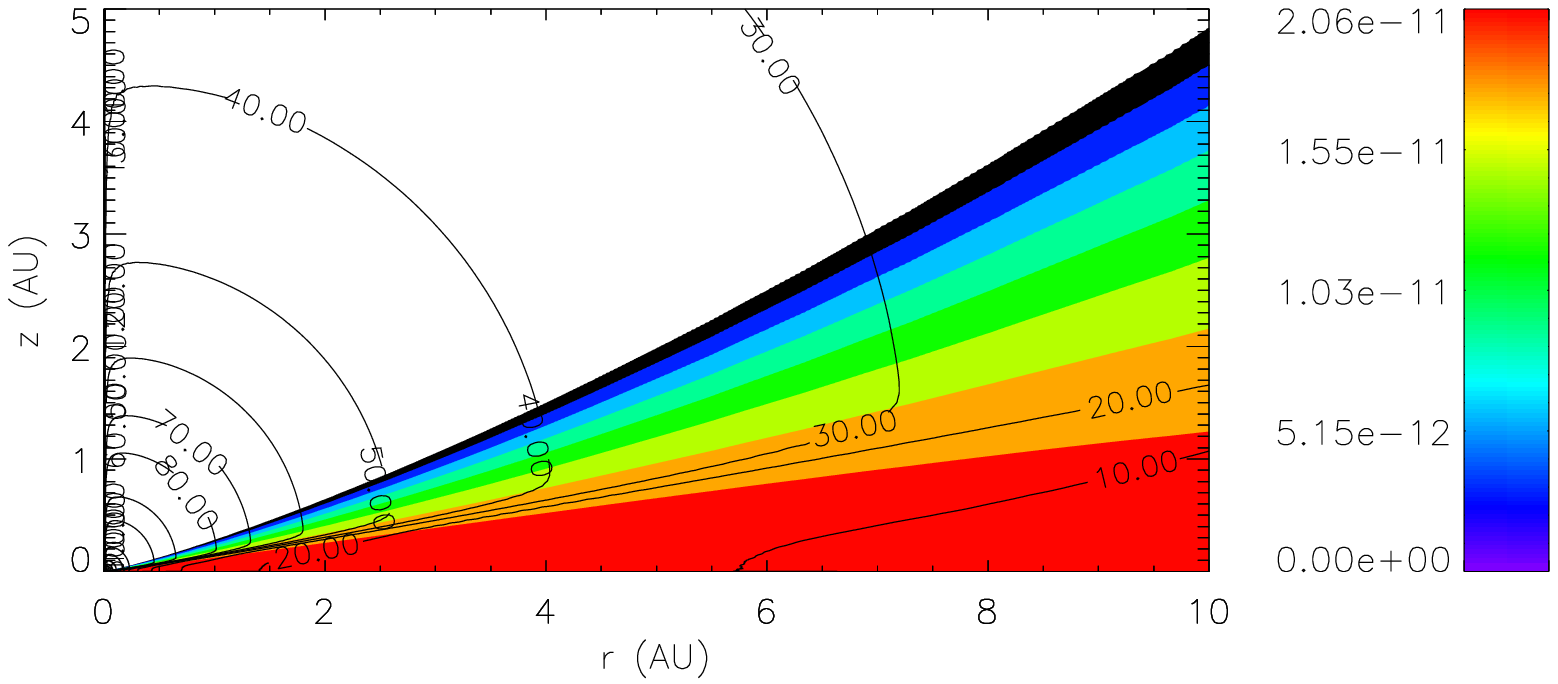}
\includegraphics[width=8.4cm]{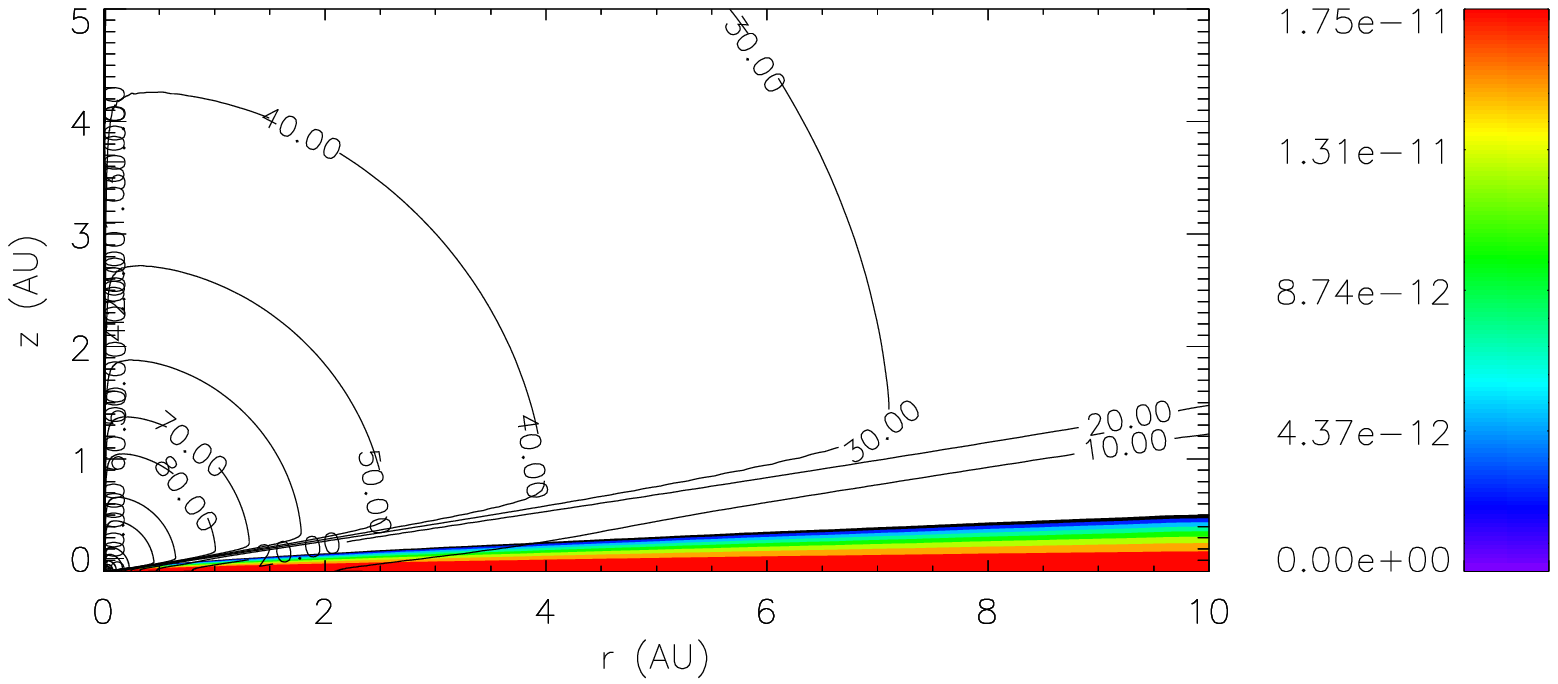}
\caption{The density and temperature structures of dust with its size $a$ without and with dust settling 
on the left and right columns, respectively. The dust density is denoted by the colors. The scale is 
shown in the color bar in units of [g cm$^{-3}$]. Top: $a=0.01 \umu$m. Middle: $a=3.16 \umu$m. 
Bottom: $a=1$mm. The dust temperature is a decreasing function of $a$. Dust settling provides a flatter 
shape for the distribution of larger grains and a lower temperature in the mid-plane region.}
\label{fig2}
\end{center}
\end{minipage}
\end{figure*}

In this subsection, we focus on the effect of dust settling on the density and thermal structures of 
discs. Fig. \ref{fig2} shows the results of our simulations of the thermal and density structures of 
each dust grain without and with of dust settling (the left and right columns, respectively). Grain 
size increases from 0.01 $\umu$m, 3.16 $\umu$m to 1 mm on the top to bottom panels. They show that the 
temperature of a dust grain decreases as its size grows and that dust settling provides a lower temperature 
in the mid-plane region. Also, larger grains settle into much flatter layers because of their lower 
friction with the turbulent gas.

\begin{figure}
\begin{center}
\includegraphics[width=8.4cm]{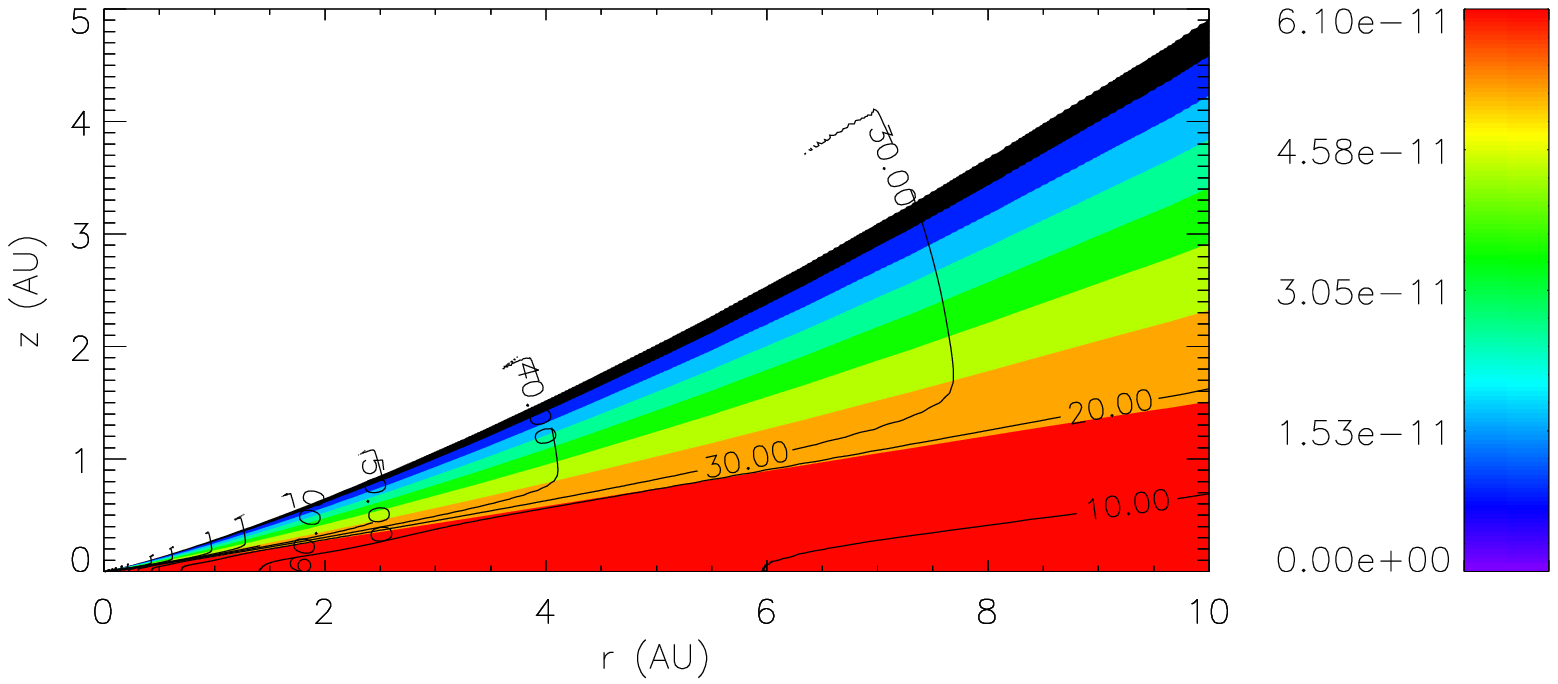}
\includegraphics[width=8.4cm]{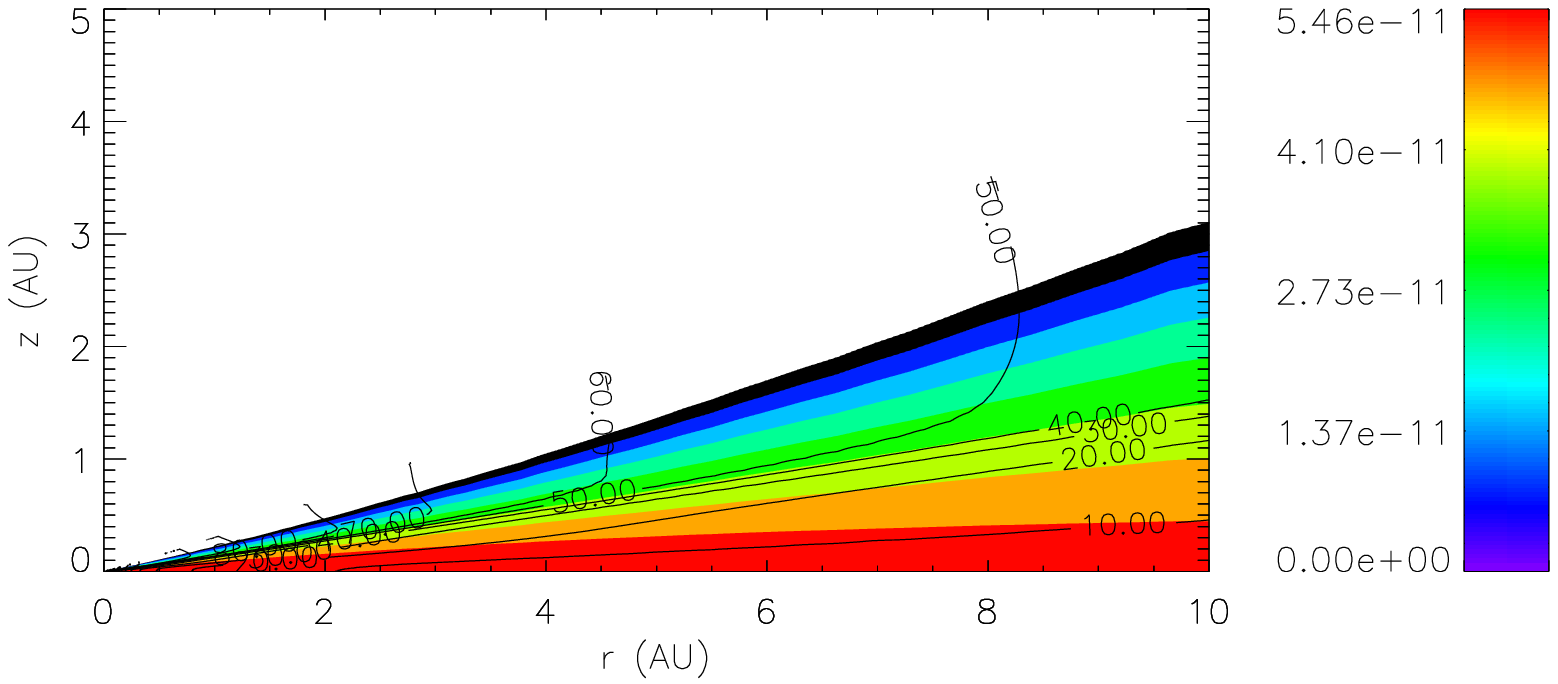}
\caption{The density structure of dust and the temperature structure of disc. The dust density is 
denoted by the colors. The scale is shown in the color bar in units of [g cm$^{-3}$]. The 
temperature is denoted by the contours in the unit of [Kelvin]. Top: dust is well-mixed with gas. 
Bottom: dust settling included. Dust settling makes the surface layer higher temperature and lower 
density and makes the mid-plane region lower temperature and higher density. It also makes the dust 
distribution geometrically flatter (also see Fig. \ref{fig4}).}
\label{fig3}
\end{center}
\end{figure}

Fig. \ref{fig3} shows the results of our simulations of the disc's temperature and density structure of 
the dust component for disc models with and without of dust settling (the bottom and top panels, 
respectively). The figures present the mass-averaged disc temperature given by equation (\ref{def_diskT}). 
The common feature of both cases is that the surface layer has a higher temperature than 
the disc mid-plane. This is because the surface layer is directly heated by the central star while the 
mid-plane is heated mainly by the thermal emission of dust as first noted by \citet[also see CG97]{kh87}. 
This is confirmed by the structure of the temperature contours. For the optically thin surface layer, 
the contours have spherical shapes, meaning that the stellar radiation dominates the thermal emission of 
dust. For the mid-plane region, the contours show straight lines that are parallel to the density 
structure of the dust denoted by colors. In this regime, dust emission dominates stellar radiation. 
The physical picture that the mid-plane region is enclosed by the superheated layers is consistent 
with the analytical models proposed by CG97 although their disc models have only two characteristic 
temperatures, one of which is for the superheated layer, the other of which characterizes the mid-plane. 
The transition region between these two regimes is characterised by a steep vertical temperature gradient.    

\begin{figure}
\begin{center}
\includegraphics[width=8.4cm]{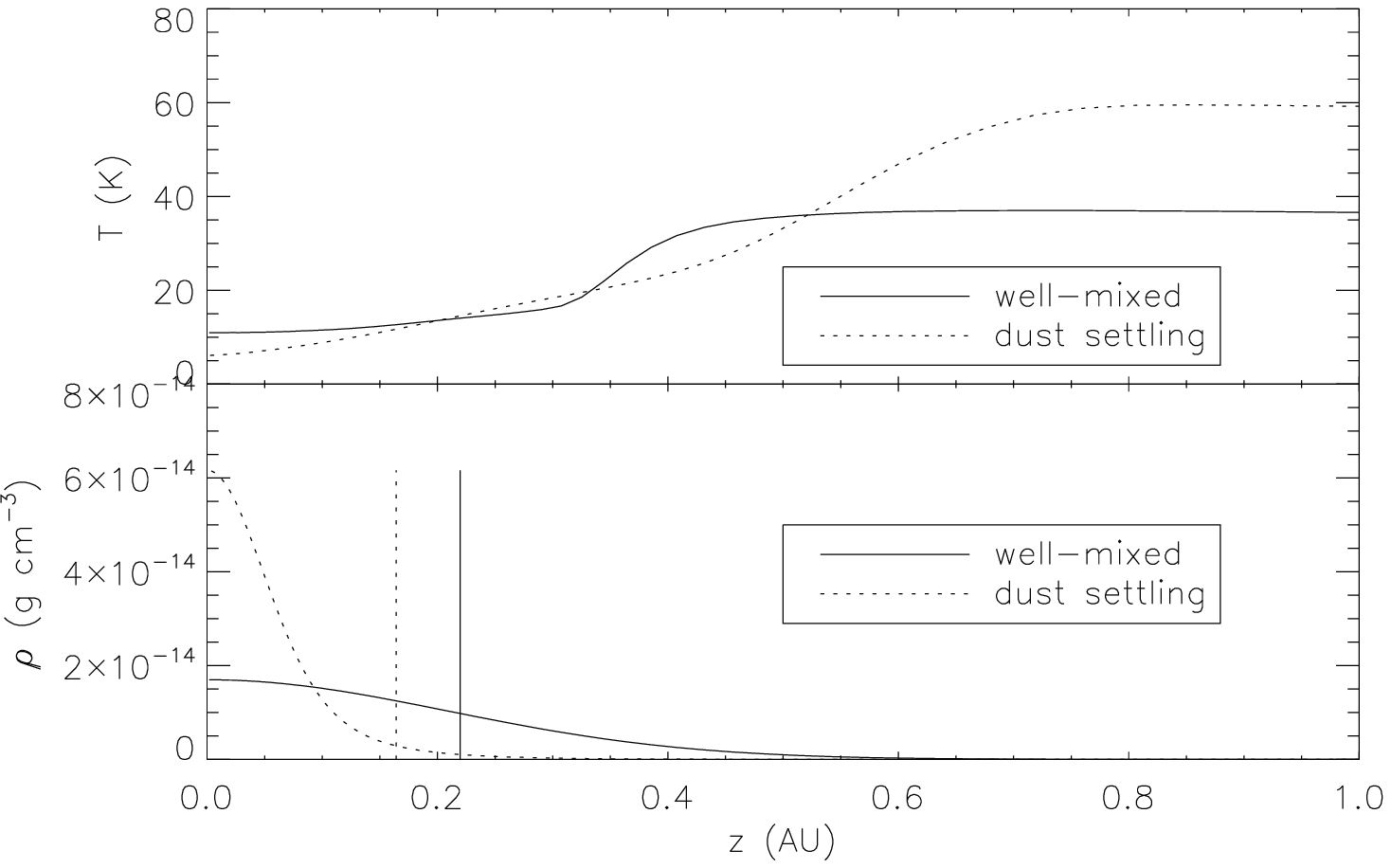}
\caption{The temperature and dust density structure at $r=5$ au as a function of distance from the 
mid-plane. Top: the temperature behaviour. Bottom: the dust density profiles. The vertical lines denote 
the scale height of gas. In both panels, any solid lines denote the case of well-mixed dust, and any 
dotted lines denote the case of dust settling. The reduced scale height of gas for the dust settling 
case also indicates flatter disc shapes.}
\label{fig4}
%\end{center}
%\end{figure}

%\begin{figure}
%\begin{center}
\includegraphics[width=8.4cm]{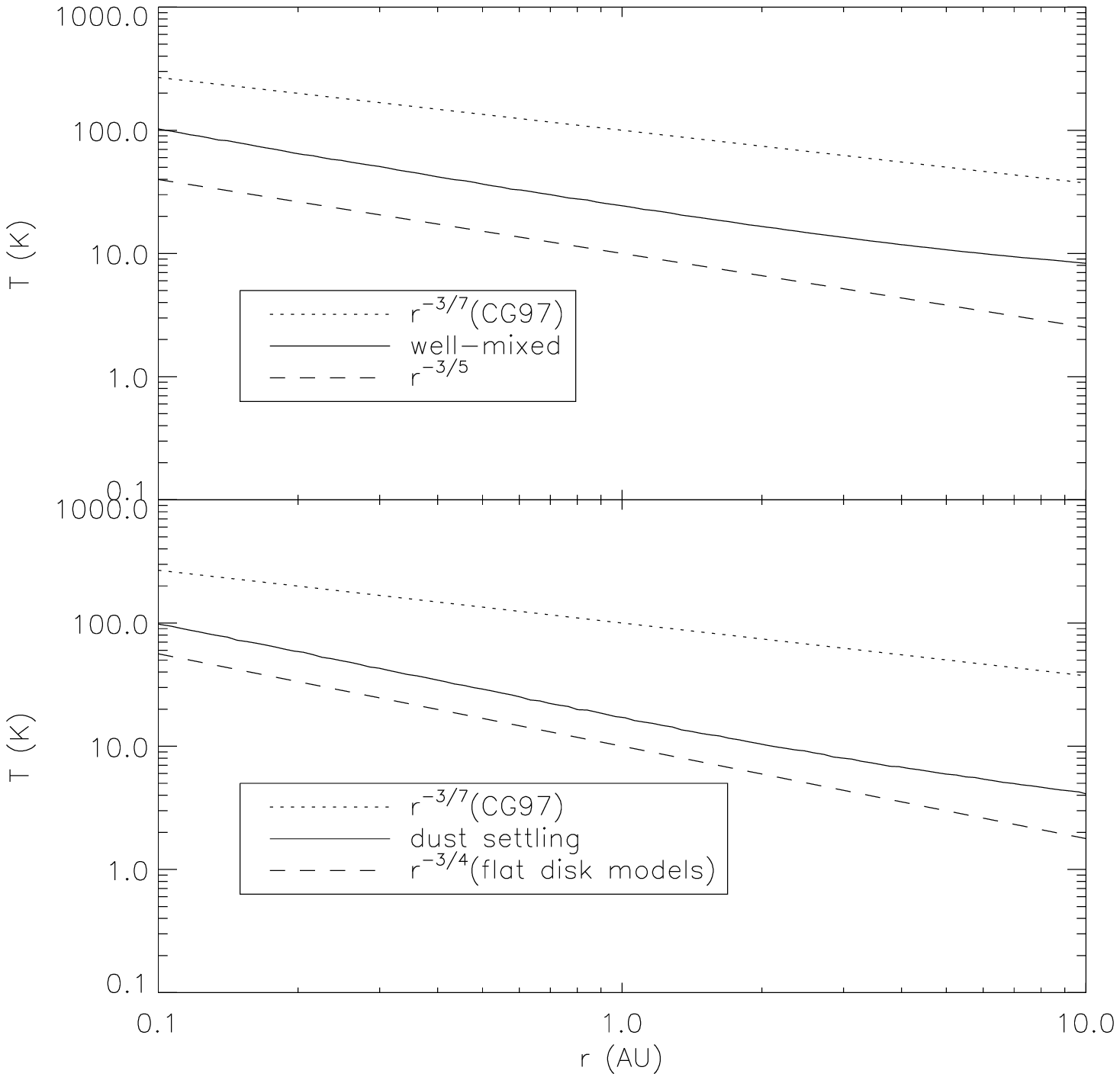}
\caption{The temperature structure as a function of disc radius. Top: the well-mixed case. The solid 
line denotes the temperature of our calculations. The dotted line denotes the temperature of flared 
disc models (CG97). The dashed line denotes the best fit for our calculations. Bottom: the dust 
settling case. The solid line denotes the temperature of our calculations. The dotted line denotes 
the temperature of flared disc models (CG97). The dashed line denotes the best fit for our calculations, 
which is also derived as flat disc models by CG97.}
\label{fig5}
\end{center}
\end{figure}

Dust settling changes the temperature structure as well as the density structure of dust as is expected 
by equation (\ref{h_d}) (see Fig. \ref{fig3} and \ref{fig4}). The main effect of dust settling is that 
the surface region has a higher temperature while the mid-plane region has a lower temperature. This 
behaviour is explained by the fact that different sizes of dust have different scale heights. In the 
surface layer, larger dust grains are depleted, but smaller grains still exist. These are more important 
for absorbing stellar radiation. Thus, photons emitted by the star can be efficiently absorbed by the 
surface layer. On the other hand, the mid-plane region has many sizes of dust due to settling, resulting 
in a higher density compared with the case of well-mixed case. Thus, the optical depth is increased and 
photons emitted from the star have more difficulty in passing through the region. Eventually, the 
energy absorbed in the region becomes smaller, and the temperature is decreased. This difference of the 
temperature, in turn, affects the density distribution of the gas by reducing its scale height (the 
vertical lines in Fig. \ref{fig4}). Thus, dust settling drives the structure of discs from flared to 
flatter shapes (see Fig. \ref{fig3}). This is consistent with the results of time-dependent dust settling 
models \citep{dd04b} while we assumed dust settling to be in the steady state.

Fig. \ref{fig5} shows the temperature at the mid-plane as a function of disc radius. The top panel 
shows the well-mixed model in which the temperature profile is well represented by $r^{-3/5}$. Thus, 
our resultant discs are less flared relative to CG97 because of the increase of disc masses. The 
bottom panel shows the case of dust settling in which the temperature profile is well represented by 
$r^{-3/4}$. Interestingly, this power-law can be derived from flat disc models (CG97). This is another 
indicator of flatter disc shapes.   

\subsection{The effects of planets}

\begin{figure*}
\begin{minipage}{17cm}
\begin{center}
\includegraphics[height=4.2cm]{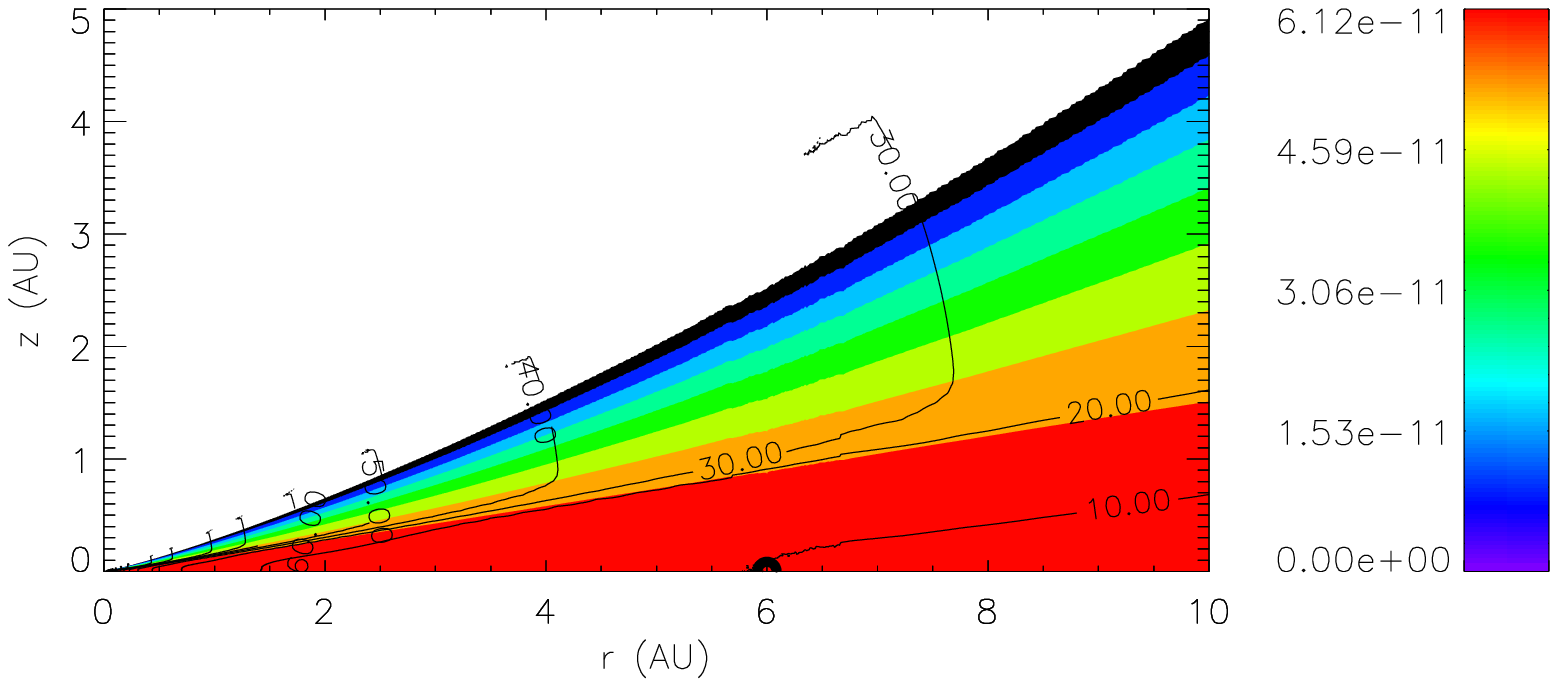}
\includegraphics[height=4.2cm]{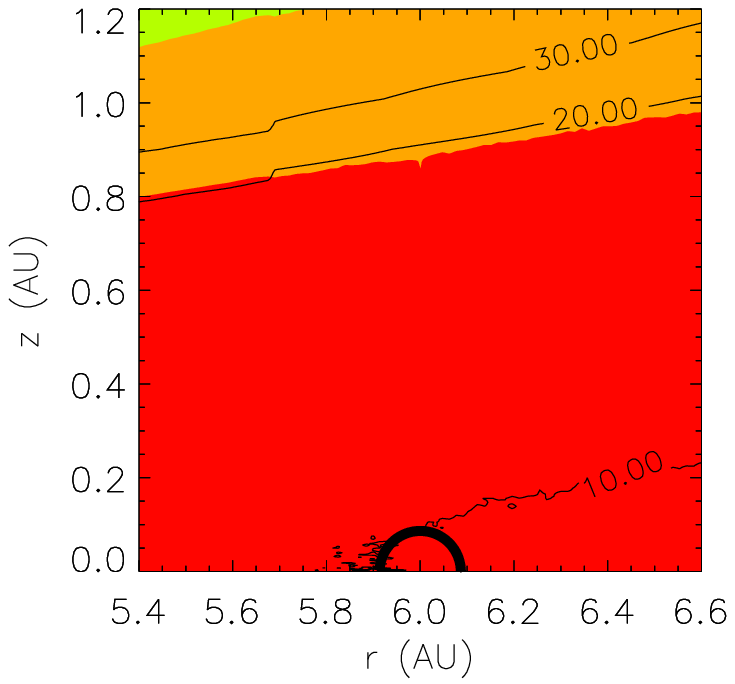}
\includegraphics[height=4.2cm]{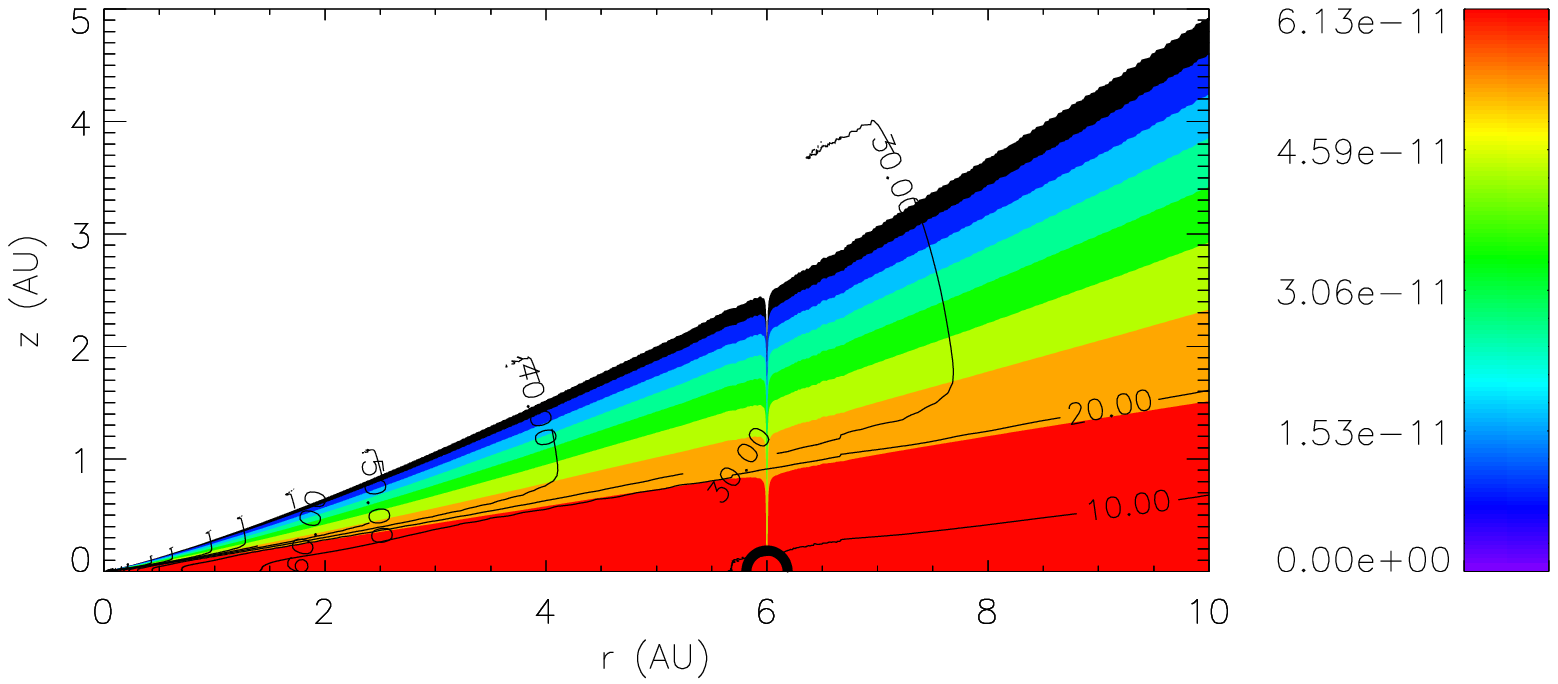}
\includegraphics[height=4.2cm]{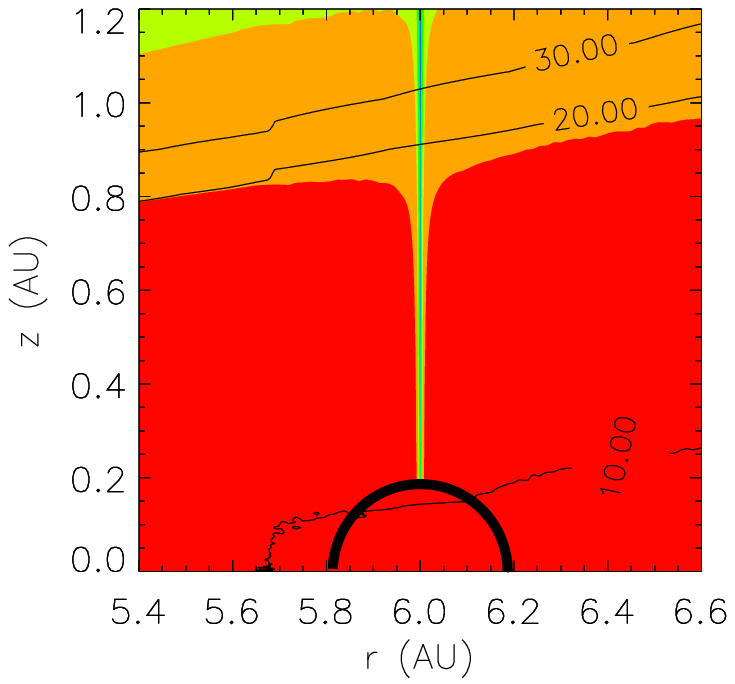}
\includegraphics[height=4.2cm]{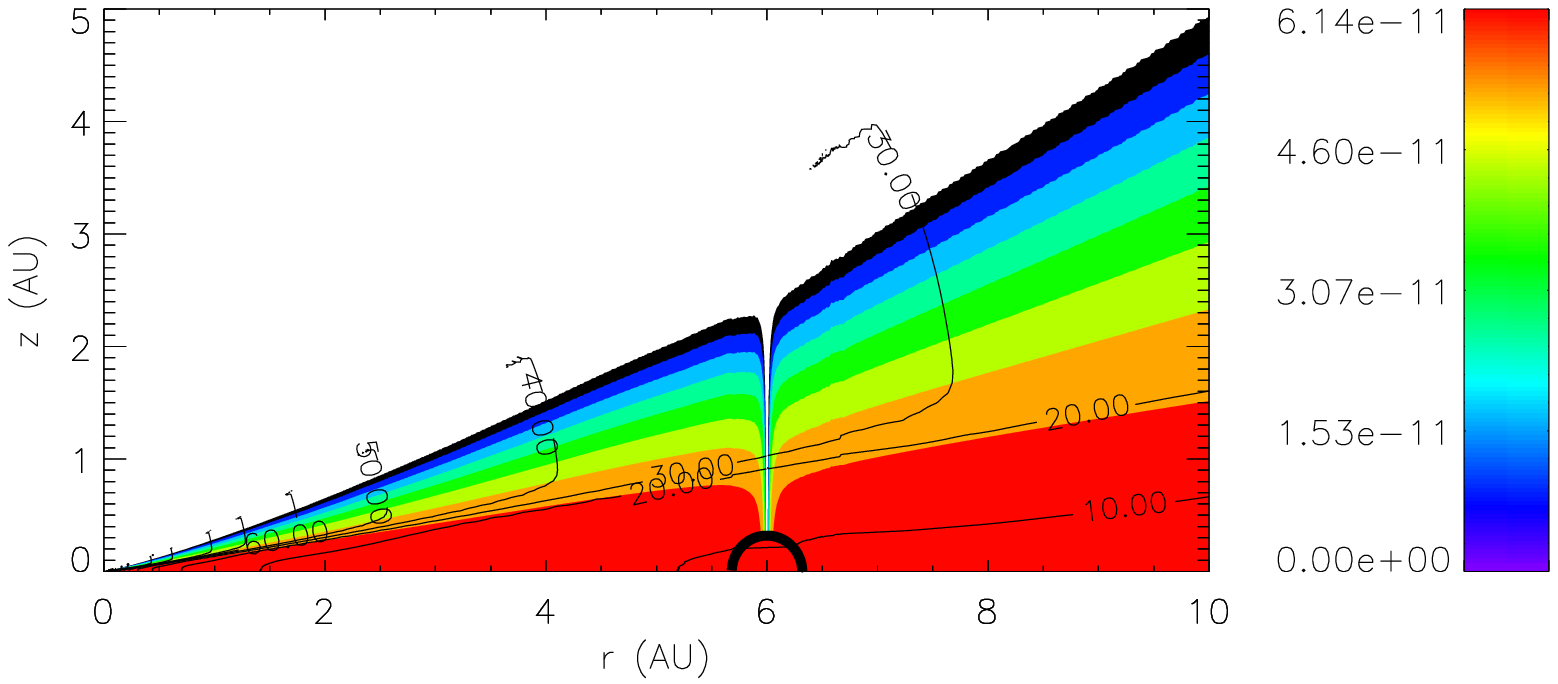}
\includegraphics[height=4.2cm]{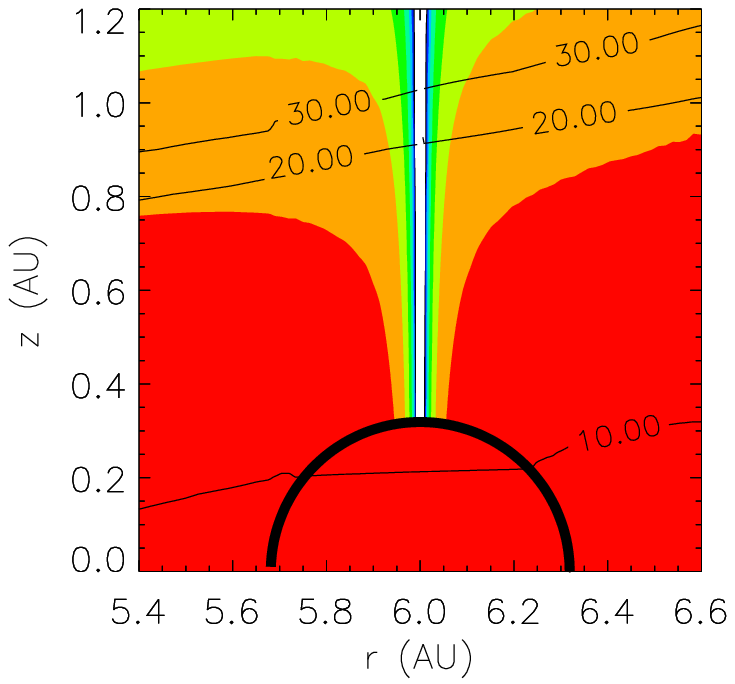}
\includegraphics[height=4.2cm]{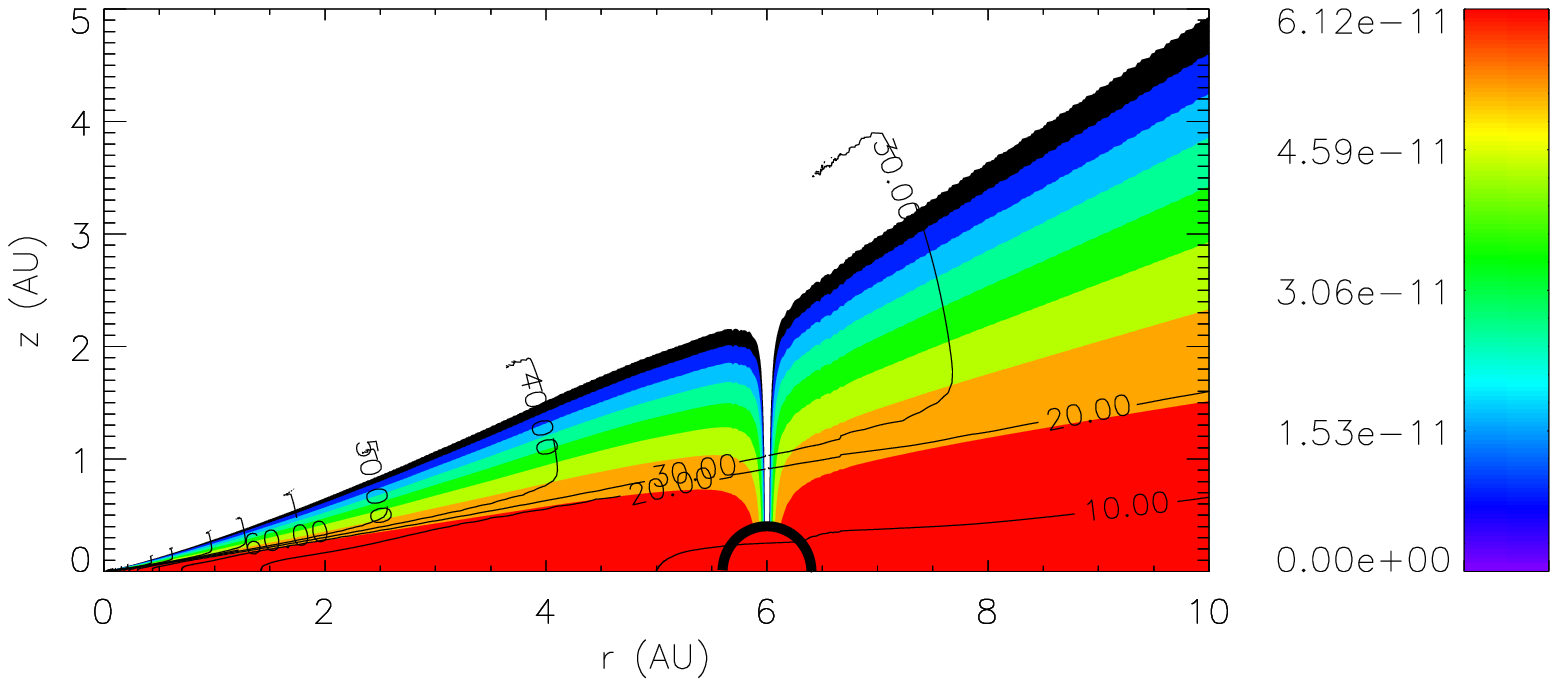}
\includegraphics[height=4.2cm]{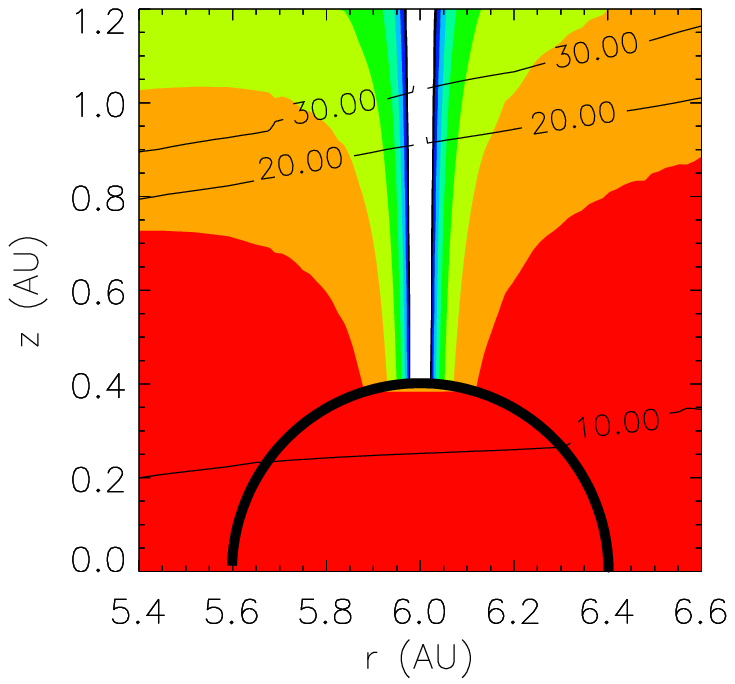}
\caption{The density structure of dust and the temperature structure of disc with well-mixed dust 
(as Fig. \ref{fig3}). The blow-up versions near the planet are shown in the right column. The thick 
solid lines denote the Hill radius of the planets. Top: 0.1 $M_{\oplus }$. Second: 1 $M_{\oplus }$. 
Third: 5 $M_{\oplus }$ Bottom: 10 $M_{\oplus }$. The presence of the planets produces a lower temperature 
of the mid-plane region in front of them due to the compression of material above the planets 
(see the contour of 10 K).}
\label{fig6}
\end{center}
\end{minipage}
\end{figure*}

In this subsection, we focus on the effect of the gravitational force of planets with various masses. 
The fiducial location of a planet is fixed as 6 au. Fig. \ref{fig6} shows the results of our simulations 
of the thermal structure of discs and the density structure of dust with planets. The masses of the 
planets are 0.1, 1, 5, and 10 $M_{\oplus }$ on the top to bottom panels, respectively. The main 
difference in the density structure of dust is the decrement of the density above the planet which 
is expected by equation (\ref{rho_w_pl}). This effect increases as the masses of planet increase. 
Another difference is the decrement of the temperature of the mid-plane region on the dayside of the 
planet as shown by the contour of 10 K. This effect is also enhanced with planetary masses since it 
arises from the gravitational force of the planet to compress the material above them, resulting in the 
increase of the density of dust in the mid-plane region (also see Fig. \ref{figB1}). This makes it 
more difficult for photons to penetrate and heat the region. The lower temperature was not found by 
JS03, JS04, and J08 because they included viscous heating although it is dominant only within a few au. 

\begin{figure}
%\begin{minipage}{17cm}
\begin{center}
\includegraphics[width=8.4cm]{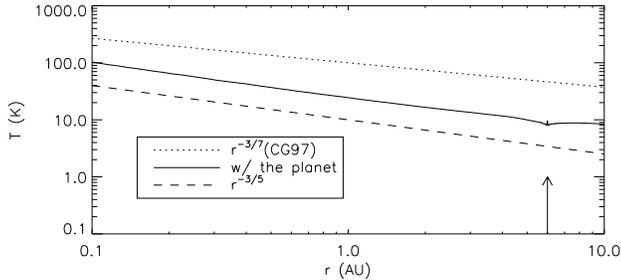}
\caption{The temperature structure with a 10 $M_{\oplus }$ planet as a function of disc radius 
(as Fig. \ref{fig5}). The location of the planet is indicated by the arrow. The planet produces a 
slightly lower temperature region in the vicinity of the planet and a sharp peak at its location due 
to the compression of material.}
\label{fig7}
\end{center}
\end{figure}

Fig. \ref{fig7} shows the temperature at the mid-plane for the case of 10 $M_{\oplus }$. As discussed 
above, the temperature on the dayside of the planet becomes lower than that of the unperturbed case 
(also see Fig. \ref{fig5}). Furthermore, the temperature on the nightside of the planet also decreases 
for the same reason. Interestingly, the temperature at the location of the planet has a sharp peak. 
This arises from the compression of material above the planet, so that mean free paths of photons 
become longer. Consequently, more photons with higher energy can reach to the mid-plane regions. 
Further detailed analyses and discussion are tossed into Appendix B.

\begin{figure*}
\begin{minipage}{17cm}
\begin{center}
\includegraphics[height=4.2cm]{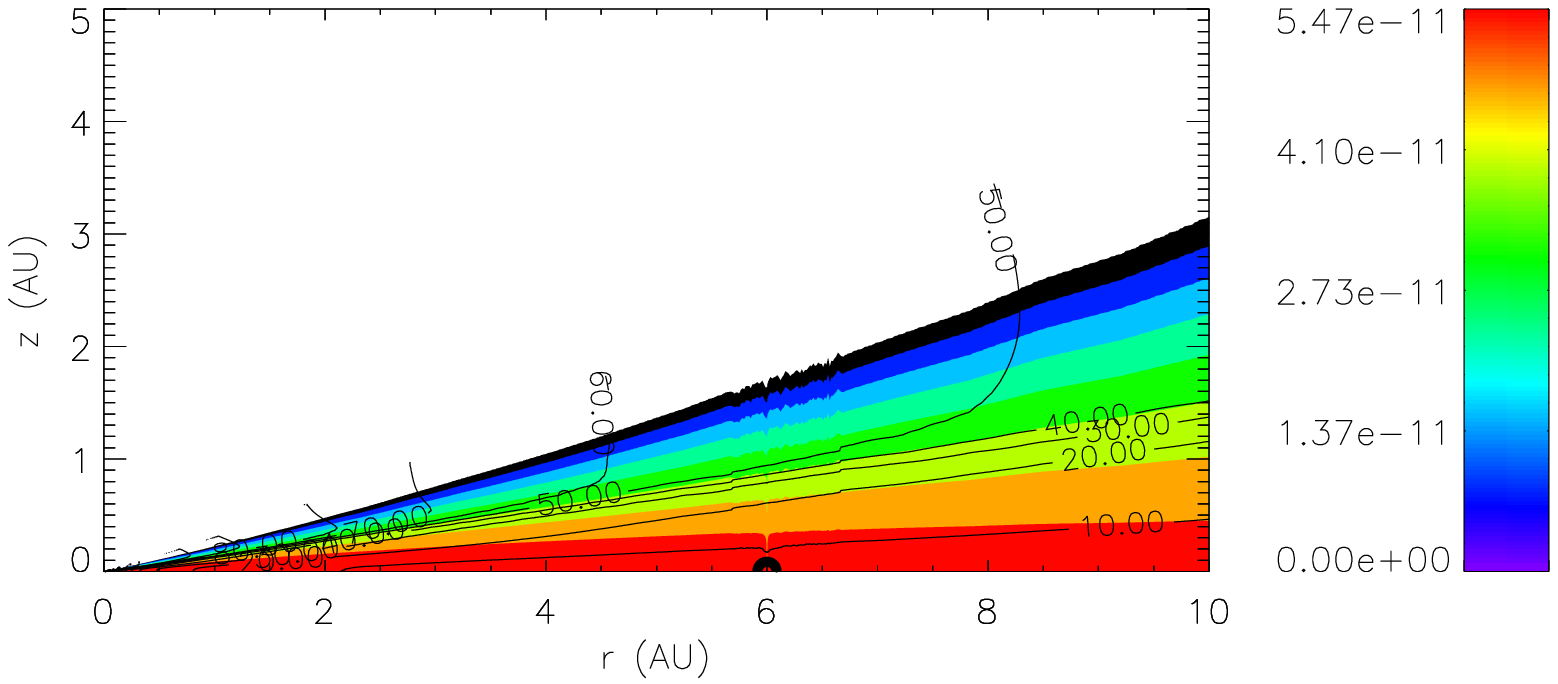}
\includegraphics[height=4.2cm]{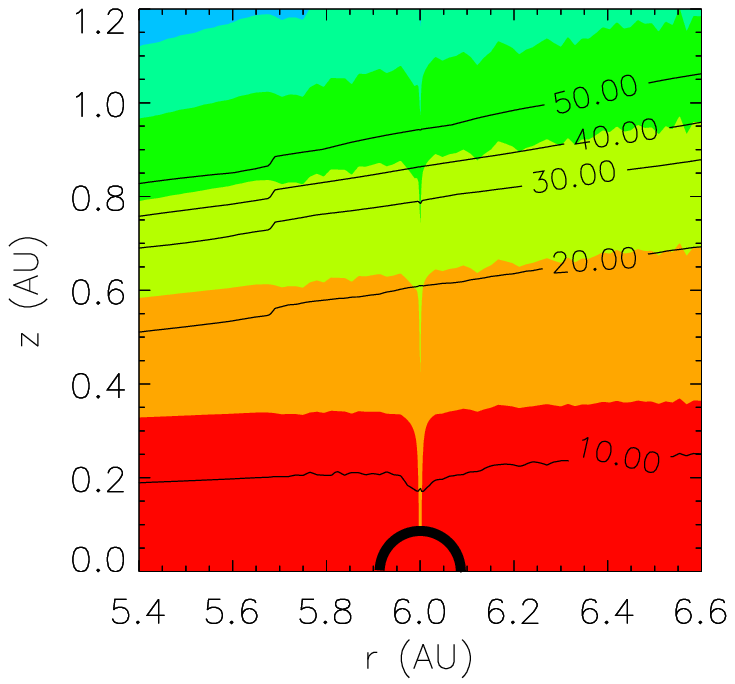}
\includegraphics[height=4.2cm]{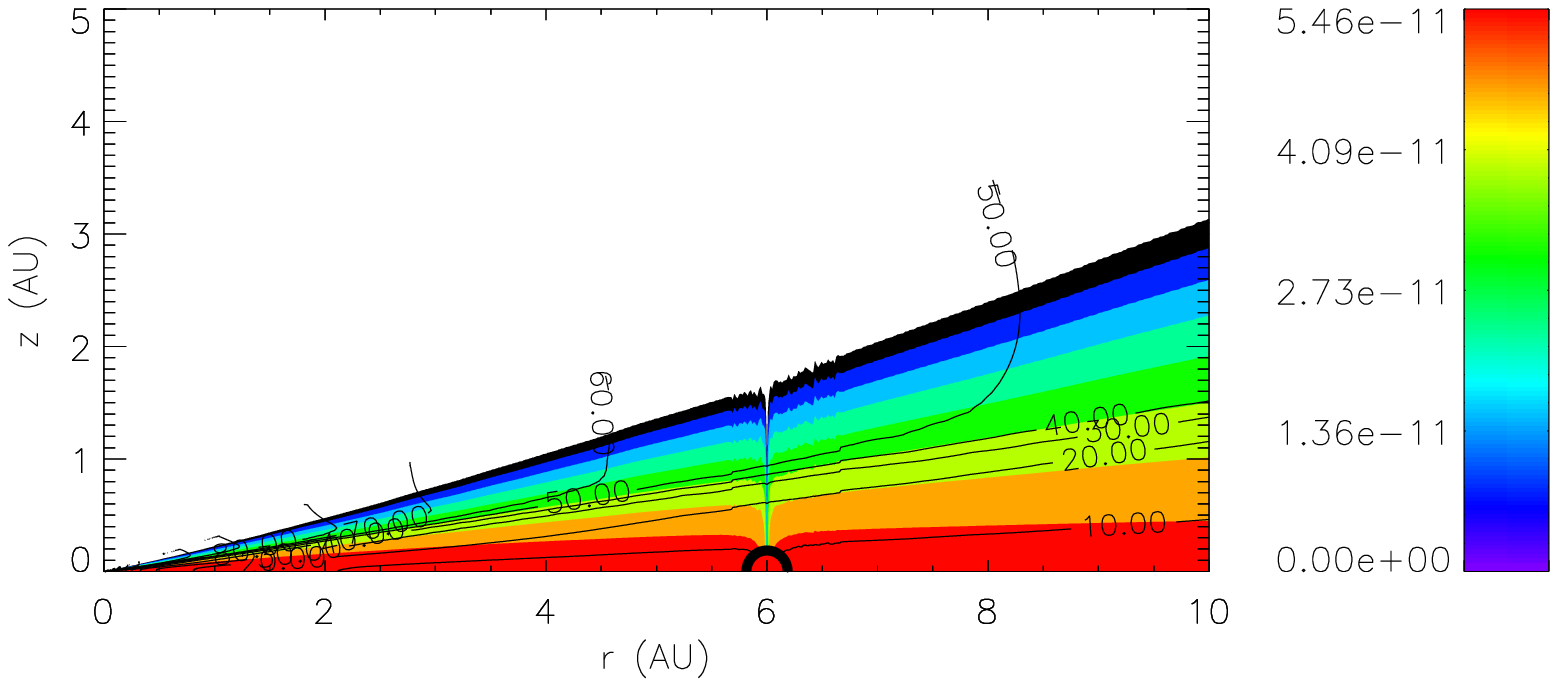}
\includegraphics[height=4.2cm]{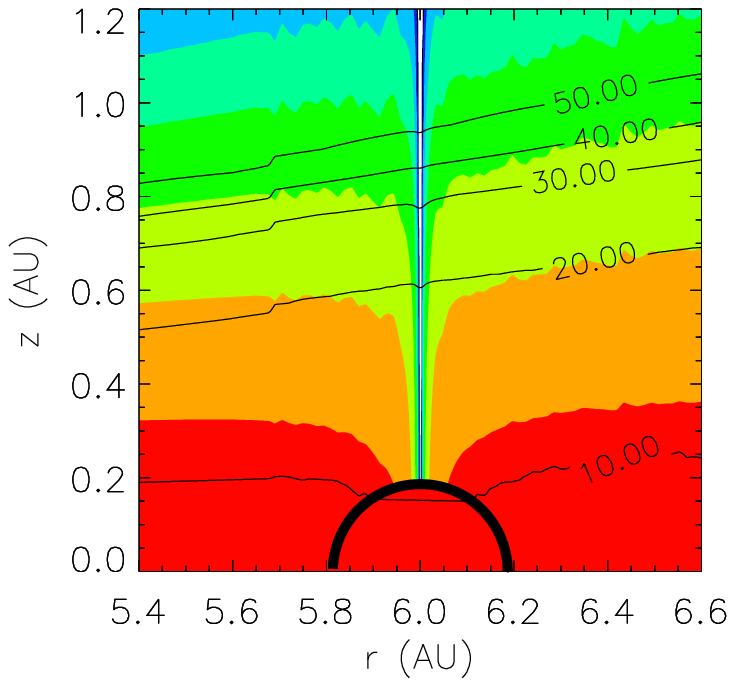}
\includegraphics[height=4.2cm]{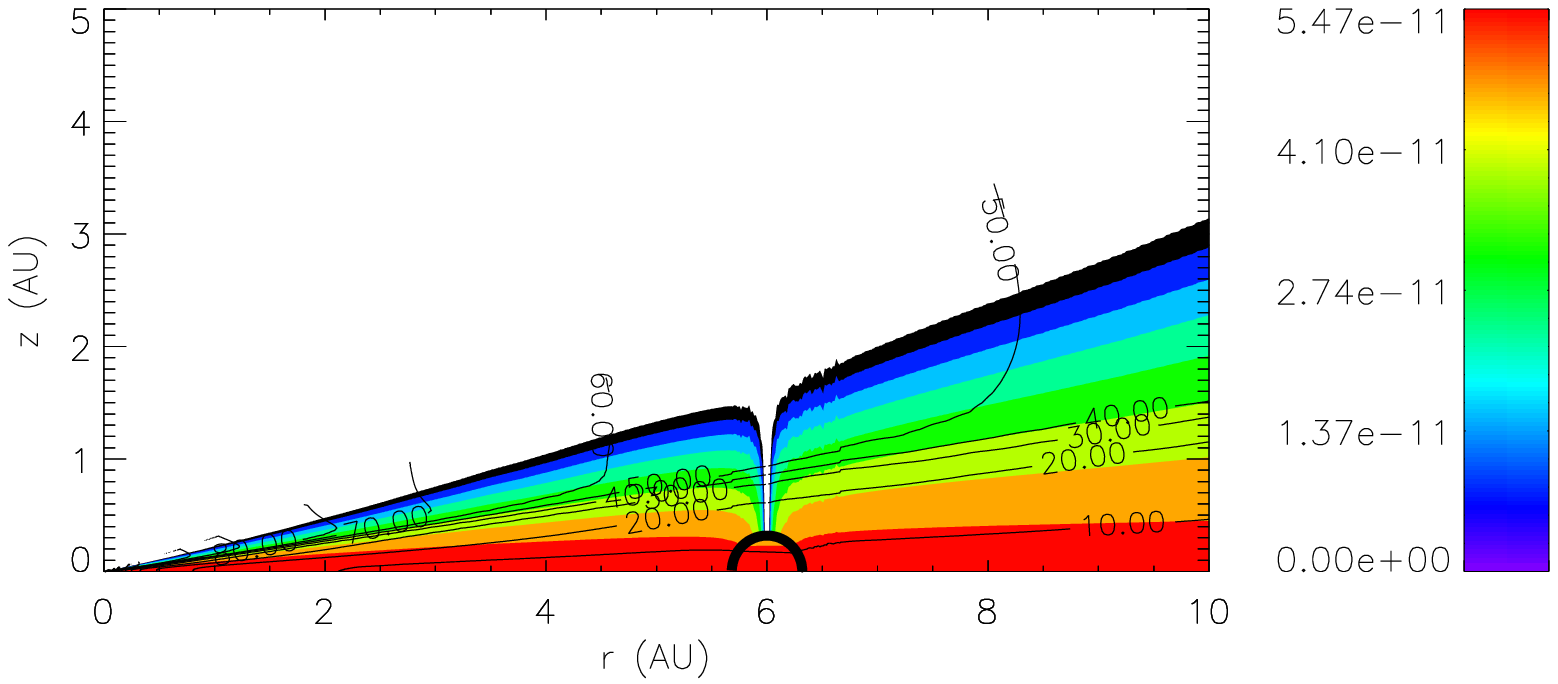}
\includegraphics[height=4.2cm]{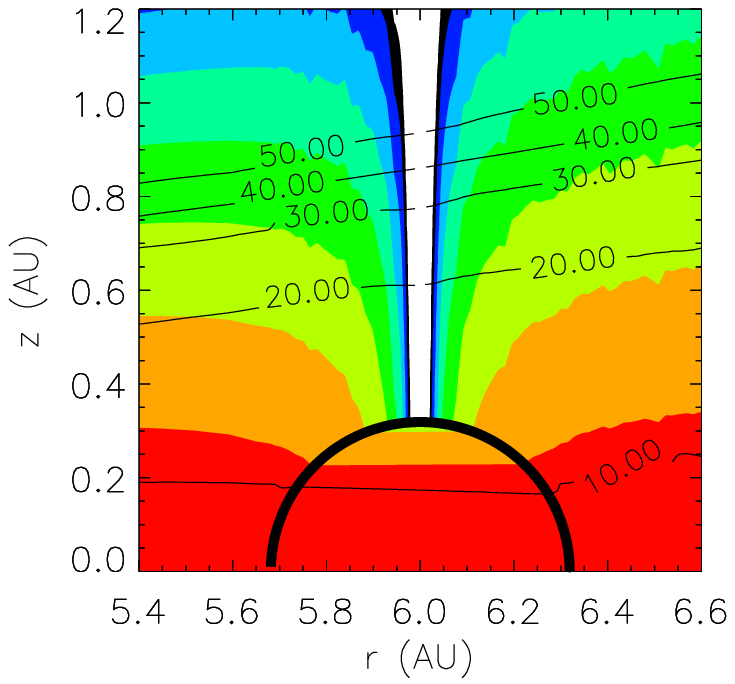}
\includegraphics[height=4.2cm]{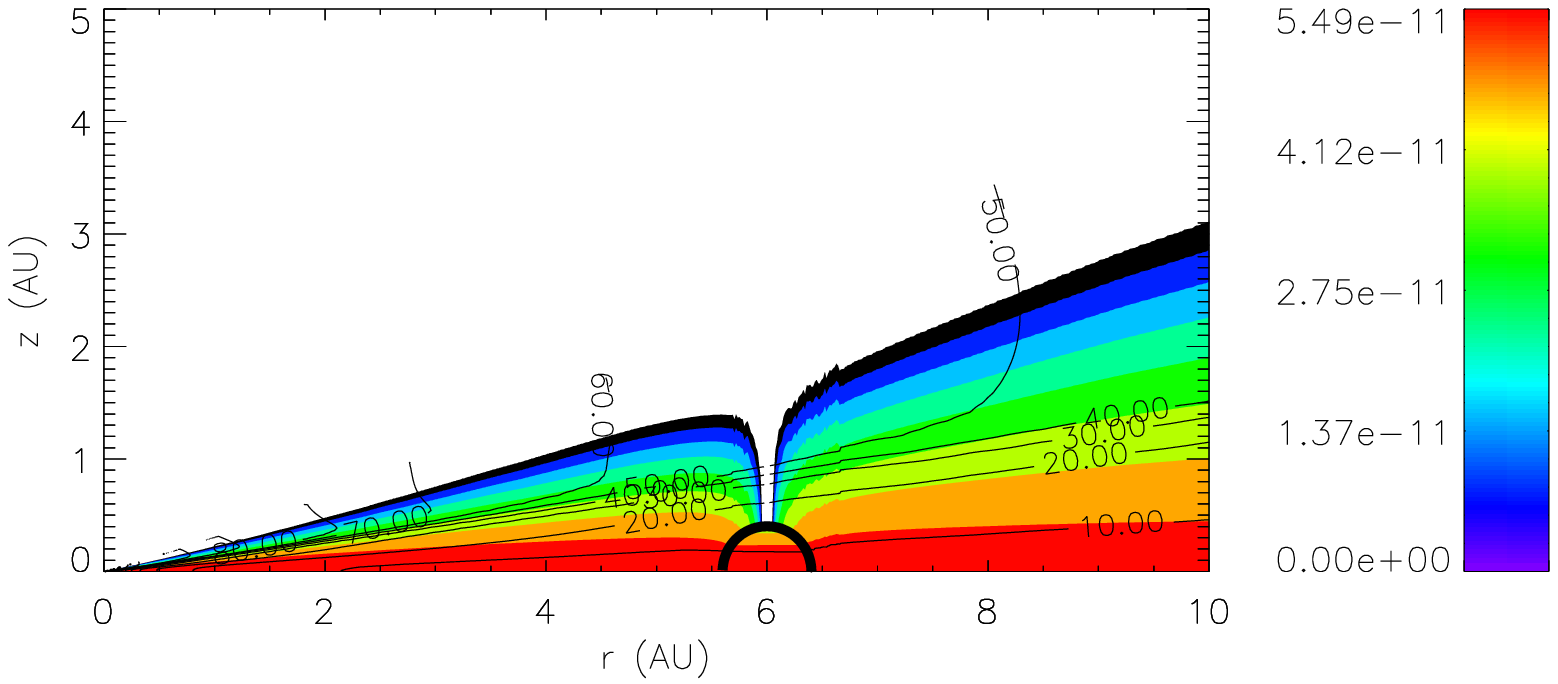}
\includegraphics[height=4.2cm]{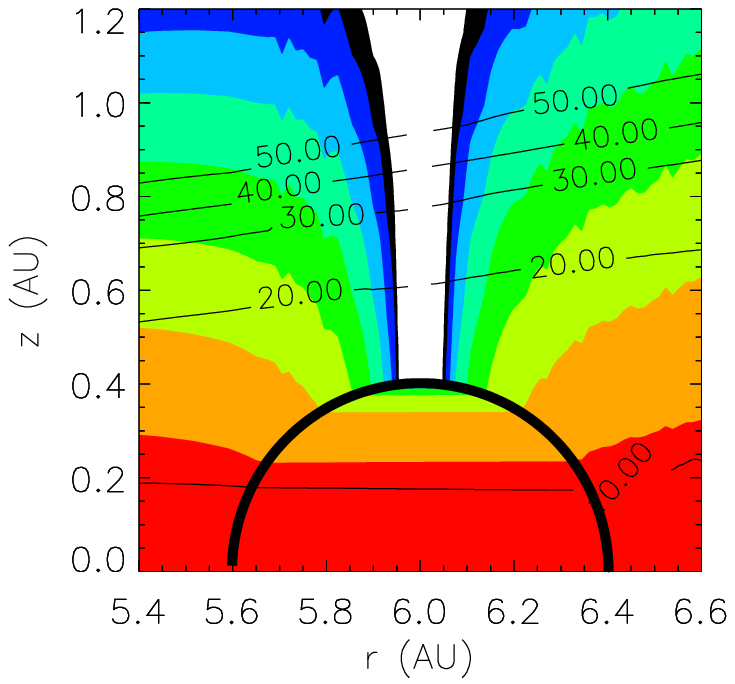}
\caption{The density structure of dust and the temperature structure of disc with dust settling as well 
as a planet (as Fig. \ref{fig6}). The combination of dust settling and planets produces higher temperature 
above the planets due to the longer mean free paths of photons in the region (see the contour of 10 K).}
\label{fig8}
\end{center}
\end{minipage}
\end{figure*}

We now focus on the combined effect of dust settling and planets. Fig. \ref{fig8} shows the results of 
our simulations of the thermal structure of discs and the density structure of dust with dust settling 
and planets. Planetary masses are changed from 0.1, 1, 5, to 10$M_{\oplus }$ on the top to bottom 
panels, respectively. The important consequence of the combination is a dip of the temperature contours 
at the location of the planets, which is clearly seen by the contour of 10 K for the case of the planet 
with 0.1 $M_{\oplus }$. The existence of this higher temperature region above the planets is explained 
by the same reason of the effect of dust settling. In other words, the lower density of dust produced 
by the combination of the planet and dust settling increases the mean free paths of photons. This allows 
more photons penetrate deeper into the mid-plane region. Consequently, the region above the planets has 
a higher temperature. 

\begin{figure}
\begin{center}
\includegraphics[width=8.4cm]{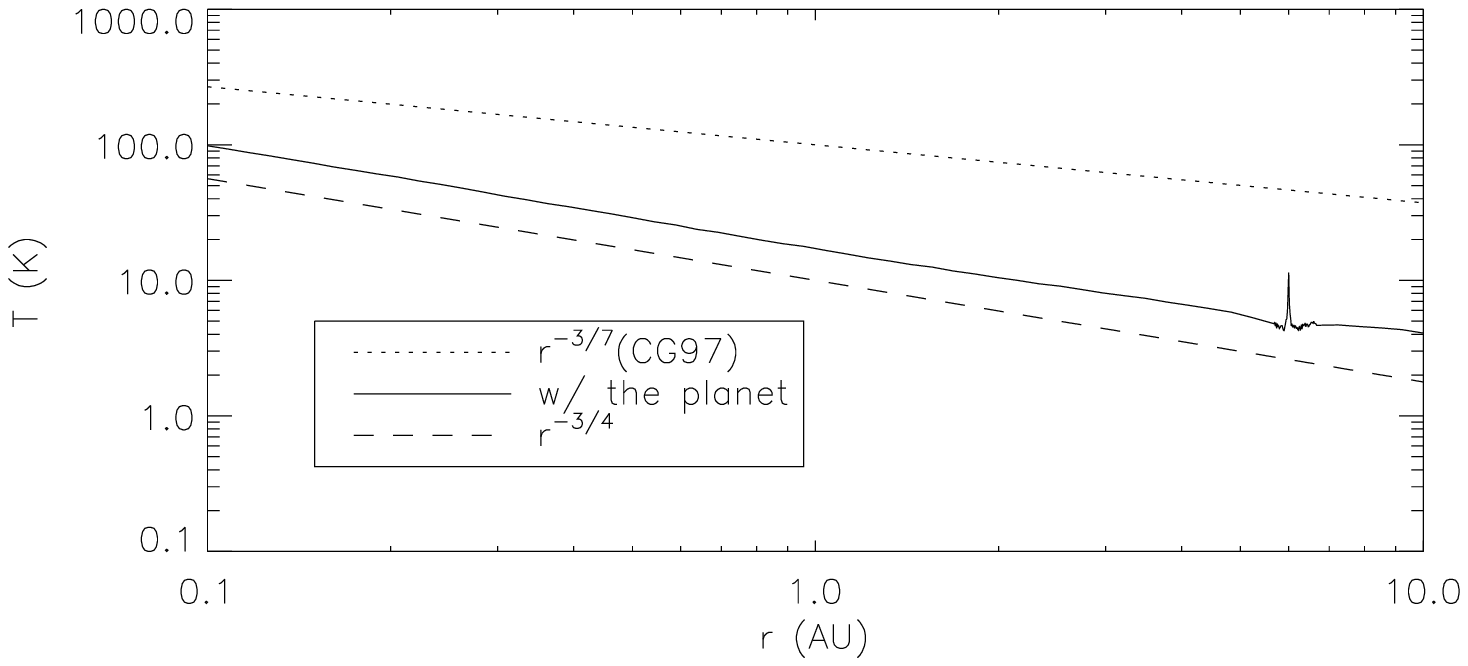}
\caption{The temperature structure with a 10 $M_{\oplus }$ planet and dust settling as a function of 
distance from the central star (as Fig. \ref{fig5}). The combination of the planet and dust settling 
enhances a sharp peak at the location of the planet.}
\label{fig9}
\end{center}
\end{figure}

Fig. \ref{fig9} shows the temperature distribution with 10 $M_{\oplus }$ planet at the mid-plane. A 
temperature peak at the co-orbital radii of the planet is enhanced due to the combined effect of the 
planet and dust settling, as discussed above. It is likely that this effect may drastically change 
the formation of gas giants. This is because accretion of gas envelops surrounding planets is sensitive 
to the gas temperature. Further detailed analyses and discussion of the dust distribution around a planet 
are presented in Appendix B.

\subsection{Dead zones in models with and without planets}

\begin{figure}
\begin{center}
\includegraphics[width=8.4cm]{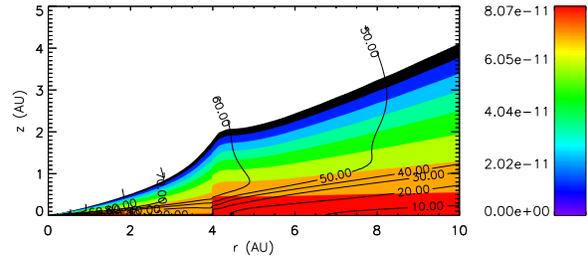}
\caption{The density structure of dust and the temperature structure of disc with a dead zone and dust 
settling (as Fig. \ref{fig3}). The size of the dead zone is 4 au. The rapid settling of the dust in the 
dead zone leaves a dusty wall at the inner edge of the active zone which is {\it directly} heated by the 
central star, resulting in the strange shapes of temperature contours sandwiched by the spherical and 
straight lines.}
\label{fig10}
\end{center}
\end{figure}

The major effect on the thermal structure of a disc arises from a dead zone (which is 4 au in size in 
our model). Fig. \ref{fig10} shows the results of our simulations of the thermal structure of discs and 
the density structure of dust with dead zones and dust settling. The dead zone drastically changes the 
density distribution of dust because dust settling is enhanced due to the low turbulence there (also see 
equation \ref{h_d}). Thus, larger dust grains accumulate at the mid-plane since the scale heights of 
such dust become much smaller. The accumulation of dust into the mid-plane is prevented in the active 
zone due to the stronger turbulence. Consequently, a distinct step in the scale height of dust - or "wall" 
- is left at the boundary between the active and dead zones. This dusty wall plays a significant role in 
the thermal structure of discs since dust is the main absorber of stellar irradiation. 

The temperature structure of discs is also substantially changed by the dead zone although the basic 
physics occurring within the dead zone is the same as that of dust settling. Thus, the surface layer has 
a higher temperature and the mid-plane region has a lower temperature. Also, these two regions are 
separated by a transition region that features a tight collection of the temperature contours. The 
transition region is more declined toward the mid-plane in the dead zone. This arises from the 
enhancement of dust settling within the dead zone. Thus, in the dead zone, the superheated layer 
expands and the mid-plane region shrinks. In contrast, the turbulence in the active region keeps dust 
aloft although dust settling is inevitable because of grain growth due to agglomeration. 
The dust wall formed at the outer edge of the dead zone 
is thus {\it directly} exposed to the stellar radiation. This is one of the most important findings in 
our study. Consequently, the strange shapes of temperature contours sandwiched by the spherical shapes 
(the optically thin, surface regions) and straight lines (the optically thick, mid-plane regions) are 
produced.

\begin{figure}
\begin{center}
\includegraphics[width=8.4cm]{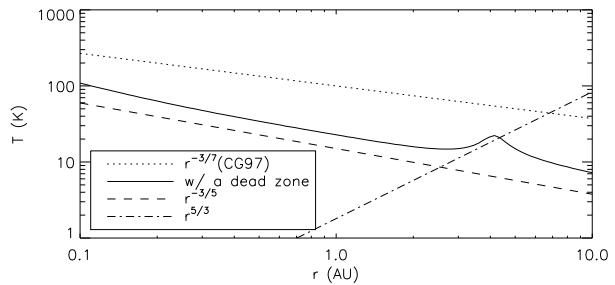}
\caption{The temperature structure with a dead zone and dust settling as a function of distance from 
the central star (as Fig. \ref{fig5}). A positive temperature gradient is formed at the boundary between 
the active and dead zones due to the {\it direct} heating by the central star. This is explained by 
radiative diffusion of the energy absorbed by the dusty wall (the dashed-dot line).}
\label{fig11}
\end{center}
\end{figure}

We found another important effect of the dusty wall. Fig. \ref{fig11} shows the temperature of the 
mid-plane. Importantly, the temperature in the vicinity of the dead zone increases toward the dusty wall. 
This unusual temperature behaviour is also explained by the {\it direct} exposure of the wall to stellar 
radiation. The direct heating provides the wall with higher energy, resulting in a higher temperature 
(otherwise the region is heated only by the thermal emission of dust, which provides much lower energy). 
Since the region is optically thick, the energy absorbed by the wall is transferred by diffusion 
processes \citep{hp09}. Thus, the presence of the hot dusty wall provides a positive temperature 
gradient. It is well-represented by an $r^{5/3}$ temperature distribution which can be also derived 
from analytical calculations. This temperature behaviour provides striking effects on planetary 
migration, which are addressed in a companion paper \citep{hp09}.     

\begin{figure*}
\begin{minipage}{17cm}
\begin{center}
\includegraphics[height=4.2cm]{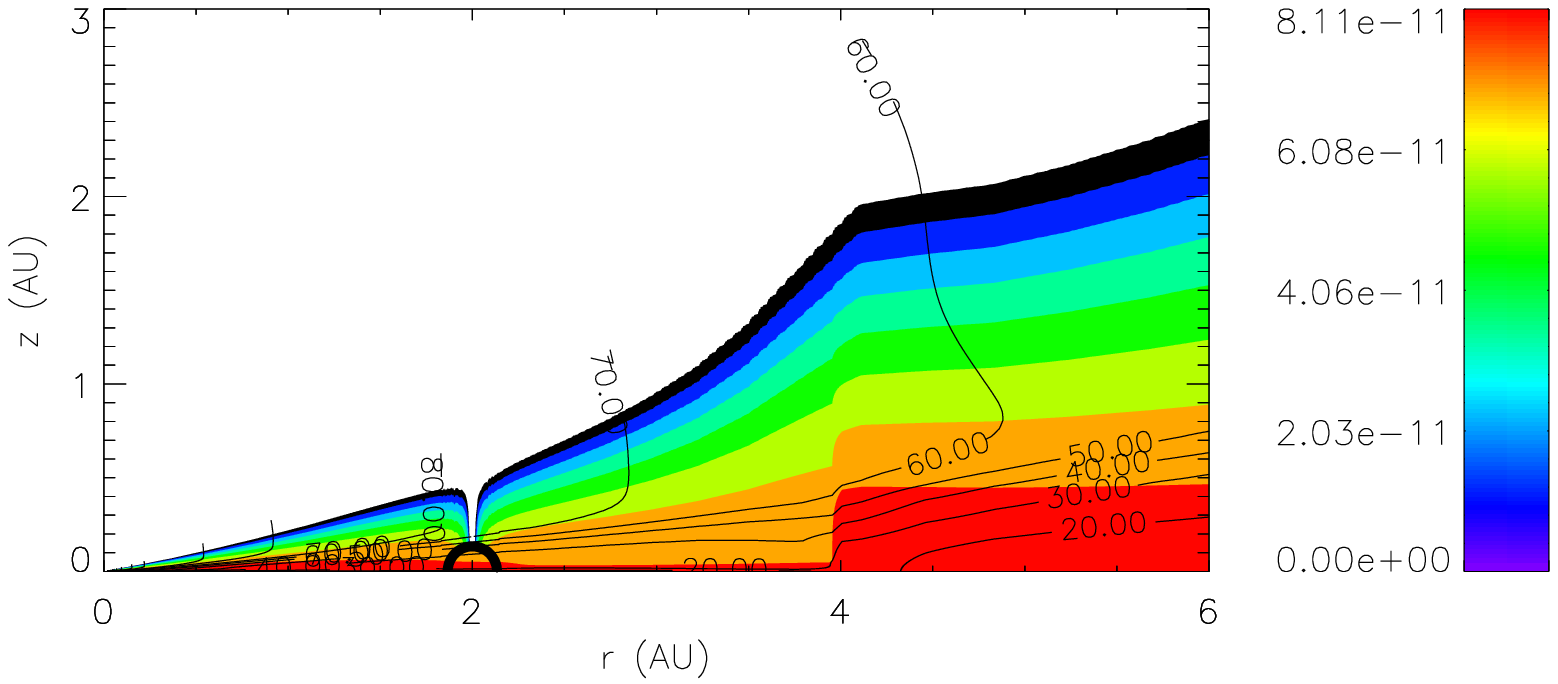}
\includegraphics[height=4.2cm]{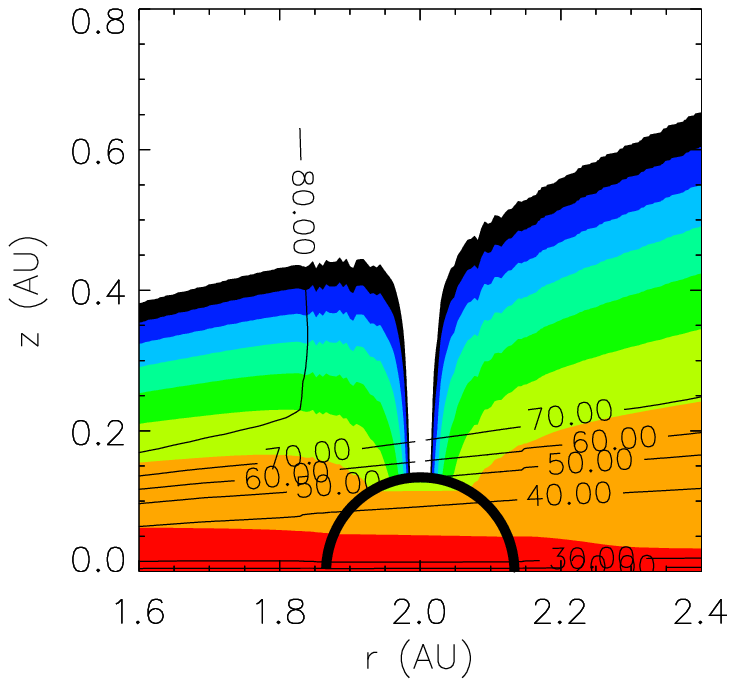}
\includegraphics[height=4.2cm]{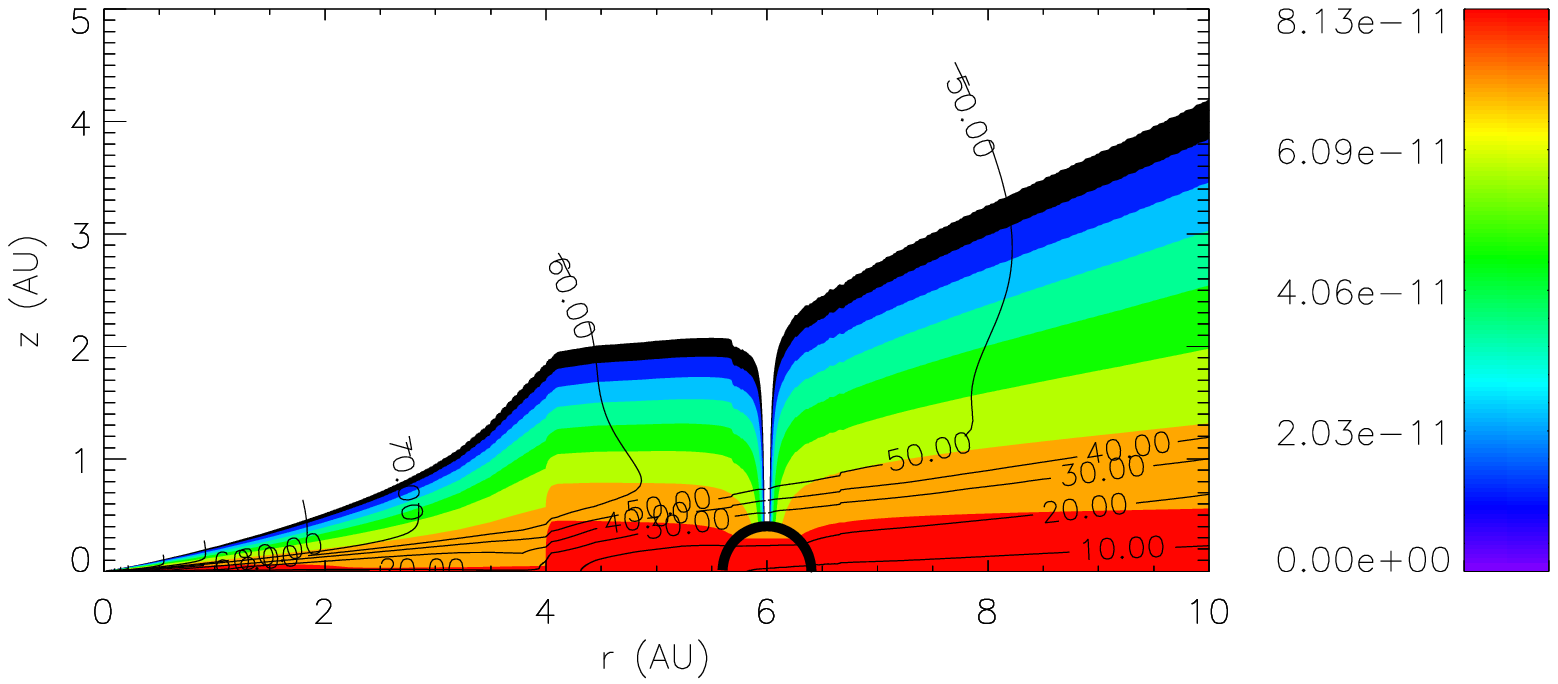}
\includegraphics[height=4.2cm]{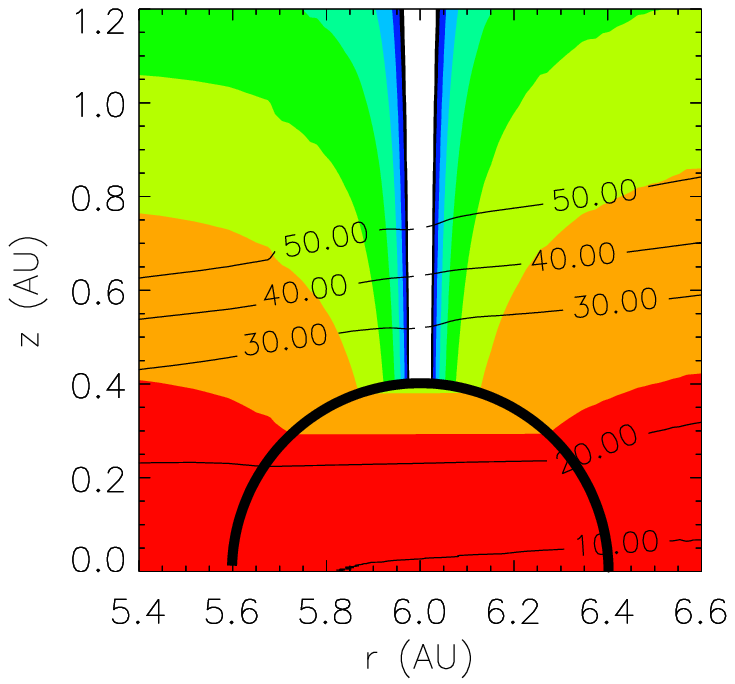}
\caption{The density structure of dust and the temperature structure of disc with a dead zone, dust 
settling, and the planet with 10 $M_{\oplus }$ (as Fig. \ref{fig6}). Top: the orbital radii is 2 au. 
Bottom: the orbital radii is 6 au. Both panels show the compression of the density of dust above the 
planet. For the bottom panel, a self-shadowing is seen clearly by the contour of 10 K (compare with 
Fig. \ref{fig10}). The dusty wall is maintained even with the presence of the planet for both cases.}
\label{fig12}
\end{center}
\end{minipage}
\end{figure*}

We now focus on the combined effects of the dead zone with dust settling and planets with 10 
$M_{\oplus }$. We examine two extreme cases; one of which is that of a planet that is located outside 
the dead zone ($r_p=6$ au), and the other in which it is located at $r_p=2$ au inside the dead zone. 
Fig. \ref{fig12} shows our results for the temperature structure of discs and the density structure of 
dust. The common feature of the density structure is the compression of the density above the planet 
although the overall density structure is similar to the case only with the dead zone. The main 
difference between these cases is the effect of the planet on the mid-plane region of discs. For the 
planet outside the dead zone, the density is further distorted by the gravitational force of the 
planet while, for the inside case, the distortion by the planet is not clear. This is explained by 
the enhancement of dust settling in the dead zone. As discussed above, dust settles much more 
compactly into the mid-plane in the dead zone. The planet also compresses dust into the mid-plane. 
Thus, both the dead zone and the planet affect the density distribution in a similar way, although 
the former is global and the latter is local. Thus, the global effect of the dead zone dominates the 
local effect of the planet. The active zone, however, allows the perturbation of the planet to be shown 
clearly due to the stronger turbulence. Thus, the combination of the dead zone and the planet still allows 
dust to settle into the mid-plane further although the dead zone surpasses the effect of the planet. 
Furthermore, the formation of the dusty wall is never affected by the inclusion of a planet, provided 
that the planets is not placed in the immediate vicinity of the dead zone. 

\begin{figure}
\begin{center}
\includegraphics[width=8.4cm]{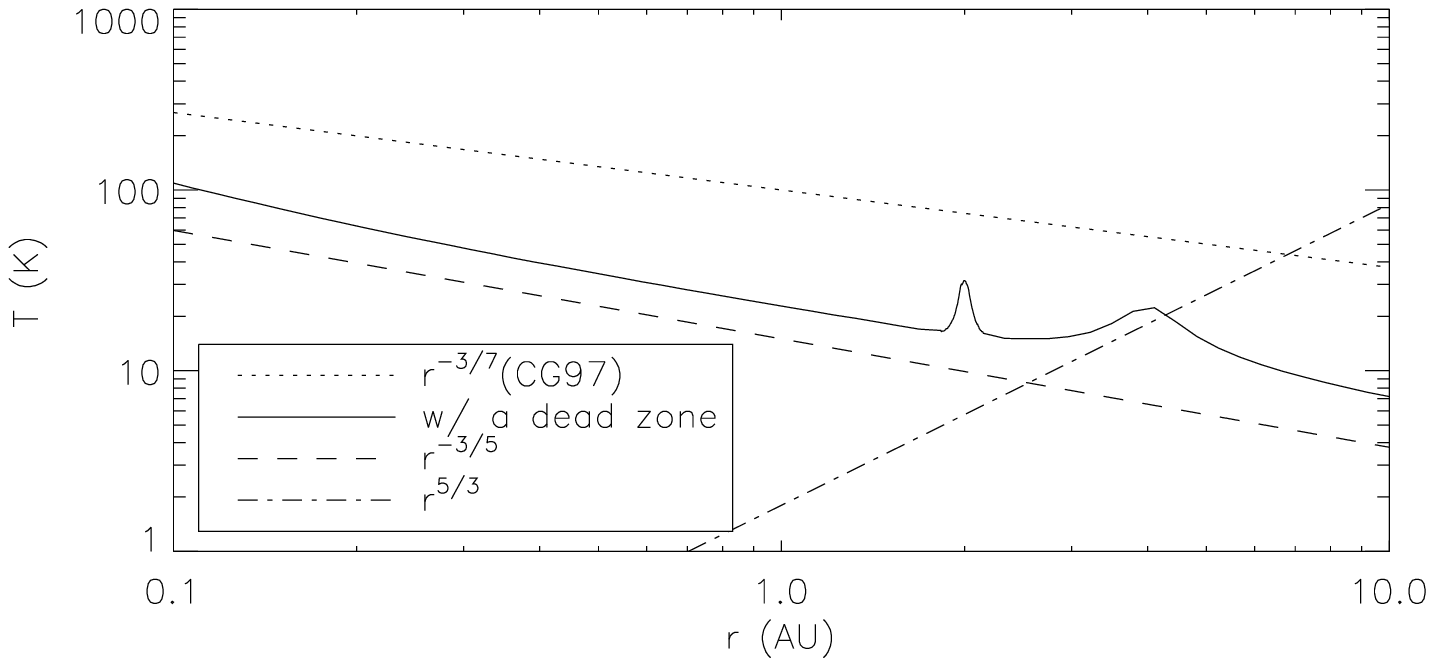}
\includegraphics[width=8.4cm]{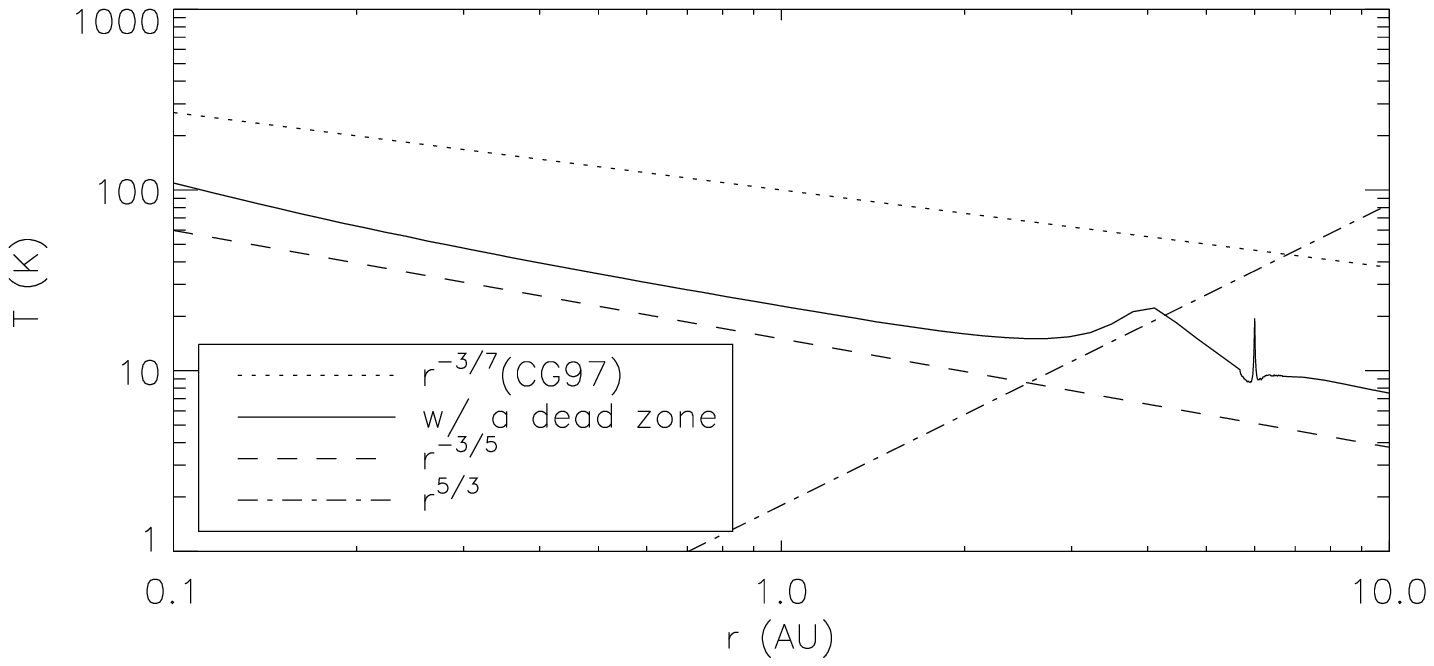}
\caption{The temperature structure with a dead zone, dust settling, and the planet with 10 
$M_{\oplus }$ as a function of disc radius (as Fig. \ref{fig11}). The combined effects produce the 
positive temperature gradient as well as the sharp peak at the location of the planet at the position 
of the planet.}
\label{fig13}
\end{center}
\end{figure}

The temperature structure is also changed by the combination of the two although it is roughly similar 
to the case with only the dead zone. The main difference arises when the planet is placed outside 
the dead zone. The compression of dust produces the self-shadowing region shown by the contours of 
10 and 20 K of Fig. \ref{fig12} (also see Fig. \ref{fig10}). There is no clear difference for the case 
of the planet inside the dead zone. This is explained by the fact that the dust distribution with the 
planet is almost identical to that without the planet since the effect of the dead zone is too strong 
relative to the gravitational force of the planet. Thus, the physical picture that the dusty wall becomes 
thermally hot due to the direct heating of the star is true even if the perturbation of the planets is 
included. This is also confirmed by Fig. \ref{fig13} which shows the temperature at the mid-plane. 

\section{The effects of disc mass and viscous heating } \label{discu}

We have demonstrated a novel thermal effect due to the presence of the dead zone, namely, the 
appearance of the dusty wall at the inner edge of the active zone and the resultant positive temperature 
gradient. In this section, we undertake parameter studies to confirm this finding for various disc 
parameters. Also, we discuss the effects of other heat sources that have been neglected so far. 

\subsection{Parameter study}  

We have used disc models and parameters that reproduce the observed SEDs around M stars (S07) in 
which $\Sigma \propto 1/r$. On the other hand, there is another famous disc model, so-called minimum 
mass solar nebula (MMSN) model in which $\Sigma \propto r^{-3/2}$ \citep{h81}. MMSN models also 
succeed in reproducing the observed SEDs (e.g., CG97). Thus, protoplanetary discs are likely to have 
the range for the surface density from $r^{-1}$ to $r^{-3/2}$. Thus, it is important to check our 
findings for MMSN models as well. Also, we have increased the disc mass by a factor of 10 which is 
favoured if massive planets are to be formed \citep{ki98,ambw05}.

\begin{figure}
\begin{center}
\includegraphics[width=8.4cm]{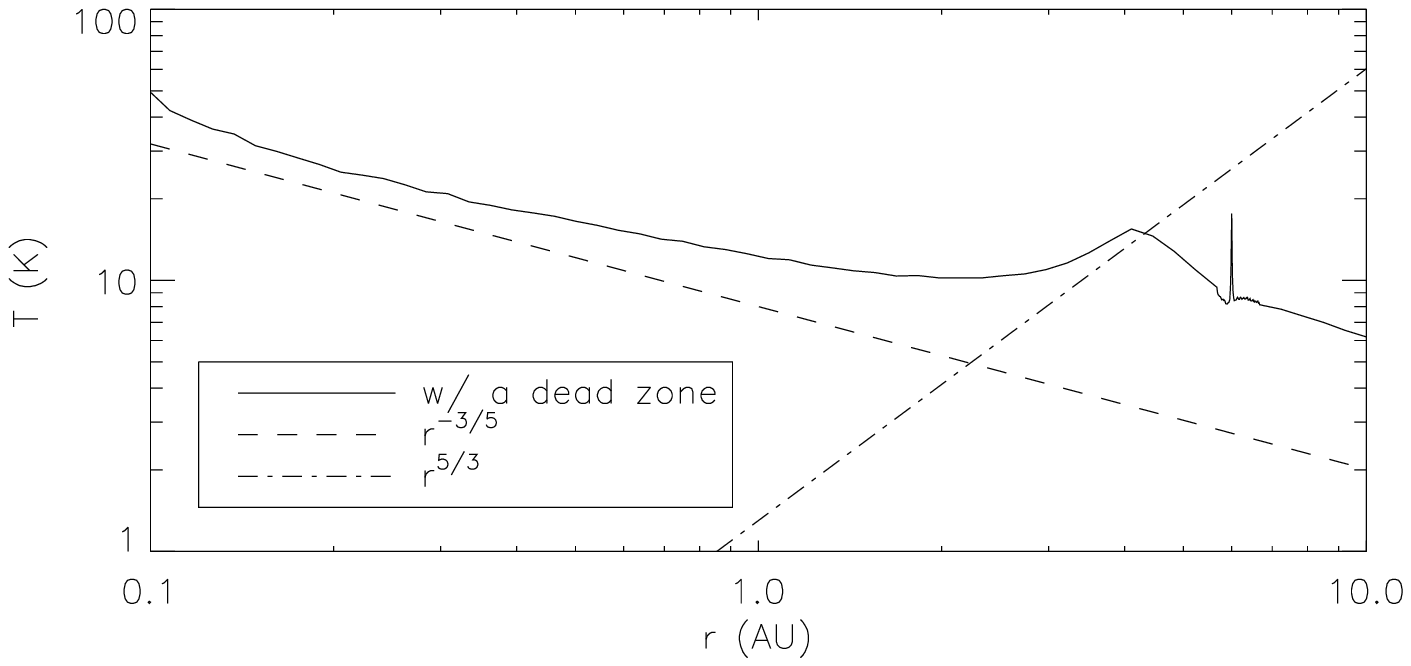}
\includegraphics[width=8.4cm]{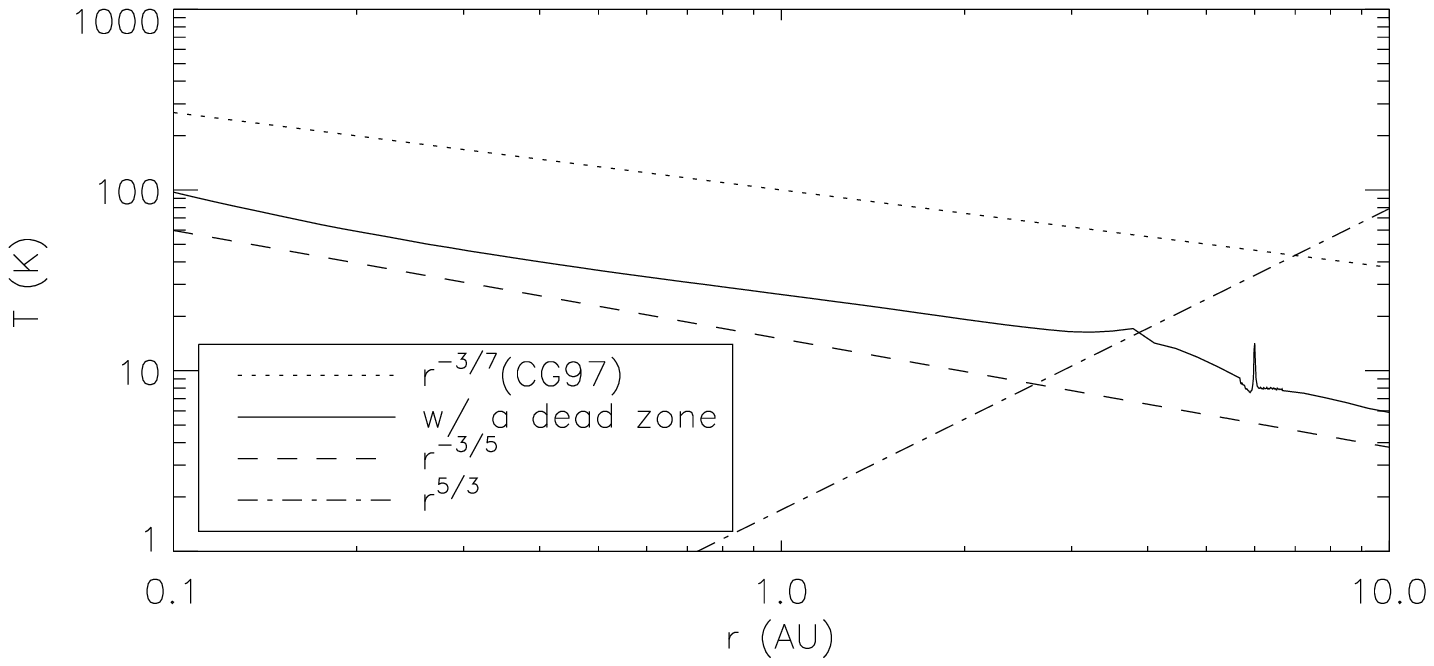}
\includegraphics[width=8.4cm]{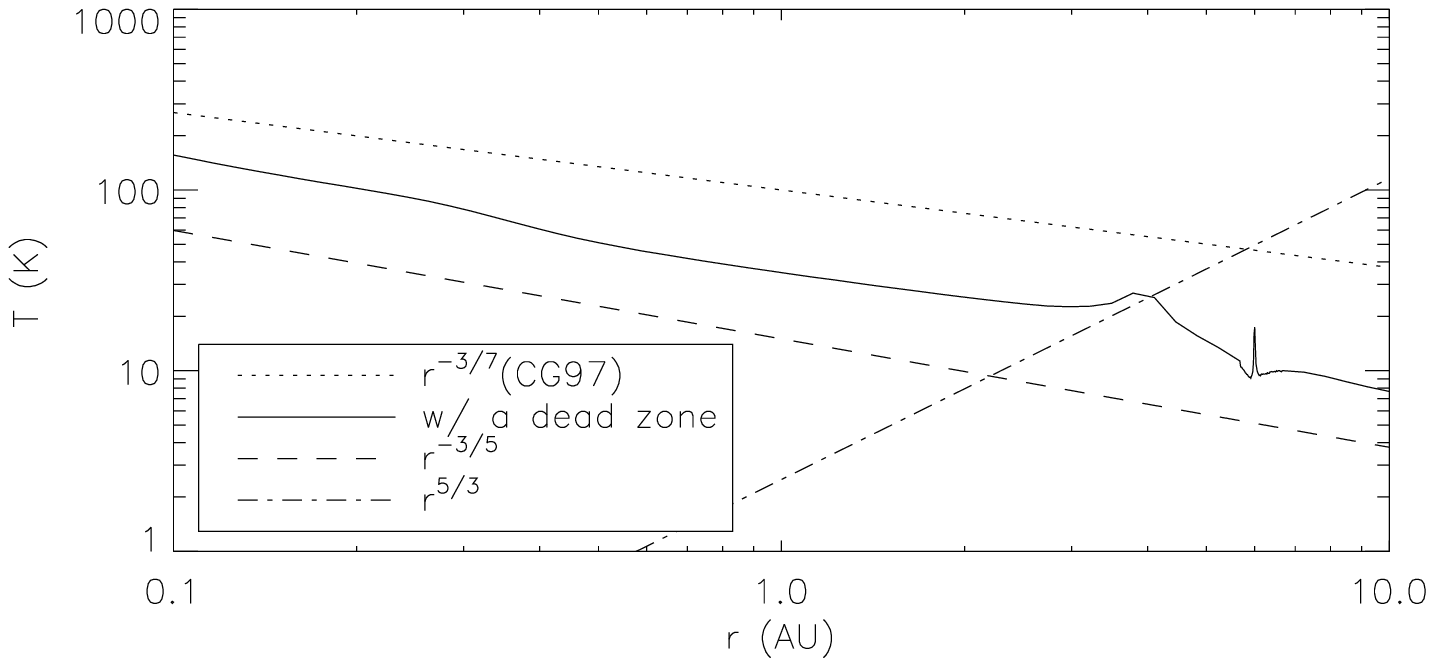}
\caption{The temperature structure with a dead zone, dust settling, and a planet with 5 
$M_{\oplus }$ as a function of disc radius (as Fig. \ref{fig11}). Top: MMSN-3. Middle: 
MMSN-4 Bottom: S07-4. Every panel shows a positive temperature gradient and a sharp peak at 
the location of the planet ($r_p=6$ au) although the slope of the temperature depends on the disc mass.}
\label{fig14}
\end{center}
\end{figure}

With this motivation, we performed a parameter study involving the slope of surface density and the 
disc mass. Table \ref{parameter_study} summarises our runs (also see Table \ref{param_disk}). In the 
simulations, we include a 5 $M_{\oplus }$ planet at 6 au to verify the effect of the gravitational 
force of the planet. Fig. \ref{fig14} shows the temperature of the mid-plane for the runs. Every case 
shows the positive temperature gradient although MMSN-4 provides a shallower slope. This shallower 
slope is explained by a higher temperature of the mid-plane for lower mass discs, resulting in 
positive temperature gradients that are somewhat diminished by the original temperature of the 
mid-plane. Thus, the formation of the positive temperature gradient depends on the mass of discs, not 
the slope of the surface density. Furthermore, sharp peaks produced by the presence of the planet are 
also sustained. Thus, our findings of the dusty wall and the positive temperature gradient are robust 
although the slope of the temperature may be changed by the mass of discs.

\begin{figure}
\begin{center}
\includegraphics[width=8.4cm]{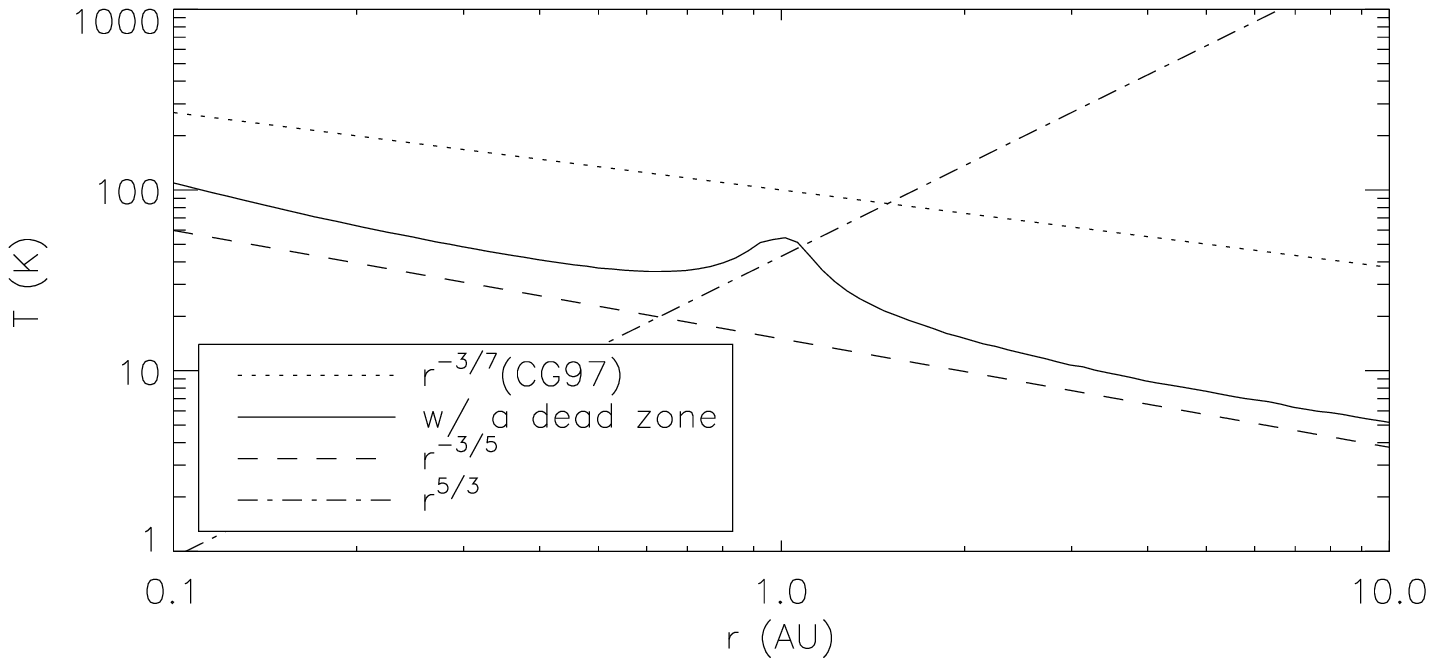}
\includegraphics[width=8.4cm]{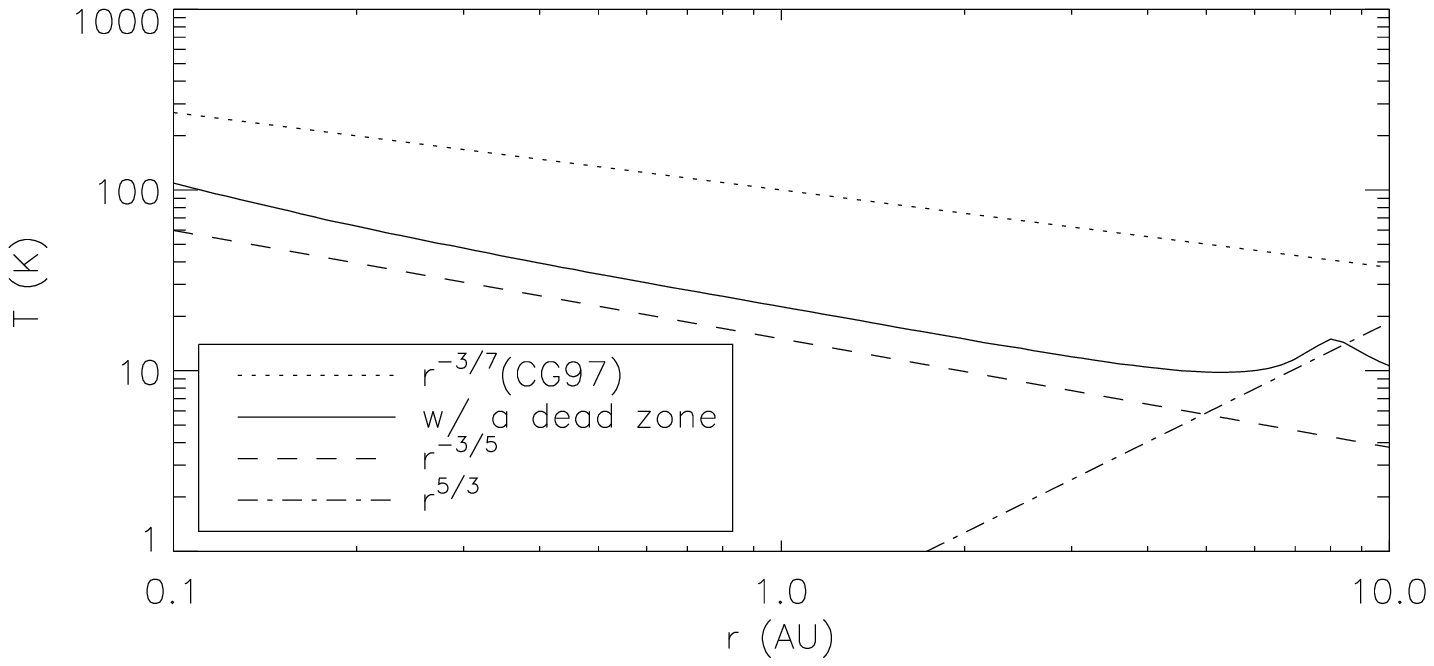}
\caption{The temperature structure with a dead zone and dust settling as a function of distance from 
the central star (as Fig. \ref{fig11}). Top: the size of the dead zone is 1 au. Bottom: the size of 
the dead zone is 8 au. Both panels show a positive temperature gradient formed at the boundary between 
the active and dead zones.}
\label{fig15}
\end{center}
\end{figure}

We now investigate the effect of the size of the dead zones. We have fixed the size of the dead zone 
as 4 au so far. Many previous studies, however, showed that the size depends on parameters they used 
although there is the agreement with each other that its size is roughly 10 au. Furthermore, 
\citet{mpt09} found that dead zones evolve with time and shrink following the accretion of gas onto 
the star. Thus, the size of the dead zones also has some uncertainties. In order to compare our findings 
for various sizes of the dead zones, we performed simulations with 1 and 8 au size of the dead zones 
with dust settling. In the simulations, we use the parameters of Table \ref{param_disk} and do not 
include the planet since we have shown that the resultant effects of the dead zone and the planet are 
robust. Fig. \ref{fig15} shows the temperature at the mid-plane. Both cases show the positive temperature 
gradient at the outer edge of the dead zones. Furthermore, the temperature behaviours are also 
well-represented by $r^{5/3}$, meaning that the energy absorbed in the dusty wall due to the direct 
heating from the star is transferred by the diffusion processes. Thus, our findings are independent 
of the size of dead zones.

\begin{figure*}
\begin{minipage}{17cm}
\begin{center}
\includegraphics[width=8.4cm]{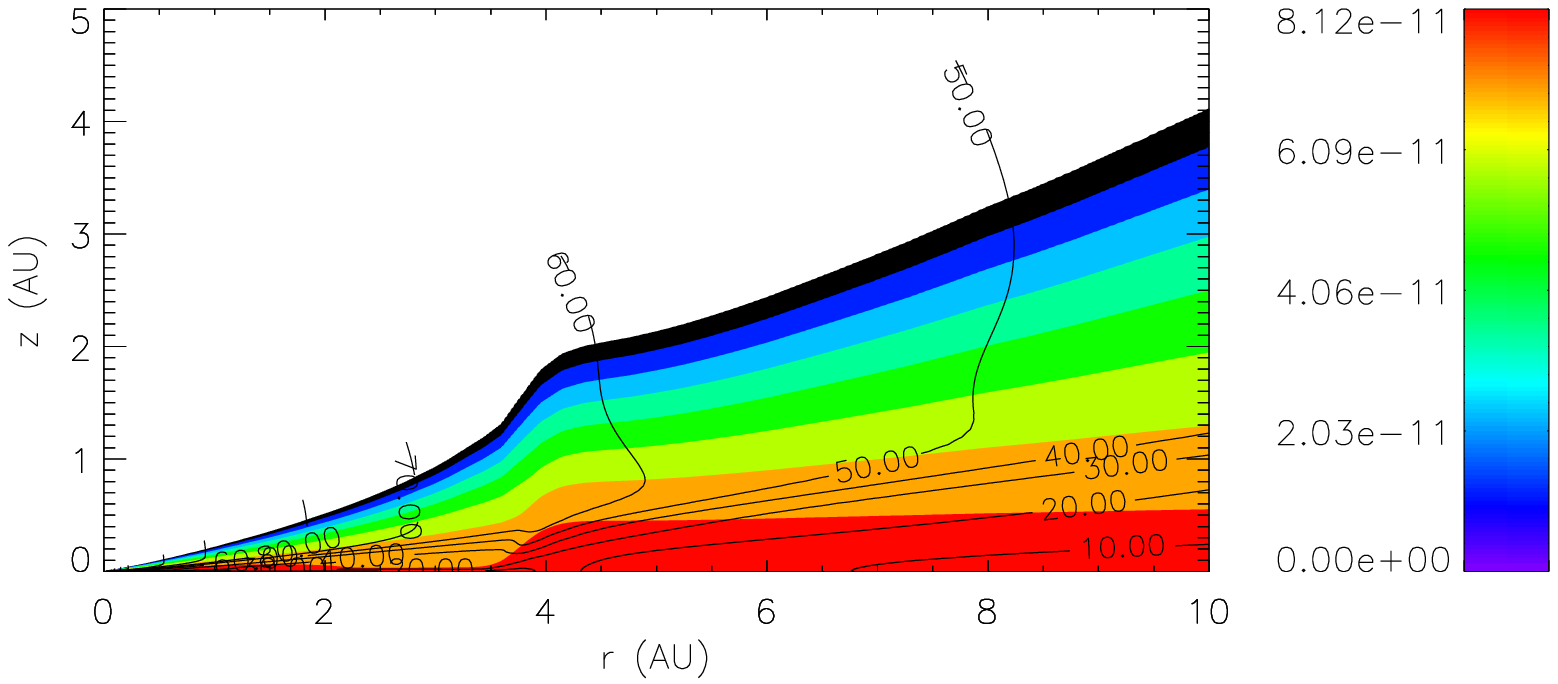}
\includegraphics[width=8.4cm]{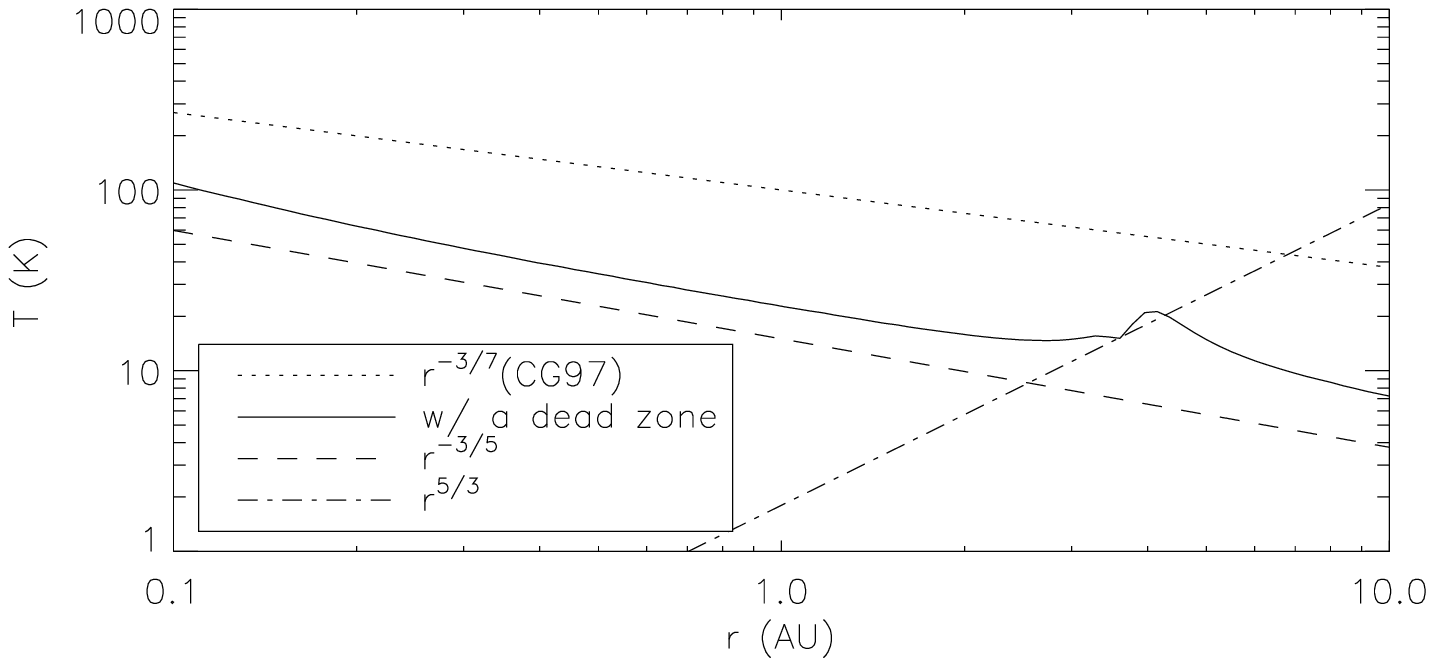}
\includegraphics[width=8.4cm]{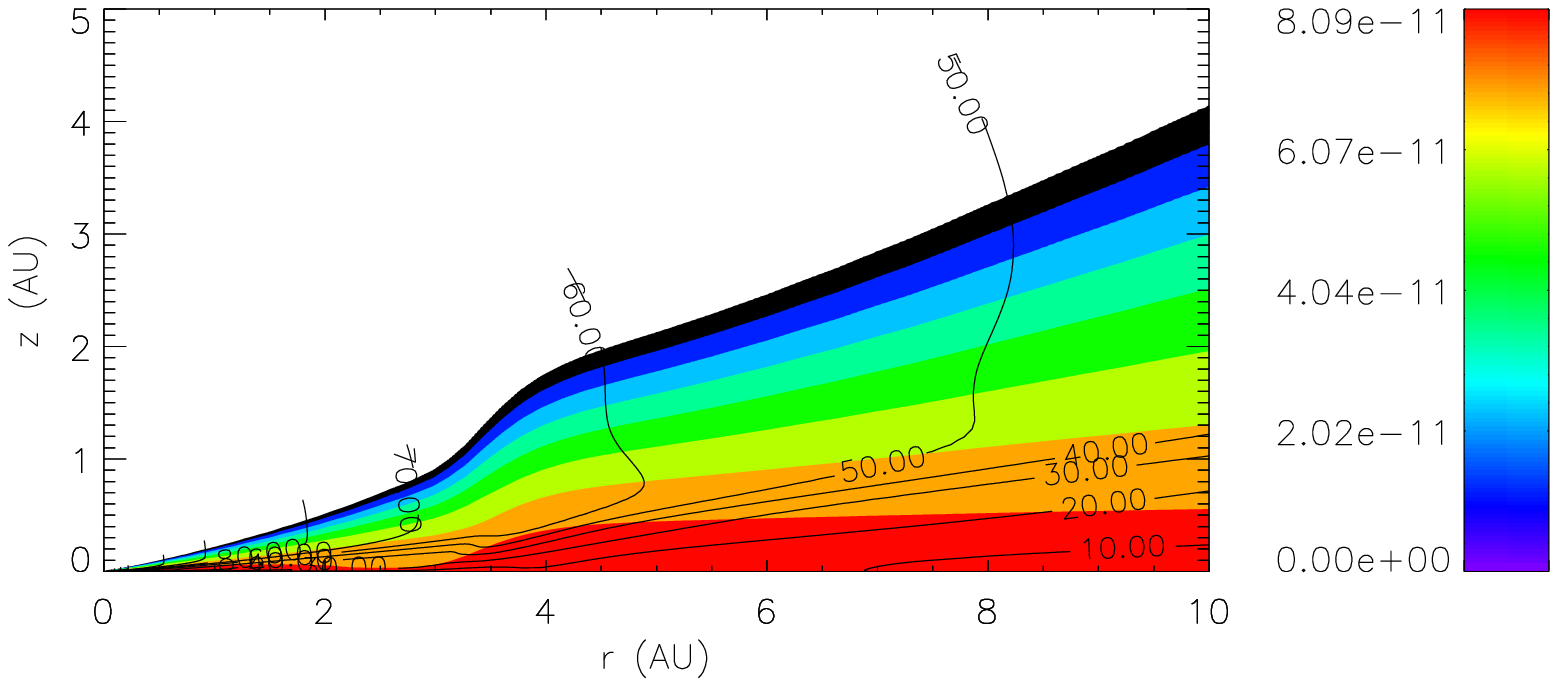}
\includegraphics[width=8.4cm]{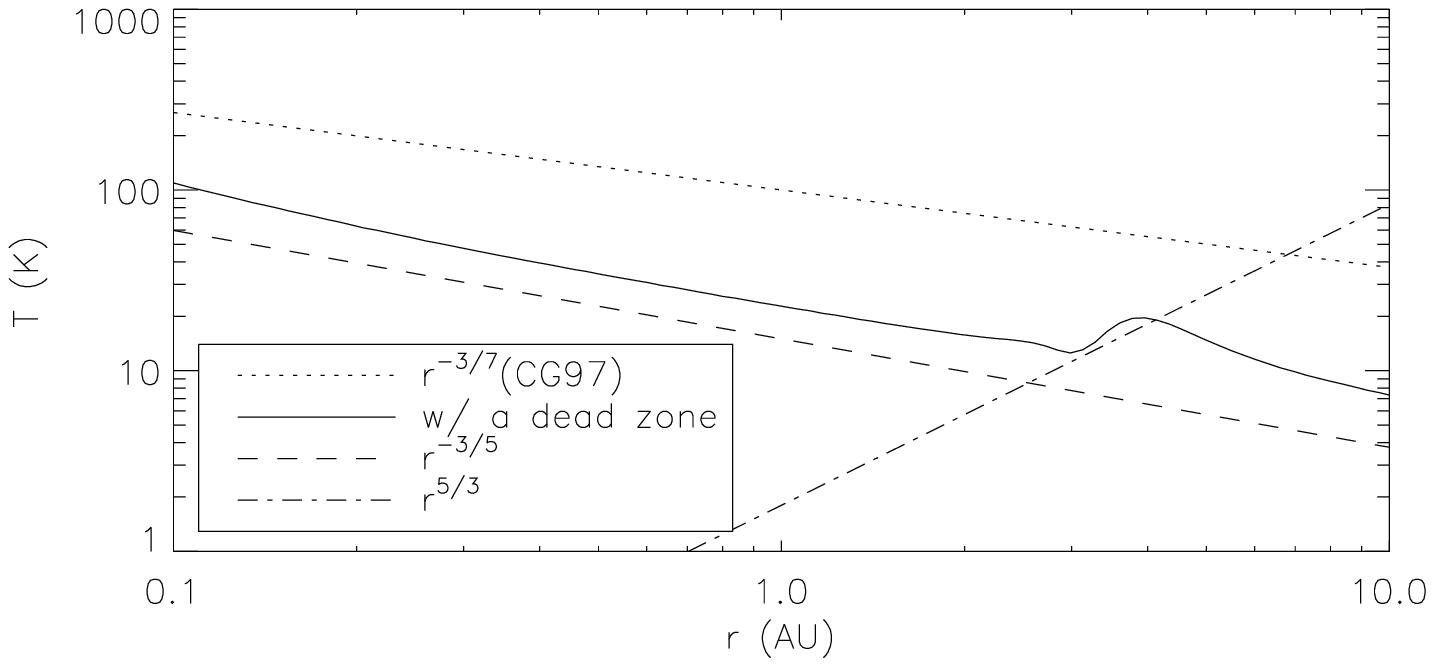}
\includegraphics[width=8.4cm]{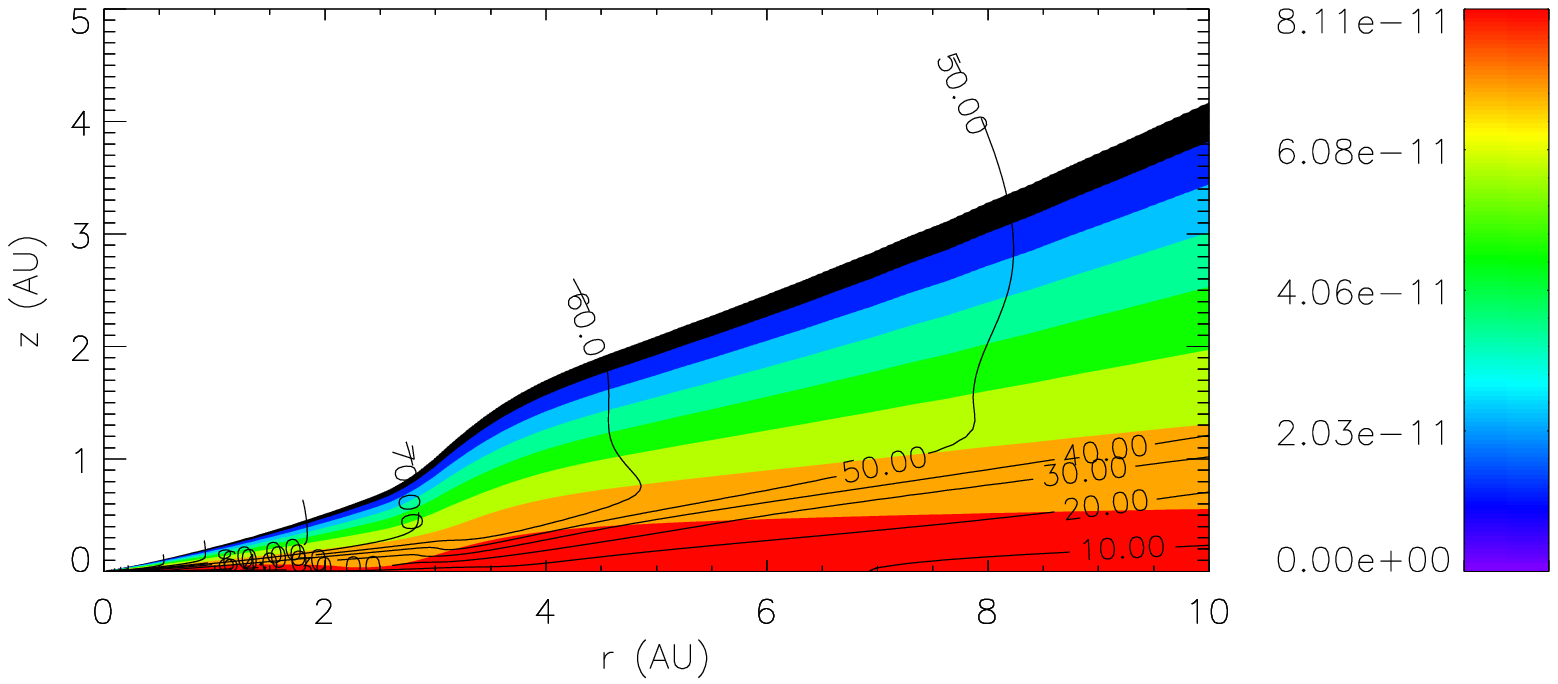}
\includegraphics[width=8.4cm]{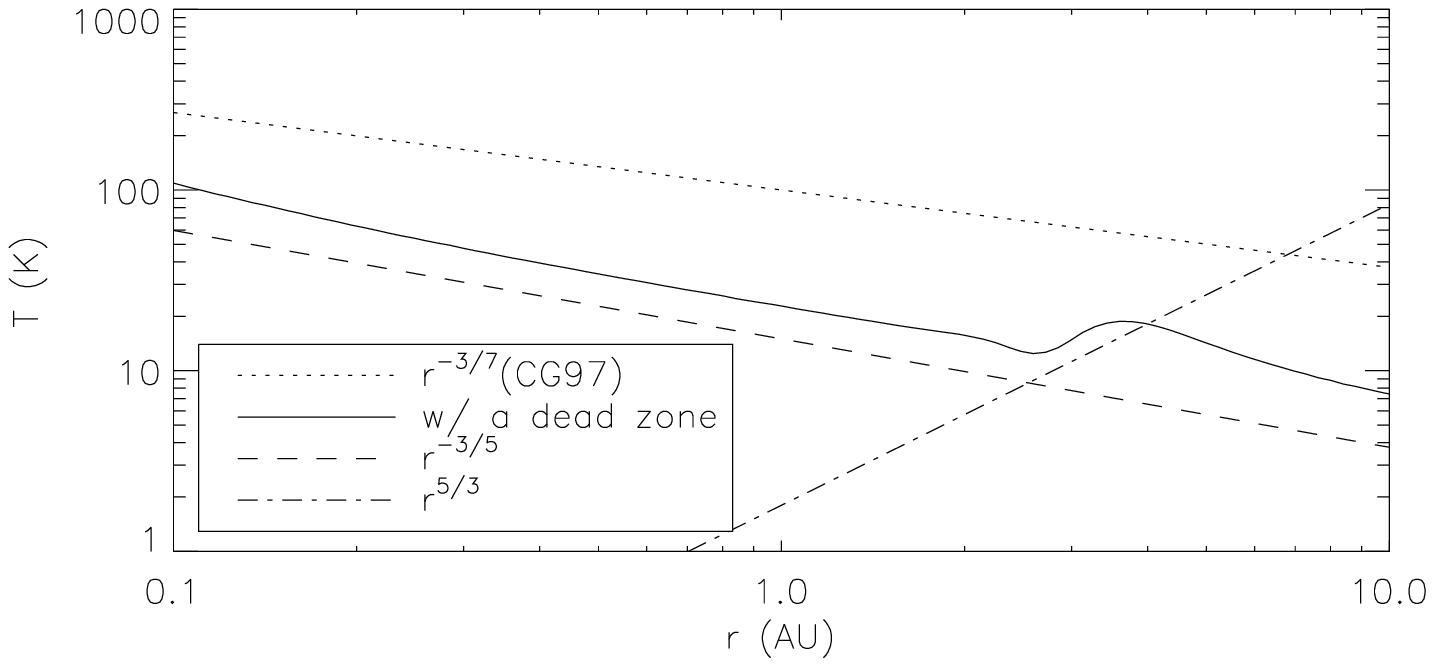}
\includegraphics[width=8.4cm]{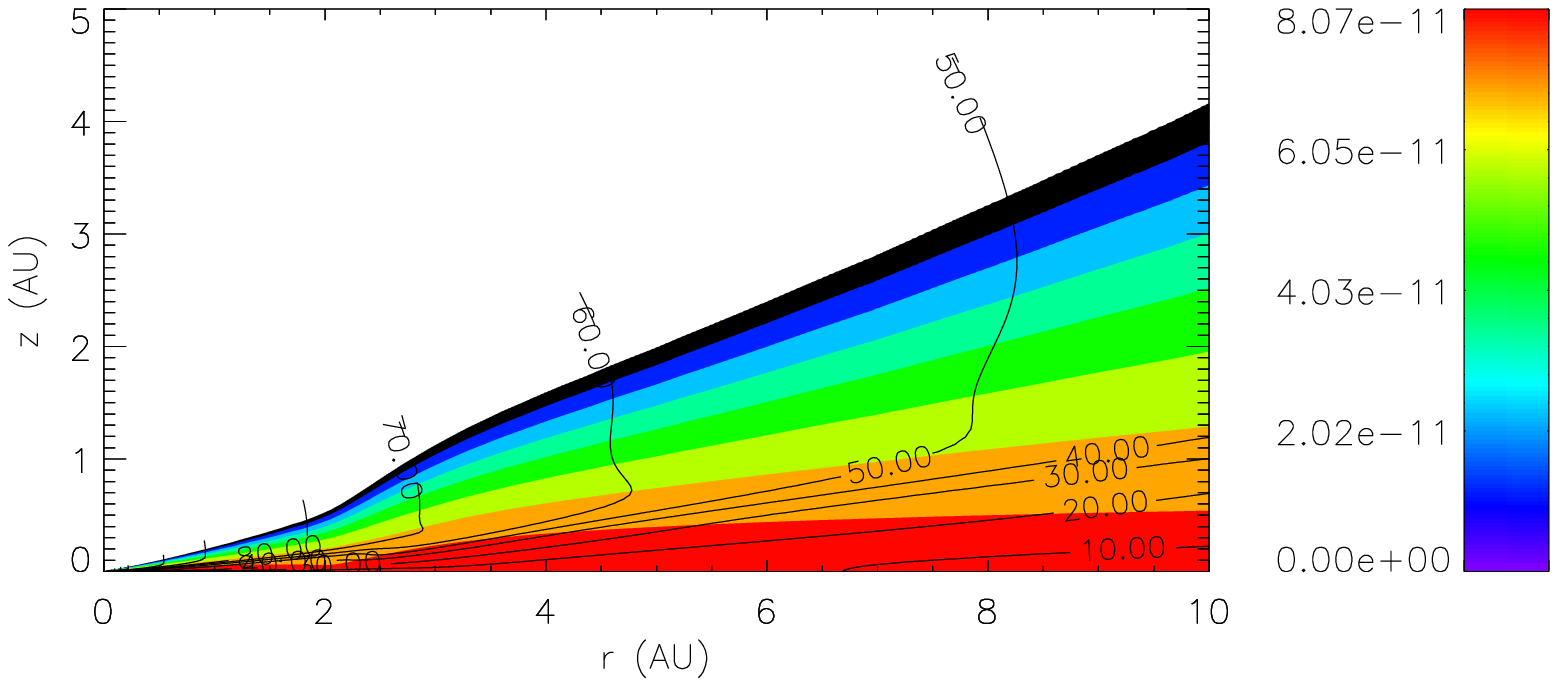}
\includegraphics[width=8.4cm]{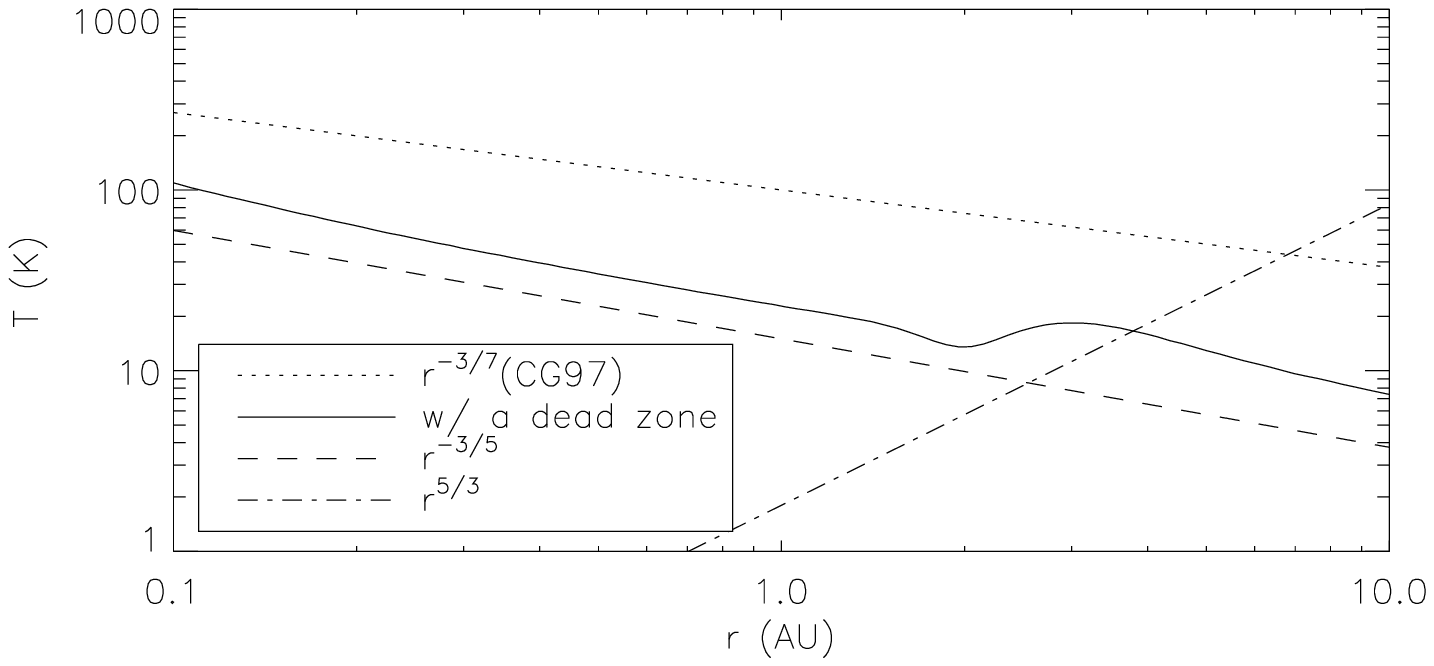}

\caption{The density structure of dust and the temperature structure of disc with the dead zone and 
dust settling with a finite size of the transition region $\bigtriangleup r$ (as Fig. \ref{fig3}) and 
the temperature of the mid-lane as a function of disc radius (as Fig. \ref{fig11}) on left and right 
columns, respectively. Top: $\bigtriangleup r=1h$. Second: $\bigtriangleup r=3h$. Third: 
$\bigtriangleup r=5h$ Bottom: $\bigtriangleup r=10h$. Every case shows a dusty wall and the resultant 
positive temperature gradient although its slope becomes shallower and its position becomes closer to 
the central star, as $\bigtriangleup r$ is increased.}
\label{fig16}
\end{center}
\end{minipage}
\end{figure*}

We next turn to examining dead zones with a finite radial transition region between the dead and 
active zones. Although we have assumed so far that the active region suddenly appears at various 
disc radii from the central star, the transition region will in general be diffused since the presence 
of the dead zones activates various instabilities as discussed below. Dead zones with a finite 
transition region can be generally written as
\begin{equation}
\alpha(r)=\alpha_{DZ}+\frac{\alpha_{active}}{2} \left[ \tanh\left( \frac{r-r_0}{\bigtriangleup r} \right) +1 \right],
\label{dead_zone_r}
\end{equation}
where $r_0=4$ au is the fiducial size of the dead zone and $\bigtriangleup r$ is the thickness of 
the transition region. Fig. \ref{fig16} shows our results - the thermal and density structures of dust 
and the temperature of the mid-plane as a function of disc radius from the star (the left and right 
columns, respectively). The parameter $\bigtriangleup r$ is changed from 1, 3, 5, and 10 $h$, where 
$h$ is the scale height of gas, on the top to bottom panels, respectively. The positive temperature 
gradient appears in every case. Interestingly, the slope of the positive temperature gradient is 
more or less represented by $r^{5/3}$ although it becomes shallower as $\bigtriangleup r$ increases. 
The difference of the slope is explained by the shape of the dusty wall. A gradual transition of the 
value of $\alpha$ for a finite value of $\bigtriangleup r$ provides smoother density distribution of 
dust, resulting in a gentle temperature gradient. Furthermore, the location of the positive temperature 
gradient is changed for these cases. The positive temperature gradient appears at the mid-plane region 
closer to the star, as $\bigtriangleup r$ increases.  This is also understood by the structure of the 
dusty wall. For the dead zone with a thicker transition region, the dusty wall presents at smaller disc 
radii from the host star.  

\begin{table}
%\begin{center}
\caption{The parameter study}
\label{parameter_study}
\begin{tabular}{ccc}
\hline
Run         & $\Sigma$    & $M_d$                \\ \hline
MMSN-3      & $r^{-3/2}$  & 4.5$\times 10^{-3}$  \\
MMSN-4      & $r^{-3/2}$  & 4.5$\times 10^{-4}$  \\
S07-4       & $r^{-1}$    & 4.5$\times 10^{-4}$  \\
\hline
\end{tabular} 
%\end{center}
\end{table}

\subsection{The addition of viscous heating}

In our simulations, we have considered stellar irradiation as the main heat source for discs, so that 
other potential heat sources such as viscous heating and the accretion luminosity of planets have been 
neglected. This is because we have assumed discs to be passive following CG97. This assumption is likely 
to be appropriate since viscous heating is dominant only within 0.1 au for M star systems. Furthermore, 
the resultant SEDs of passive discs agree with the observed ones very well \citep[CG97;][]{cg99,cjc01}. 
On the other hand, the SEDs produced from the models including viscous heating as well as stellar 
irradiation are also able to reproduce the observations \citep{dccl98,dch99}. Thus, it is worth 
analysing the effects of viscous heating on the dusty wall and the resultant temperature structures since 
the constraints arising from SEDs are necessary, but not sufficient conditions.

The positive temperature gradient formed in front of the dust wall is maintained even if viscous 
heating is included. This is because the temperature around the dust wall determined by the direct 
exposure of stellar irradiation is much higher. Viscous heating cannot wash out this effect since the 
temperature due to viscous heating is dominant only within 0.1 au.

Furthermore, the low turbulence within the dead zones implies that viscous heating is strongly reduced. 
Weaker viscous heating strongly supports our assumption of passive discs. Therefore, our assumption is 
valid throughout the disc beyond 0.1 au and our findings are robust features of the dead zone even if 
viscous heating is included.  

In addition to viscous heating, the presence of planets provides another heat source due to accretion 
processes. During planet formation, protoplanets accrete gas, dust and planetesimals, resulting in 
the radiation of the extra gravitational energy. The temperature structure produced by the accretion 
luminosity of the planet is represented by $r^{-1/2}$ due to the inverse square law. Thus, the 
temperature peak produced by the gravitational force of the planet becomes wider due to the accretion 
effect. Thus, although the accretion luminosity provides slight modification, the basic picture of our 
results is still valid. Planetary migration, however, may be affected by the accretion luminosity since 
it is determined by the quantities in the vicinity of planets. We leave this study in the 
future publication. 

\section{Inducing dead zone turbulence by dust settling} \label{khi}

The presence of a dead zone is known to activate various instabilities in protoplanetary discs. In 
this section, we focus on the Kelvin-Helmholtz instability (KHI) that may arise.

The KHI is triggered by dust settling because the vertical shear is induced by the vertical 
difference of dust density \citep{s98,jhk06}. The increase of the density of dust in the mid-plane 
causes the gas to be in Keplerian motion since the collisional interactions with dust become more 
efficient while the gas above the mid-plane is still in sub-Keplerian due to lower density of dust. 
This vertical difference in the gas velocity induces a vertical shear flow, resulting in the KHI. 
Such regions can therefore become turbulent. 

This dust-induced turbulence is local and well-represented by the so-called Richardson number $Ri$,
\begin{equation}
Ri=\frac{g_z\partial \ln (\rho + \rho_d)/\partial z}{(\partial u_y/\partial z)^2},
\end{equation} 
where $g_z$ is the vertical gravitational force, $\rho$ is the gas density, $\rho_d$ is the dust 
density and $u_z$ is the vertical component of the gas velocity. The classical theory finds that 
$Ri <$ 1/4 is the unstable condition \citep{c61} although it is the necessary, but not sufficient 
condition. Recent numerical simulations have shown that the critical value is 1 \citep{jhk06}. 
Furthermore, the critical $Ri$ allows one to estimate the critical scale height of dust $h_{d,crit}$,
\begin{equation}
h_{d,crit}=\sqrt{Ri}\frac{| \zeta  |}{2\gamma_{ad}}h,
\label{h_d_crit}
\end{equation}
where $\zeta =(h/r)(\partial \ln \rho/\partial \ln r)$ and $\gamma_{ad}=5/3$ is the adiabatic index 
of gas \citep{s98}.

As discussed in $\S$ \ref{diskmodel}, dust settling is compensated by the turbulence of discs. In our 
disc models, the strength of turbulence is controlled by $\alpha$ which is a useful prescription for 
any global turbulence. Thus, we have treated the active and dead zones by adjusting $\alpha$. In the 
active well coupled zone, the origin of the turbulence is obvious, that is, the MRI. The origin of 
the turbulence in the dead zone, however, is ambiguous since the MRI cannot be active due to the low 
ionization. Thus, we examine whether or not our assumption for the dead zones is consistent with the KHI.

\begin{figure}
\begin{center}
\includegraphics[width=8.4cm]{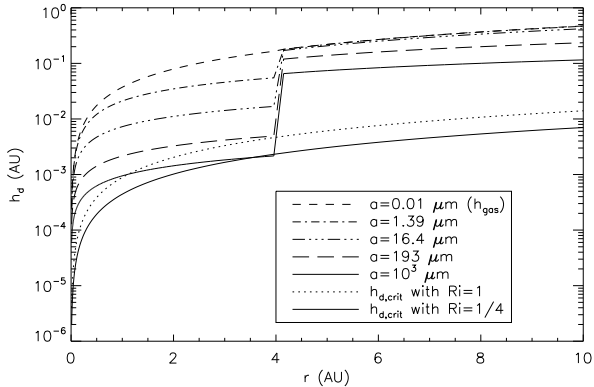}
\caption{The scale heights of dust as a function of distance from the central star with a 4 au sized 
dead zone. The scale heights of dust with various grain sizes in our disc models are shown. In addition, 
the critical scale heights with $Ri=1/4$ and $Ri=1$ are denoted by the solid and dotted lines, 
respectively. The KHI becomes active for dust grains with 1 mm and with 193 $\umu$m when $Ri=1/4$ and 
$Ri=1$, respectively.}
\label{fig17}
\end{center}
\end{figure}

Fig. \ref{fig17} shows the scale height of dust as a function of the distance from the star with a 4 au 
sized dead zone. Since the scale height of dust depends on the strength of turbulence $\alpha$ as well 
as its size $a$ in our disc models (see equation (\ref{h_d})), it drastically changes at the boundary 
of the active and dead zones. Also, the critical scale heights of dust with $Ri=1/4$ and $Ri=1$ are 
plotted. In the active zone, both of the critical scale heights of dust are much smaller than those of 
our scale heights of dust. This indicates that the MRI turbulence is so strong that the KHI cannot be 
active and is not therefore the primary driver of disc turbulence. 

The situation in the dead zone is significantly different. The critical scale height of dust with 
$Ri=1/4$ occurs for dust grains with 1mm size. For the case of $Ri=1$, the critical case is that for 
dust with 193 $\umu$m size. For these grains, the KHI becomes active. This implies that the origin of 
turbulence in the dead zones may be due to the KHI. Thus, it is worth estimating the value of $\alpha$ 
arising from the KHI following \citet{jhk06}. From equation (\ref{h_d}), the ratio of the scale height 
of dust to that of gas becomes
\begin{equation}
\frac{h_d}{h}={\bar H}
\end{equation}  
since $\bar H \ll 1$. In addition, the scale height of dust estimated from the KHI is given by 
equation (\ref{h_d_crit}). Equating these two equations, the $\alpha_{KHI}$ arising from the KHI is
\begin{equation}
\alpha_{KHI}=\sqrt{2\pi ( 1+\gamma_{turb} )}\left( \frac{\zeta }{2\gamma_{ad}} \right)^2 \frac{\rho_s a}{\Sigma}Ri.
\end{equation}

\begin{figure}
\begin{center}
\includegraphics[width=8.4cm]{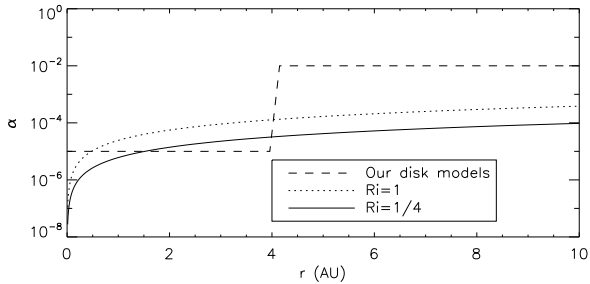}
\caption{The values of $\alpha$ as a function of disc radius. The $\alpha_{KHI}$ estimated from 
the KHI with $Ri=1/4$ and $Ri=1$ are denoted by the solid and dotted lines, respectively. The value of 
$\alpha$ in our disc models is denoted by the dashed line. The value of $\alpha$ within the dead 
zone lies the range between $10^{-4}$ and $10^{-6}$ for both values of $Ri$. This is consistent with 
the literature.}
\label{fig18}
\end{center}
\end{figure}

Fig. \ref{fig18} shows the value of $\alpha_{KHI}$ as a function of disc radius. As discussed above, 
the $\alpha_{KHI}$ in the active zone is about two orders of magnitude smaller than that of the MRI. 
Within the dead zone, the $\alpha_{KHI}$ lies in the range between $10^{-6}$ and $10^{-4}$. This supports 
the value of $\alpha$ in the dead zone used in the literature \citep[e.g.,][]{is05,mp06}. The origin of 
turbulence in the dead zone is probably governed by the KHI, although the derived $\alpha$ is a kind of 
the maximum value because the recent study including the radial Keplerian shear has shown that the KHI can 
be active only when a mode of the instability grows faster than it is sheared out \citep{is03}. 
Furthermore, \citet{is03} have shown that although the equilibrium state of the dust distribution can 
be established, the critical conditions are different from the Richardson number because it does not 
include the effect of the Keplerian shear.  Further study such as global simulations is needed.      

\section{Conclusions} \label{conc}

In this paper, we have investigated the detailed thermal structure of protoplanetary discs around 
M stars by solving the radiative transfer equation in 2D discs by means of the Monte Carlo method. 
In our simulations, stellar irradiation is absorbed by dust in discs and is their main heat source. 
Thus, the properties of dust such as its composition, size and density distributions are crucial in 
this study. For the composition, we have adopted the models that are derived from the solar abundance. 
For the size distribution, we have adopted the power-law distribution in which the size of dust is 
from 0.1 $\umu$m to 1 mm. Since we have used elaborated models for the dust distribution, we had to 
sample the size of dust within this range. Our convergence study has shown that 15 sizes of dust chosen 
from this range are sufficient to describe the temperature of discs to an accuracy of better 
than 10 per cent. Our results of the dust temperatures are more accurate than those derived from the 
mean opacity for the ensemble of larger dust grains (e.g., S07). Furthermore, we have included dust 
settling, the gravitational force of the planet and the presence of dead zone.

The most important findings of this study are the followings;

\begin{enumerate}
\item The temperature of the mid-plane is well-represented by $r^{-3/4}$ when dust settling is 
included. This power-law was derived by CG97 by assuming discs to be flat. Well mixed dust models 
result in an $r^{-3/5}$ temperature distribution. We have confirmed that dust settling makes discs 
geometrically flatter.

\item The presence of the dead zone in discs implies the presence of a dust wall at the boundary of 
the active and dead zones. This wall will be directly exposed  to stellar irradiation. Consequently, 
a positive temperature gradient is formed in the temperature profile of the mid-plane. This is explained 
by the higher energy at the wall that is propagated by radiative diffusion since the mid-plane is 
optically thick. We have found that even if the planets are present in front of the wall, this 
thermally hot dusty wall and the positive temperature gradient are maintained.

\item The dead zones excite the Kelvin-Helmholtz instability because of enhanced dust settling 
within them. Consequently, we have demonstrated that a self-consistent value of $\alpha \simeq 10^{-5}$ 
in dead zones is expected. This result backs up the assumptions of many previous models for dead zones.

\end{enumerate} 

Furthermore, we have confirmed earlier results in the literature that the surface layer has a higher 
temperature than the mid-plane region in any disc model. The surface layer is directly heated by the 
central star while the mid-plane region is only heated by the thermal emission of dust. Our simulations 
with dust settling have shown that this process leaves the surface layer hotter and the mid-plane 
region cooler. This arises from the deficit of larger dust grains in the surface layer and the 
increment of them in the mid-plane region. This can be also understood by the mean free paths of 
photons that are determined by the density of dust. Therefore, dust settling makes the mean free paths 
in the surface layer longer and those in the mid-plane region shorter. Also, we have shown that the 
inclusion of the planets triggers the compression of the material above them, resulting in longer mean 
free paths of photons. The region above the planets has a higher temperature. 

We have also performed a parameter study for our disc models by varying the total disc masses and 
the surface density distributions. It is generally thought that the disc masses are about ten times 
more massive than the observed ones in order for gas giants to be formed. Also the observations have 
indicated that the surface density has some range from $r^{-1}$ to $r^{-3/2}$. We have confirmed that 
our findings are basically unaffected for such parameter spaces, although the lower disc masses result in 
a higher temperature at the mid-plane, causing the positive temperature gradient to be somewhat 
diminished. In addition, the formation of the positive temperature gradient is independent of the 
size of dead zones since the dusty wall is present for any size of the dead zone. Furthermore, 
the positive temperature gradient is sustained for dead zones with a finite transition for $\alpha$.

We have also examined other heat sources for discs such as viscous heating and the accretion 
luminosity of the planets. We have estimated viscous heating is only important within about 0.1 au 
for M stars although it is known to be dominant within about 1 au for the classical T Tauri stars. Thus, 
we have investigated the detailed effects of dust settling, dead zones, and planets on the 
thermal structure of irradiated discs. 

In subsequent papers, we will discuss the effect of our 
results on planetary migration because Lindblad torques, the main driving force of type I migration, 
are affected by the gas pressure. In addition, we will address the effects of the dead zones on the 
spectral energy distributions (SEDs) and images.  It is well known that the presence of embedded 
planets in discs provide some interesting observational signatures on the SEDs and/or images by forming 
a gap for massive ones \citep{vbfq06} and by distorting the density distribution for low mass 
ones \citep{jc09}. We anticipate that the presence of the dead zones will also have an interesting 
contribution to the SEDs arising from the dusty wall and resultant positive temperature gradient.

\section*{Acknowledgments}
The authors are indebted to Kees Dullemond for the use of his excellent radiative transfer code and 
for several very useful discussions during a sabbatical stay at the Institute for Theoretical 
Astrophysics in Heidelberg.  We also thank Thomas Henning, Hubert Klahr, Kristen Menou, Soko Matsumura 
and Richard Nelson for stimulating discussions and an anonymous referee for useful comments on our 
manuscript. Our simulations were carried out on computer 
clusters of the SHARCNET HPC Consortium at McMaster University. YH is supported by SHARCNET Graduate 
Fellowship as well as McMaster University and REP by a Discovery Grant from the Natural Sciences 
and Engineering Research Council (NSERC) of Canada.

\bibliographystyle{mn2e}

\bibliography{mn-jour,adsbibliography}

\appendix

\section{The trajectory of photons projected into 2D planes in cylindrical coordinate systems}

Here, we obtain solutions of the trajectory of photons projected into 2D planes $(r,z)$ in cylindrical 
coordinate systems. First of all, we setup a global cylindrical $(r,z,\Phi)$ and local spherical 
$(\theta,\phi)$ coordinate system by following \citet{dt00}. We assume discs to be axisymmetrical, 
so that any dependence on $\Phi$ is dropped out. The north pole is aligned with the z-axis of the 
global system. Consequently, $\theta$ is measured from this axis toward the mid-plane and follows 
photons as they move upward or downward compared with the mid-plane. $\phi$ is measured from the local 
x-axis which is parallel to the global x-axis. The $\phi$ coordinate tracks photons as they move 
forward or backward compared with the local coordinate system.

The radiative transfer equation in the cylindrical system is 
\begin{equation}
\frac{dI_{\nu}}{ds}=\sqrt{1-\mu^2}\sin \phi \frac{\partial I_{\nu}}{\partial r} + \mu \frac{\partial I_{\mu}}{\partial z} + \frac{\sqrt{1-\mu^2}\cos \phi}{r}\frac{\partial I_{\nu}}{\partial \phi},
\end{equation}
where $I_{\nu}$ is the specific intensity at frequency $\nu$, $ds$ is the displacement of the 
path of a photon, and $\mu=\cos \theta$ \citep{lzt06}. Thus, the variations of $r$, $z$, $\mu$, and 
$\phi$ along the path of the photon are
\begin{subequations}
\begin{eqnarray}
\frac{dr}{ds}    & = &   \sqrt{1-\mu^2}\sin\phi,            \\
\frac{dz}{ds}    & = &   \mu,                               \\
\frac{d\mu}{ds}  & = &   0,                                 \\
\frac{d\phi}{ds} & = &   \frac{\sqrt{1-\mu^2}\cos \phi}{r}.
\end{eqnarray}
\end{subequations}
A set of solutions we found are
\begin{subequations}
\begin{eqnarray}
r^2       & = &    (1-\mu^2)s^2+b^2, \label{sol_r_c}       \\
z         & = &    \mu_0 s + z_0,  \label{sol_z_c}         \\
\mu_0     & = &    \mu,                                    \\
\sin \phi & = &    \frac{\sqrt{1-\mu^2}}{r}s,
\end{eqnarray}
\end{subequations}
where $b$ is the impact parameter of the ray with respect to the origin, $z_0$ is the height from 
the mid-plane of the closest point to the north pole of the global system.

We use the solutions to follow the path of photons. When a photon is located at $P=(r_k,z_l)$ and its 
new direction is determined as $(\mu,\phi)=(\mu_i,\phi_j)$ by a random number, the constants are 
\begin{subequations}
\begin{eqnarray}
b^2  & = &   r_k^2\cos^2 \phi_j, \\
z_0  & = &   z_l- \frac{\mu_0}{\sqrt{1-\mu^2_0}}r_k\sin \phi_j,
\end{eqnarray}
\end{subequations}
where $k$, $l$, $i$, and $j$ represent the grid number at which the photon is placed since although the 
Monte Carlo method is a non grid-based code, all physical quantities such as the density and temperature 
are defined on the grid. Thus, the next point it will reach is determined by the intersection between 
grid points and the solutions. 

Equations (\ref{sol_r_c}) and (\ref{sol_z_c}) give us two and one values for $s$, respectively. Since 
possible grid points the photon can take are $r_{k-1}$, $r_{k}$, and $r_{k+1}$ ($r_{k-1}<r_{k}<r_{k+1}$), 
and $z_{l-1}$, $z_{l}$, and $z_{l+1}$ ($z_{l-1}<z_{l}<z_{l+1}$ ). Thus, the total number of possible 
values of $s$ is 9 and are given by 
\begin{subequations}
\begin{eqnarray}
s_n  & = &  \pm \sqrt{\frac{r_K^2-b^2}{1-\mu^2}}, \label{s_n_c} \\
s_m  & = &  \frac{z_L-z_0}{\mu_0}, \label{s_m_c}
\end{eqnarray}
\end{subequations}
where $n=1,2,...,6$, $K=\{k-1,k,k+1 \}$, $m=7,8,9$, and $L=\{l-1,l,l+1 \}$. The actual path is determined 
so that the absolute value of $s$ is a minimum and not the same as $s_P$ which is the path evaluated at 
$P=(r_k,z_l)$.

Thus, the motion of photons projected into 2D plane in cylindrical coordinate systems is determined by 
equations (\ref{s_n_c}) and (\ref{s_m_c}).

\section{Temperature and dust density distribution around a planet}

Here, we carefully analyse and discuss the thermal and dust density structures of discs with a planet 
for the well mixed and dust settling cases.   

\subsection{The well-mixed case}

\begin{figure}
\begin{center}
\includegraphics[width=8.4cm]{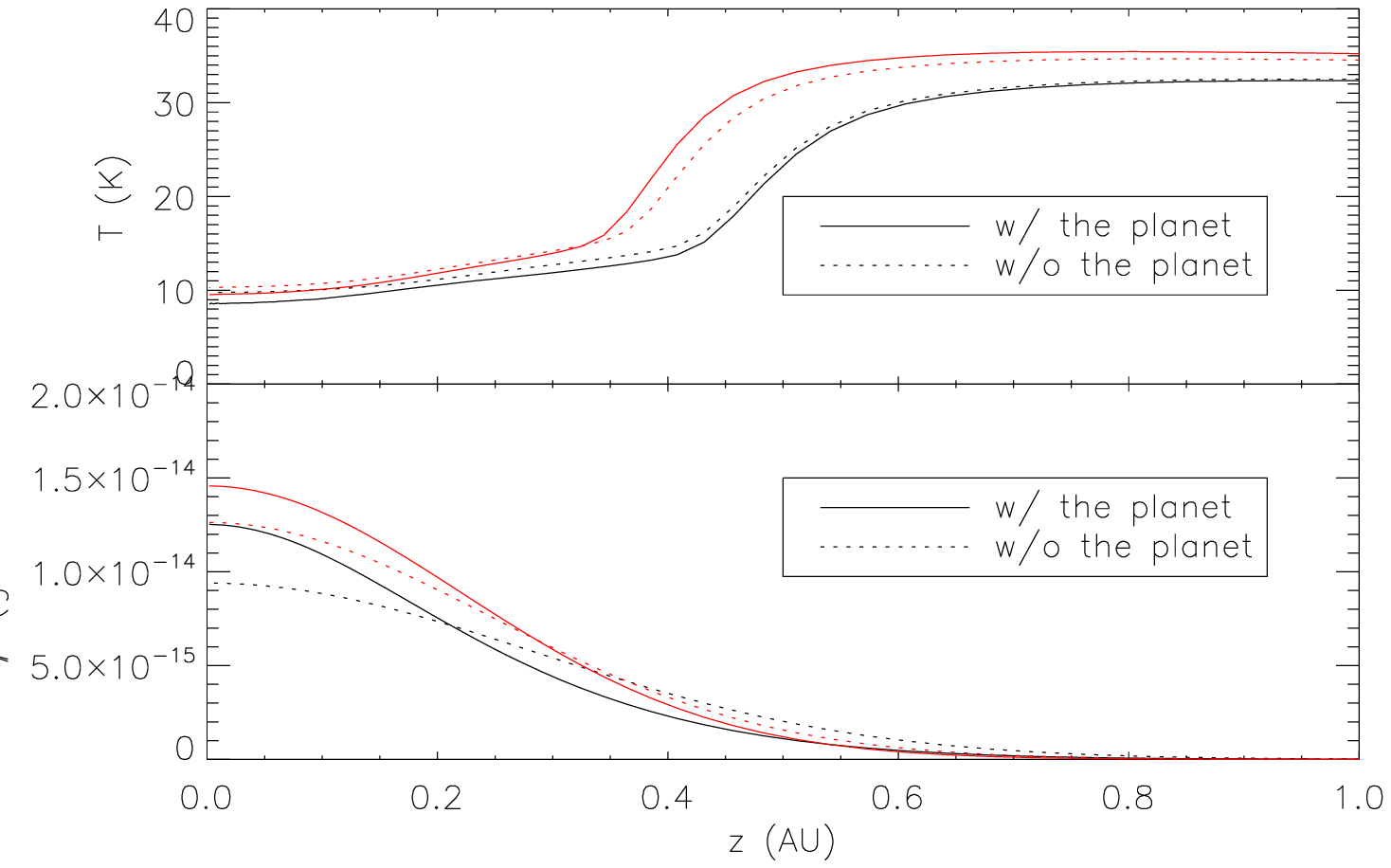}
\caption{The temperature and dust density structure in a disc with well mixed dust and a 10 
$M_{\oplus }$ planet as a function of distance from the mid-plane. Top: the temperature behaviours. 
Bottom: the dust density profiles. In both panels, solid lines denote the case with the planet, and 
dotted lines denote the case without the planet for comparison. The red lines denote quantities at 
$r_p-r_H$, and the black lines denote quantities at $r_p+r_H$. The perturbation due to the planet 
accumulates the material near the mid-plane region, and produces the self-shadowing region in the 
vicinity of the planet.}
\label{figB1}
\end{center}
%\end{minipage}
\end{figure}

Fig. \ref{figB1} shows the vertical temperature and density of dust, in a model with well mixed dust, in 
the vicinity of the 10 $M_{\oplus }$ planet (Solid lines). For comparison, the unperturbed case is 
denoted by the dotted lines. Quantities at $r_p-r_H$ are denoted in red while those at $r_p+r_H$ are in 
black. The presence of a planet increases the density at the mid-plane region due to the gravitational 
force of the planet. This results in the compression of dust (and gas) at $r_p \pm r_H$ (compare solid 
lines with dotted lines). For the temperature structure, the inclusion of the planets causes the surface 
layer to be hotter at $r_p-r_H$ than the case without the planet (see two red lines). This is explained 
by similar effects of dust settling. The compression of material reduces the optical depth, resulting in 
a higher temperature. On the other hand, the surface layer at $r_p+r_H$ becomes slightly cooler by the 
planets (see two black lines). In other words, there is no self-illumination region in our simulations. 
For the mid-plane region, the temperature becomes cooler due to the gravitational force of the planets 
at $r_p \pm r_H$ (compare solid lines with dotted lines). This is a self-shadowing effect, as found by 
JS03, JS04, J08.

Thus, we could not find the self-illumination regions clearly, differing from the findings of JS03, JS04 
and J08. We did find the self-shadowing regions, however, although its effects are relatively small. 
This is because of the increment of disc masses and the resultant relatively less flaring disc 
structure. JS04 found that the temperature variation due to the planets becomes smaller when the slope 
of the surface density is steeper because higher density prevents photons from penetrating it deeply. 
Also, J08 found that the self-shadowing and illumination regions are very sensitive to the incident 
angle of stellar radiation. The increase of mass allowing massive planets to be formed results in a 
relatively flatter disc structure (see Fig. \ref{fig5}). This prevents the regions perturbed by the 
planet from being exposed to stellar radiation. Thus, the results of our simulations are not precisely 
the same, although consistent with those of JS03, JS04 and J08 due to the increase of the disc mass.

\subsection{The dust settling case}

\begin{figure}
\begin{center}
\includegraphics[width=8.4cm]{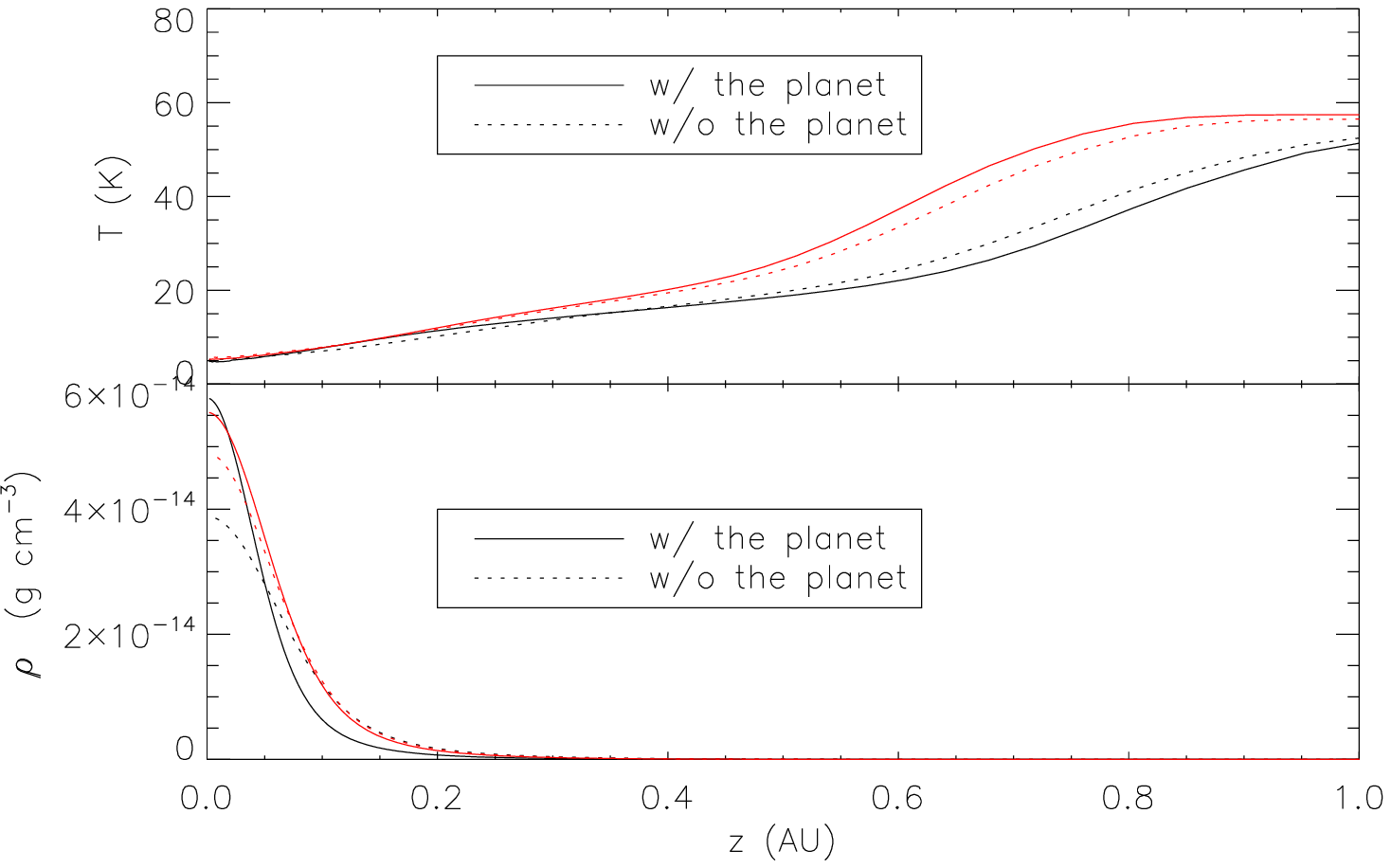}
\caption{The temperature and density structure in a disc with dust settling and a 10 $M_{\oplus }$ 
planet, as a function of distance from the mid-plane (as Fig. \ref{figB1}). Higher dust density in 
the mid-plane region is due to the combined effect of the planet and dust settling. This results in 
the much smaller temperature difference between $r_p\pm r_H$.}
\label{figB2}
\end{center}
\end{figure}

Fig. \ref{figB2} shows the vertical temperature and density of dust, in a model with dust settling,  
in the vicinity of the planet with 10 $M_{\oplus }$. The main trend is the same as the case including 
the planets only. Higher density in the mid-plane regions results in much smaller temperature difference 
between $r_p\pm r_H$. For the surface region, the temperature at $r_p-r_H$ becomes higher while the one 
at $r_p+r_H$ is lower when the planets are included.  

\bsp

\label{lastpage}

\end{document}